\documentclass[useAMS,usenatbib]{mn2e}

\usepackage[dvips]{graphicx}
\usepackage{subfigure}
\usepackage{float}
\usepackage{amsmath}
\usepackage{color}

\newcommand{\apj}{ApJ}
\newcommand{\apjl}{ApJL}
\newcommand{\pasj}{PASJ}

\newcommand{\aj}{AJ} 
\newcommand{\mnras}{MNRAS}

\title[Do GMCs Care About the Galactic Structure?]{Do Giant Molecular Clouds Care About the Galactic Structure?}
\author[Fujimoto et al.]{Yusuke Fujimoto\thanks{E-mail:
yusuke@astro1.sci.hokudai.ac.jp}, Elizabeth J. Tasker, Mariko Wakayama and Asao Habe\\
Department of Physics, Faculty of Science, Hokkaido University, Kita 10 Nishi 8 Kita-ku, Sapporo 060-0810, Japan}
\begin{document}

\date{}
\maketitle

\pagerange{\pageref{firstpage}--\pageref{lastpage}} \pubyear{2013}

\label{firstpage}

\begin{abstract}
We investigate the impact of galactic environment on the properties of simulated giant molecular clouds (GMCs) formed in a M83-type barred spiral galaxy. Our simulation uses a rotating stellar potential to create the grand design features and resolves down to 1.5\,pc. From the comparison of clouds found in the bar, spiral and disc regions, we find that the typical GMC is environment independent, with a mass of $5\times 10^5$\,M$_\odot$ and radius 11\,pc. However, the fraction of clouds in the property distribution tails varies between regions, with larger, more massive clouds with a higher velocity dispersion being found in greatest proportions in the bar, spiral and then disc. The bar clouds also show a bimodality that is not reflected in the spiral and disc clouds except in the surface density, where all three regions show  two distinct peaks. We identify these features as being due to the relative proportion of three cloud types, classified via the mass-radius scaling relation, which we label {\it A, B} and {\it C}. {\it Type A} clouds have the typical values listed above and form the largest fraction in each region. {\it Type B} clouds are massive giant molecular associations (GMAs) while {\it Type C} clouds are unbound, transient clouds that form in dense filaments and tidal tails.  The fraction of each clouds type depends on the cloud-cloud interactions, which cause mergers to build up the GMA {\it Type B}s and tidal features in which the {\it Type C} clouds are formed. The number of cloud interactions is greatest in the bar, followed by the spiral, causing a higher fraction of both cloud types compared to the disc. While the cloud types also exist in lower resolution simulations, their identification becomes more challenging as they are not well separated populations on the mass-radius relation or distribution plots. Finally, we compare the results for three star formation models to estimate the star formation rate and efficiency in each galactic region.
\end{abstract}


\section{Introduction}
\label{sec:intro}

At first glance, star formation appears to be a localised process. The coldest gas in the galactic interstellar medium (ISM) clumps into turbulent aggregations known as the giant molecular clouds (GMCs). These stellar nurseries are on average, three orders of magnitude smaller than the galactic radii and the pockets within them that collapse to form stars are another order of magnitude smaller still. Additionally, GMC properties have been observed to be remarkably similar across different galaxies, which might suggest a disregard for the structure of the galactic host \citep{Blitz2007, Bolatto2008, Heyer2009, DonovanMeyer2013}.

Yet, there is still more evidence that star formation is far from being unaware of its global environment. Observations indicate an empirical relation that relates the galaxy's gas surface density ($\Sigma_{\rm gas}$) and its star formation rate surface density ($\Sigma_{\rm SFR}$) by a simple power law \citep{Schmidt1959, Kennicutt1989, Kennicutt1998, WongBlitz2002, Kennicutt2007, Bigiel2008, Leroy2008, Schruba2011}:

\begin{equation}
\Sigma_{\rm SFR} \propto \Sigma_{\rm gas}^N
\label{eq:ks}
\end{equation}

\noindent where the measurements of the power index, $N$, vary between 1 and 2.  Generally referred to as the Kennicutt-Schmidt Relation, this link between the gas distribution in a galaxy and its star formation holds on both local and global scales. Moreover, recent work has shown that systematic variations exist in the relation and that the star formation activity may also be sensitive to global structural variations such as galaxy type \citep{Daddi2010, Leroy2013}, conditions in the galactic central region \citep{Oka2001} and the grand design \citep{Sheth2002, Momose2010}.
Notably, this is distinct from material simply being gathered to produce a higher star formation rate, since then both the gas density and star formation rate density would rise in unison and not produce a variation in the Kennicutt-Schmidt relation. However, observations of the barred galaxy NGC 4303 by \citet{Momose2010} revealed a different star formation efficiency in the spiral arms and bar region of the disc, even in locations where the gas surface density is comparable.

If star formation truly does care about its large-scale environment then this should be reflected in the properties of the GMCs, the nurseries whose conditions determine whether a star can form. Within the Milky Way, the properties of GMCs have been measured to high precision \citep{Larson1981, Solomon1987, Heyer2009, Roman-Duval2010}, yet it is difficult from within our own disc to assess the impact of global structure. A more likely source of information comes from the growing catalogue of nearby galaxy GMC properties, many of which elude to environmentally driven differences between the populations in different galaxies \citep{Hughes2013}.  However, extragalactic data is limited by resolution, making it hard to assemble large enough samples of GMC properties to explore the dependence on internal galaxy structure (a deficit that ALMA will tackle). The results we do have from such surveys strongly indicate that structure plays a key role in star formation. In the spiral arms of M51, observations by \citet{Koda2009} find evidence for giant molecular associations (GMAs) which later fragment into smaller GMCs in the interarm region. These GMAs have a significantly higher surface density, while having the same estimated volume density, as a typical galactic GMC, suggesting they are a distinct group of objects and not simply an agglomeration of overlapping clouds. In the same galaxy,  \citet{Meidt2013} observed that sheering flows and shocks driven by spiral structures can stabilise GMCs that would otherwise collapse to form stars, changing the dependence between the star formation rate and gas surface density in such environments. It is clear, therefore, that understanding environment effects on GMC properties is key to understanding star formation itself. 

On the theoretical side, a dependence on GMC properties with galactic structure has been found by \citet{Dobbs2006b} who saw a similar result to \citet{Koda2009}, with clouds leaving the spiral arms to be sheared and form interarm feathering. More compact interarm spurs were found by \citet{Renaud2013}, due to Kelvin-Helmholtz instabilities forming down the side of the spiral arms. \citet{Renaud2013} also found that the elongated gas structures in their spiral arms dictated the spacing of their GMCs, with fragmentation occurring at regularly spaced intervals. In the transient spiral galaxy models of \citet{Wada2011}, gas arms can gather material to form a GMA, but then can themselves disperse the arm and their own structure. The two-dimensional simulations of M83 by \citet{Nimori2012} found that GMCs forming in the bar region tended to be less bound than those in the spiral arms. Their findings were consistent with observations that the velocity dispersion of molecular gas in the bar region both in M83 and in Maffei 2 is high  \citep{Lundgren2004b, Sorai2012}, raising the value of the virial parameter. 

One possible reason for \citet{Nimori2012}'s findings is the increased likelihood of collisions between GMCs in regions of high gas density. Such interactions can either build the cloud via mergers, deepening its potential well and central density to boost star formation or by triggering a shock at the collisional interface to produce stars. The latter mechanism has been previously suggested as a way to unite the local star formation process with the globally observed Kennicutt-Schmidt relation \citep{Tan2000, TaskerTan2009} and also as a way to create massive stars \citep{Furukawa2009, Ohama2010, HabeOhta1992}. Therefore, the structure of the disc has both the potential to change the properties of the GMCs and increase their interactions to trigger star formation.  

In this paper, we will focus on the effect of a grand design bar and spiral on the formation and evolution of the GMCs. In section~\S 2, we present our model of the barred galaxy, M83 and discuss the details of the three-dimensional hydrodynamical simulation. Section~\S 3 details our results, discussing first the global evolution of the disc and moving on to exploring the cloud properties and estimated star formation rates. In section~\S 4, we consider the effect of resolution and the way in which GMCs are identified. Section~\S 5 presents our conclusions.

\section{Numerical methods}
\label{sec:numerics}

\subsection{The code}
\label{sec:numerics_code}

The simulations presented in this paper were run using {\it Enzo}: a three-dimensional adaptive mesh refinement (AMR) hydrodynamics code \citep{Enzo, Bryan1999, BryanNorman1997}. {\it Enzo} has previously been used to model galactic discs where it successfully produced a self-consistent atomic multiphase ISM, consisting of a wide range of densities and temperatures \citep{TaskerBryan2006, TaskerBryan2008, TaskerTan2009, Tasker2011, Benincasa2013}.

We used a three-dimensional box of side 50\,kpc with a root grid of $128^3$ cells and 8 levels of refinement, giving a limiting resolution (smallest cell size) of about 1.5\,pc. Cells were refined whenever the mass included in the cell exceeded $\rm 1000 M_{\odot}$, or whenever the Jeans' Length covered less than four cells. This latter condition is suggested by \citet{Truelove1997} as the minimum resolution required to avoid artificial fragmentation. 

The evolution of the gas in {\it Enzo} was performed using a three-dimensional version of the {\it Zeus} hydrodynamics algorithm \citep{StoneNorman1992}. {\it Zeus} uses an artificial viscosity as a shock-capturing technique with the variable associated with this, the quadratic artificial viscosity, set to the default value of 2.0. 

The gas was self-gravitating and allowed to cool radiatively down to 300\,K. The radiative cooling used rates from the analytical expression of \citet{SarazinWhite1987} for solar metallicity down to $10^4$\,K, and continued to 300\,K with rates provided by \citet{RosenBregman1995}. This allowed the gas to cool to temperatures at the upper end of the atomic cold neutral medium \citep{Wolfire2003}. Actual GMCs have temperatures of about 10\,K, an order of magnitude below our minimum radiative cooling temperature. However, we lacked the resolution to sufficiently resolve the full turbulent structure of our smaller clouds, nor did we include pressure from magnetic fields. This temperature floor therefore imposed a minimum sound speed of $1.8$\,km/s to crudely allow for these effects. In fact, the velocity dispersion within our clouds was typically higher than this by about a factor of three, implying that the floor was not having a significant impact on our cloud properties.

To prevent unresolved collapse at the finest resolution level, a pressure floor was implemented that injected energy to halt the collapse once the Jeans length became smaller than four cells. Gas in this regime followed a $\gamma = 2$ polytrope, $P \propto \rho^{\gamma}$. In order to study the evolution of the gas clouds alone, there was no star formation or stellar feedback in this simulation. 

\subsection{The structure of the galactic disc}
\label{sec:numerics_disc}

\begin{figure}
\begin{center}
	\subfigure{
	\includegraphics[width=8.0cm,bb=0 0 1024 768]{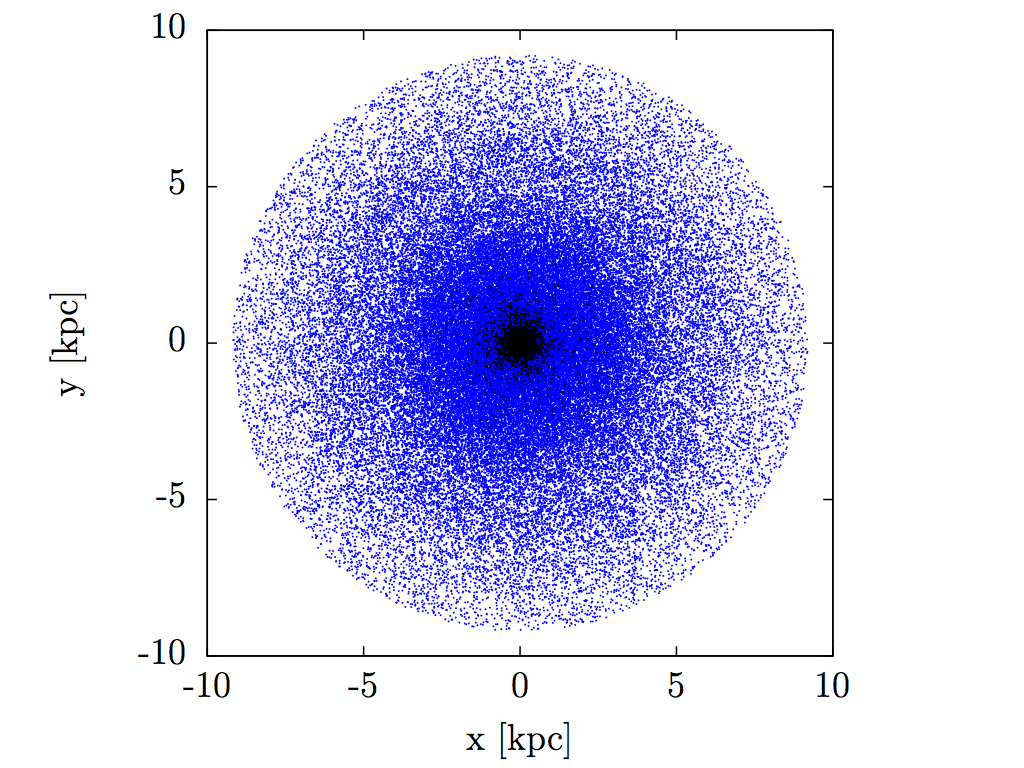}}
	\subfigure{
	\includegraphics[width=8.0cm,bb=0 0 1024 768]{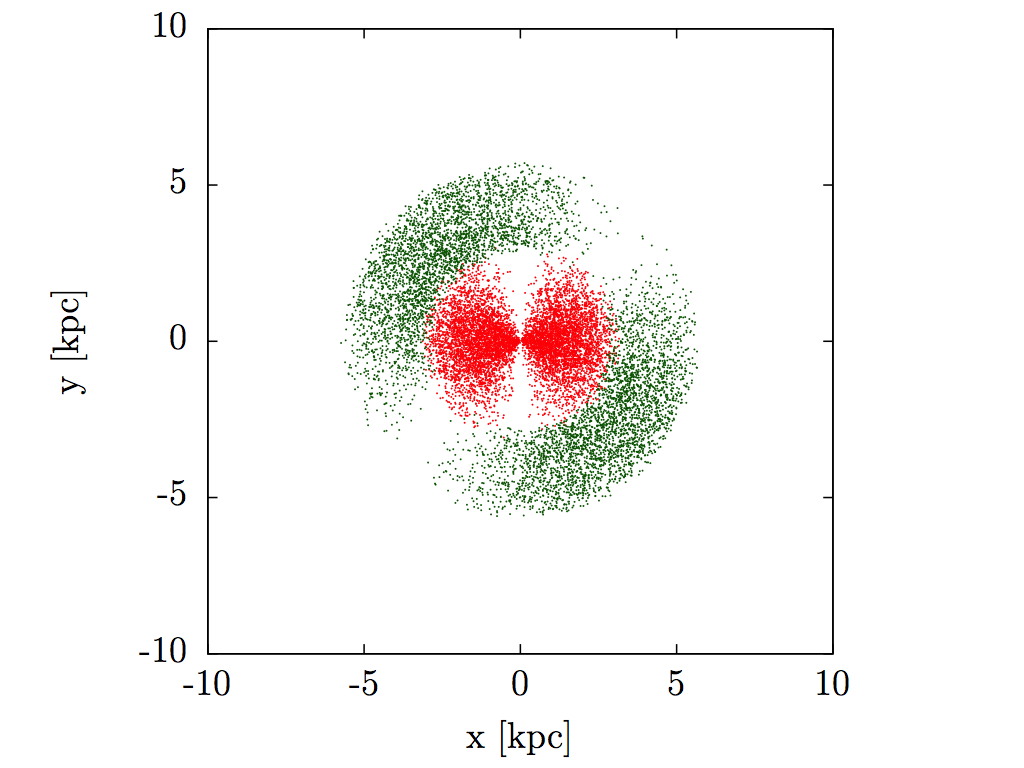}}
	\caption{The stellar component of the galactic potential. Top panel shows the axisymmetric star particle distribution, where blue dots denote disc particles and black form the bulge. Bottom panel shows the non-axisymmetric star particle distribution (red dots are the bar particles and green are the spiral arms) which are rotated at a constant pattern speed. }
	\label{star particles}
\end{center}
\end{figure}

Our galaxy was modelled on the barred spiral galaxy, M83, with the gas distribution and stellar potential taken from observational results (see below). At 4.5\,Mpc away, M83 is a nearby type SABc galaxy and has been observed at various wavelengths to measure its atomic \citep{HuchtmeierBohnenstengel1981} and molecular gas \citep{Lundgren2004a, Lundgren2004b, Sakamoto2004, Muraoka2007} as well as optical emission lines \citep{Dopita2010} and X-ray \citep{SoriaWu2003}. Its GMC properties are also being observed by ALMA in Cycle 0 and Cycle 1 (Hirota et al., in prep.).

\subsubsection{Initial gas distribution}

For the galaxy's radial gas distribution, we assumed an initial exponential density profile with a radial scale length of 2265 pc, based on the observations of \citet{Lundgren2004a}. The initial vertical distribution was assumed to be proportional to $\rm sech^2 (z/z_{\rm h})$ with a vertical scale height of $z_{\rm h} = 100$\,pc. The total gas mass in the simulation was taken again from the observations of \citet{Lundgren2004a}, where the $\rm H_2$ gas mass was recorded as $3.9\times10^9M_{\odot}$. This gave an initial gas distribution: 
 
\begin{equation}
\rho_{\rm gas}(r, z) = 0.67\ \exp\left(-\frac{r}{2265{\rm pc}}\right){\rm sech}^2\left(\frac{z}{100{\rm pc}}\right) M_{\odot}/{\rm pc}^3
\end{equation}

\noindent The gas was set in circular motion as calculated via $V_{cir}(r) = (GM_{tot}/r)^{1/2}$, where $M_{tot}$ is the enclosed mass of stars, dark matter and gas within the radius $r$.

\subsubsection{Stellar potential}

We used $10^5$ fixed-motion star particles to create a stellar potential model in keeping with the observed global characteristics of the stellar distribution in M83. This model was from the work of \citet{Hirota2009}, who analysed the 2Mass K-band image of M83 \citep{Jarrett2003}. The stellar density, consisting of the disc, bulge, bar and spiral arms was given by:

\begin{eqnarray}
\rho_{\rm star}(r, \theta, z) &=& \Sigma(r, \theta)h(z)\nonumber\\
&=& \{\Sigma_{\rm disc}(r) + \Sigma_{\rm bulge}(r) +\nonumber\\
&& \Sigma_{\rm bar}(r) \cos(2\theta) + \Sigma_{\rm spiral}(r)\sin(2\theta)\}h(z),\nonumber\\
\end{eqnarray}

\noindent where $\Sigma_{i}(r)$ is the radial distribution of each component and $h(z)$ is the vertical distribution. Each of these were given by:

\begin{eqnarray}
&&\Sigma_{\rm disc}(r) = 20\exp(-r/2230{\rm pc}), \label{eq:ic_disc}\\
&&\hspace{120pt} (0\ {\rm pc}\lid r \lid 9200\ {\rm pc})\nonumber\\[12pt]
&&\Sigma_{\rm bulge}(r) = \dfrac{1000}{\{1+(r/140{\rm pc})^2\}^{1.5}},\\
&&\hspace{120pt} (0\ {\rm pc}\lid r \lid 9200\ {\rm pc})\nonumber\\[12pt]
&&\Sigma_{\rm bar}(r) = \dfrac{3}{100}(1150{\rm pc}-r) + 3.0,\\
&&\hspace{120pt} (0\ {\rm pc}\lid r \lid 3220\ {\rm pc})\nonumber\\[12pt]
&&\Sigma_{\rm spiral}(r) = -\dfrac{0.7}{900}(r-3450{\rm pc})^2 + 0.7\nonumber \\
&&\hspace{50pt} - \dfrac{0.6}{2500}(r - 3450\ {\rm pc})(r - 5750\ {\rm pc}).\label{eq:ic_spiral}\\
&&\hspace{120pt} (2760\ {\rm pc} \lid r \lid 5750\ {\rm pc})\nonumber
\end{eqnarray}

\noindent In this model, \citet{Hirota2009} assumed that the mass to light ratio was constant between bulge and outer disc region. However, M83 has a starburst at the nucleus and so the actual total stellar mass of the nucleus has to be less than what was assumed. Owing to this, we decreased the influence of the bulge potential by 35\% by distributing the star particles between Equations~\ref{eq:ic_disc} - \ref{eq:ic_spiral} in a {\tt 2:0.7:2:2} ratio. 

We assumed the vertical stellar distribution $h(z)$ was:

\begin{eqnarray}
h(z) &=& \frac{1}{890}\ {\rm sech}^2\left(\frac{z}{450{\rm pc}}\right) \hspace{10pt} (-4600\ {\rm pc} \leq z \leq 4600\ {\rm pc})\nonumber\\
&=&\frac{1}{890} \left\{\frac{2}{\exp(z/450{\rm pc})+\exp(-z/450{\rm pc})}\right\}^2.
\end{eqnarray}

\noindent Figure~\ref{star particles} shows the resulting distribution of the star particles. The top panel shows the axisymmetric star particles distribution where blue dots mark disc particles and black show the bulge. The bottom panel shows the non-axisymmetric star particle distribution (where red dots are the bar particles and green dots denote spiral arms) which is rotated at 54 km/s/kpc, the estimated pattern speed for M83 \citep{Hirota2009}. 

Each star particle has a mass of $5.0 \times 10^{5} M_{\odot}$, giving a total stellar mass of $M_* = 5.0 \times 10^{10} M_{\odot}$, in agreement with observational results (see section~\ref{sec:numerics_masses}).

To remove the discreteness effects of the star particles, we smoothed the particles' gravitational contribution by adding the mass onto the grid at AMR level 4, with a cell size of 50\,pc.

\subsubsection{Dark matter potential}

In additional to the stellar potential, the galaxy sits in a static dark matter halo with an NFW profile \citep{Navarro1997}. The halo concentration parameter was set to $c = 10$, while the virial mass (within which the density is 200 times the cosmological critical value), $M_{200}=1.0 \times 10^{10} M_{\odot}$, a value obtained by comparison with the observational results (see section~\ref{sec:numerics_masses}).

\subsubsection{Stellar and dark matter mass ratio}
\label{sec:numerics_masses}

The stellar and dark matter masses of $M_* = 5.0 \times 10^{10}$\,M$_{\odot}$ and $M_{200} = 1.0 \times 10^{10}$\, M$_{\odot}$, were selected via comparisons between our model's rotation curve and that from the observational results from M83 \citep{Lundgren2004b}. We also compared and matched the size of the bar obtained from the M83 $\rm ^{12} CO$(J=1-0) observations \citep{Lundgren2004a} with the size of the bar structure formed in the simulation at 240\,Myr. In both cases, the position of the bar-end from the galactic centre was approximately 2.3\,kpc. Within the radius of the galaxy disc, the stellar mass dominated over the dark matter to ensure a grand design spiral.

\subsection{Cloud definition and tracking}
\label{sec:numerics_cloud}

The giant molecular clouds in our simulation were identified as coherent structures contained within contours at the threshold density of $n_{\rm H,c} = 100$\,cm$^{-3}$, similar to the observed mean volume densities of typical galactic GMCs. Note that since the formation of molecules was not being followed in our simulation, the gas is purely atomic. However, at the threshold density, it is reasonable to assume the cloud would consist of both a molecular core and atomic envelope.  We refer to this method as the {\it `contour method'} for cloud identification. 

We also used an additional method for defining GMCs that builds clouds around density local maxima\citep{TaskerTan2009}. Here, peaks are found in the baryon density field that have $n_{\rm H} \ge n_{\rm H,c} = 100 \rm cm^{-3}$.  Neighbouring cells are then recursively searched and assigned to the cloud if they also have densities $n_{\rm H} \ge 100 \rm cm^{-3}$. Density peaks further 20\,pc apart are identified as separate clouds. We refer to this method as the {\it `peak method'}.

The main difference between these two methods is that in the second case, multiple clouds may exist within the same continuous density structure if it contains more than one well-separated peak.

In this paper, we mainly focus on the results of the {\it contour method} due to its ability to identify large bodies that (visually) appear to be a single entity. This allows us to assess more easily the difference the environment was having on the cloud properties. Notably, however, the overall results from these two methods are very similar. We discuss this in a quantitative way in section~\ref{sec:cloud_comparison}.
 
To follow the evolution of the clouds, simulation outputs were analysed every 1\,Myr and the clouds were mapped between outputs with a tag number assigned to each cloud. The algorithm of this cloud tracking is described in \citet{TaskerTan2009}. A merger is said to have happened when a single cloud is at the predicted position for two other clouds after 1\,Myr of evolution.

\section{Results}
\label{sec:results}

\subsection{Global structure and disc evolution}
\label{sec:results_global}

\begin{figure*}
\begin{center}
	\includegraphics[width=18.5cm,bb=0 0 979 288]{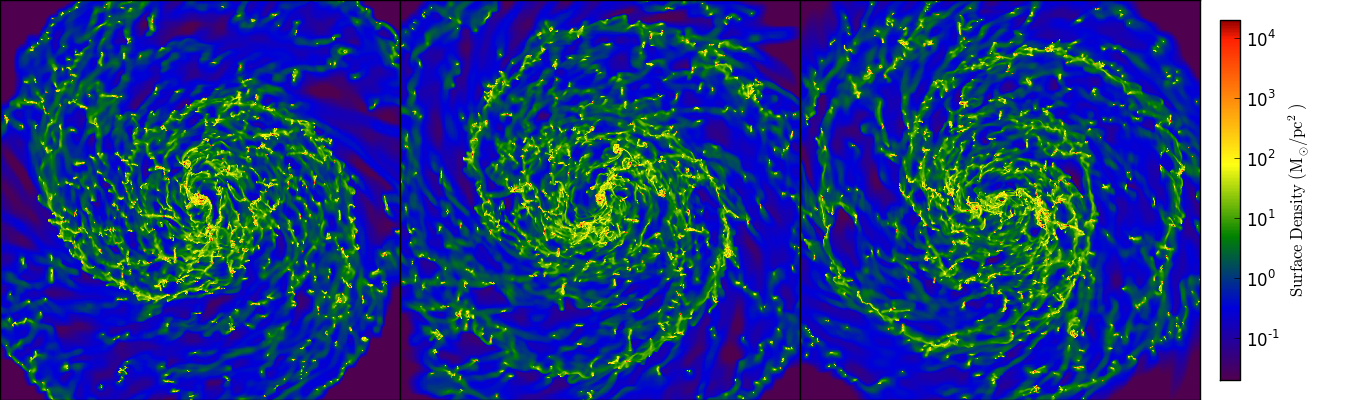}
	\caption{Evolution of the galactic disc. Images show the surface gas density at times, t = 200, 240, and 280\,Myr. Each image is 15\,kpc across.}
	\label{fig:disc_images}
\end{center}
\end{figure*}

\begin{figure*}
\begin{center}
	\hspace{-80pt}
	\subfigure{
	\includegraphics[width=80mm]{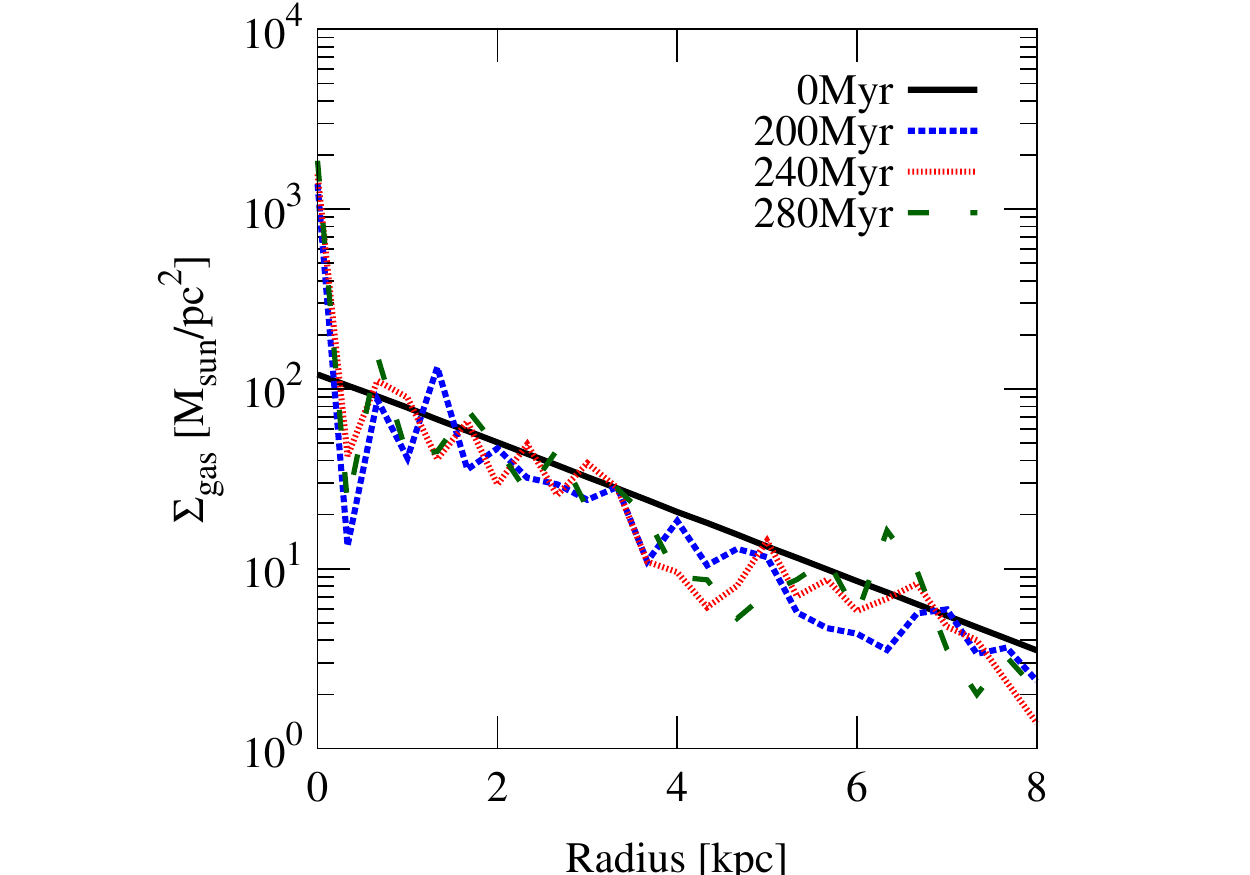}}
	\hspace{-60pt}
	\subfigure{ 
	\includegraphics[width=80mm]{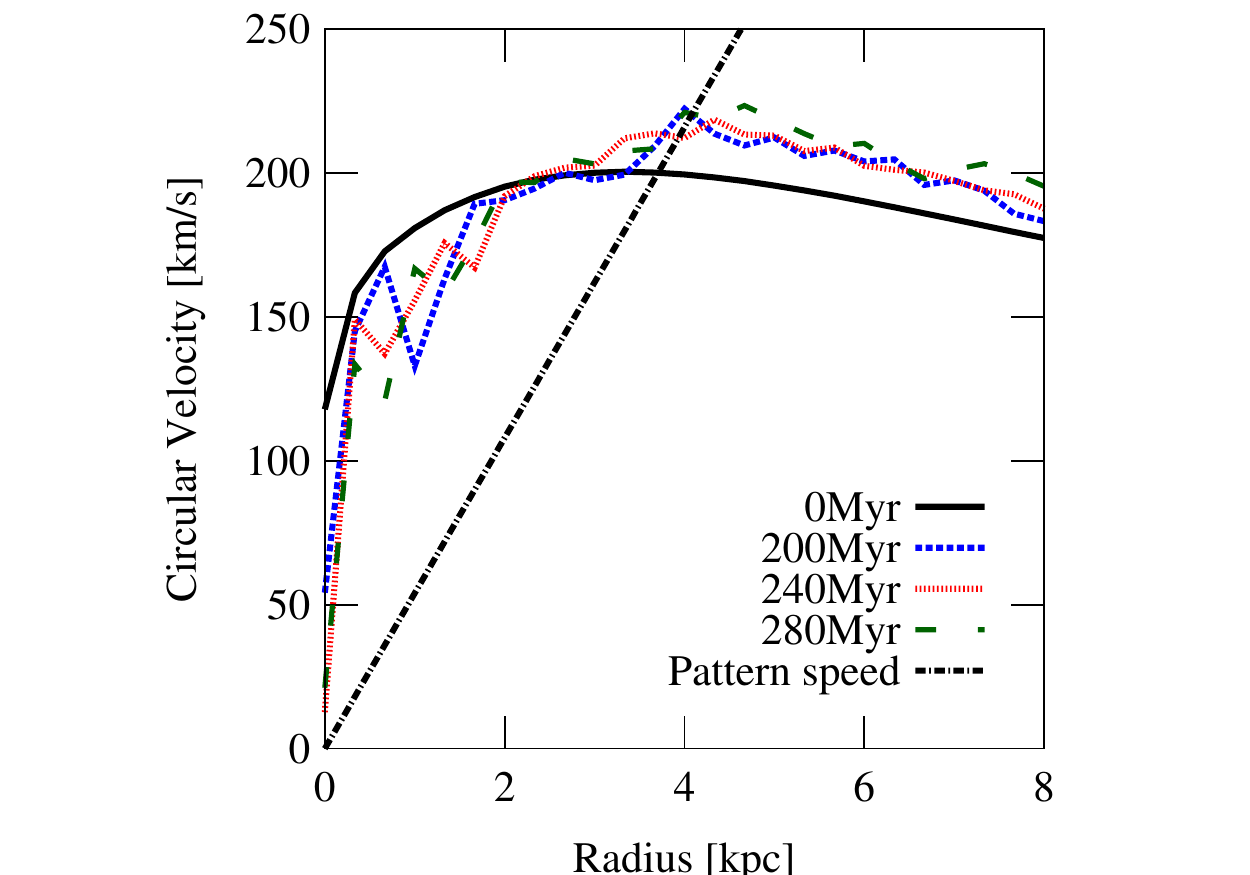}}
	\hspace{-60pt} 
	\subfigure{
	\includegraphics[width=80mm]{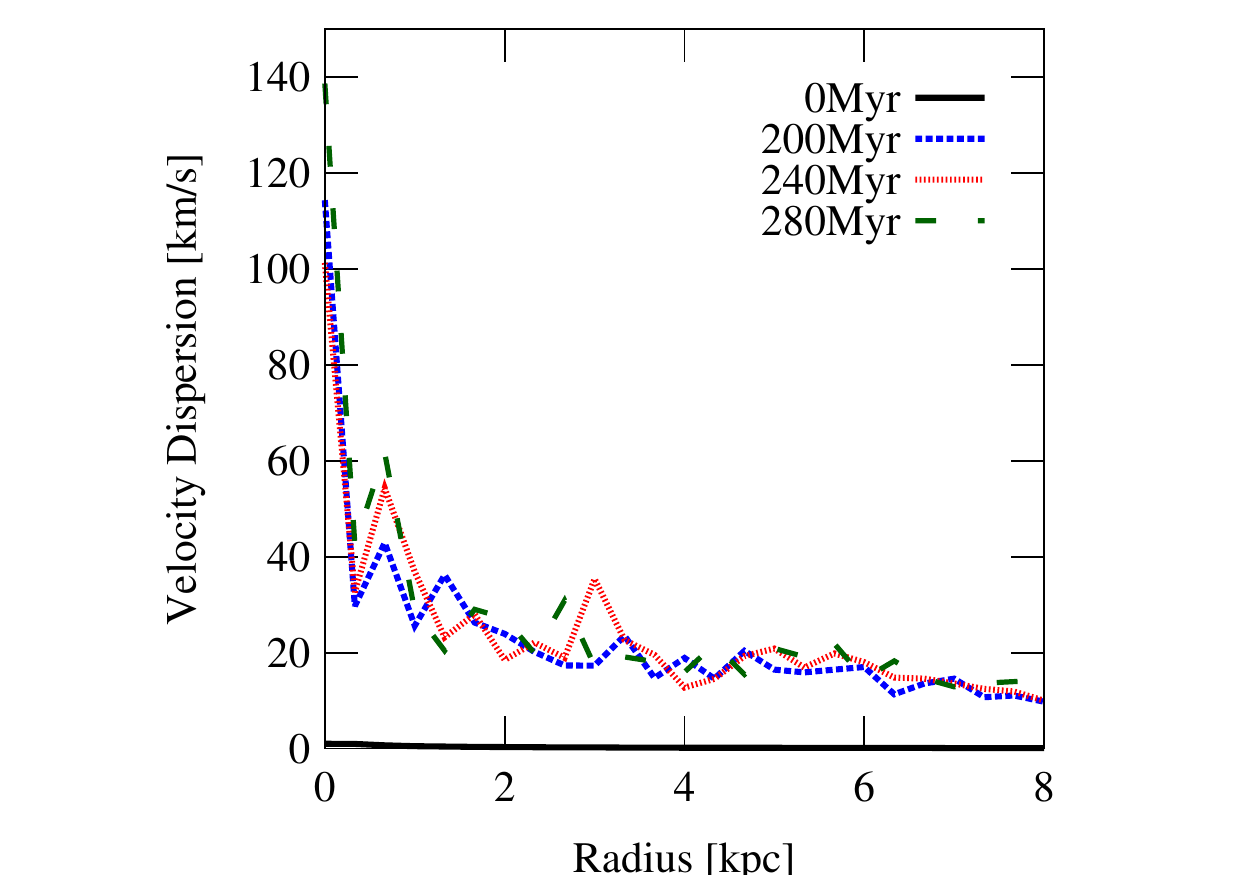}}
	\hspace{-80pt} 
	\caption{Azimuthally averaged (bin size 333\,pc) radial gas profiles for the galactic disc at t = 0, 200, 240 and 280\,Myr. From left to right: (1) gas surface density, $\Sigma_{\rm gas} = \int_{-1kpc}^{+1kpc}\rho(z)dz$, (2) gas circular velocity (mass-weighted average over -1kpc $<$ z $<$ 1kpc) and (3) 1D gas velocity dispersion (also mass-weighted average over -1kpc $<$ z $<$ 1kpc).}
	\label{Galactic disc properties}
\end{center}
\end{figure*}

In the initial stages of the simulation, the gas profile is smoothly exponential as described in section~\ref{sec:numerics_disc}. As the simulation begins and the gas feels the impact of the stellar potential, two shock waves are formed at the point of co-rotation between the stellar potential's pattern speed and the gas circular velocity. As they move in opposite directions through the disc, the gas falls into the grand design pattern. Self-gravity then begins to act, fragmenting the gas into knots and filaments. After 120\,Myr (one pattern rotation period), the gas is fully fragmented, and between 200\,Myr and 280\,Myr, the galactic disc settles into a quasi-equilibrium with no large structural change. 

The surface density of the inner 15\,kpc of the galactic disc is shown in Figure~\ref{fig:disc_images} at three different times after the main fragmentation: t = 200, 240 and 280\,Myr. Given the period of rotation for the non-antisymmetric stellar potential (pattern speed) is about 120\,Myr, the figure shows approximately $2/3$rd of a pattern rotation. The gas circular velocity gives a orbital period of about 240\,Myr (our middle panel) at the disc's outer edge, 8\,kpc. 

The grand design of the bar and spiral arms can be clearly seen in each panel of  Figure~\ref{fig:disc_images}. These global galactic structures are consistent with the $\rm ^{12} CO$(J=1-0) image of M83 \citep{Lundgren2004a}, with the bar-end at $\rm r\sim2.3\ kpc$. The bar and spiral arms rotate counter-clockwise in pace with the non-axisymmetric stellar potential.
 
 We can see the formation of clouds as dense knots in the surface density field. These clouds are seen not only in the bar and spiral arms but also in the inter-arm region and outer disc. This is in keeping with observations, where clouds are observed both within the grand design's main features and also the inter-arm regions \citep{Koda2009}. The properties of the clouds during this quasi-equilibrium stage are the focus of this paper.

The azimuthally averaged radial profiles for the galaxy disc are shown in Figure~\ref{Galactic disc properties} at four different simulation times, t = 0, 200, 240 and 280\,Myr. Between 200\,Myr and 280\,Myr (our main analysis time period), the disc properties are settled and show little evolution in agreement with what is seen visually in Figure~\ref{fig:disc_images}. 

\begin{figure*}
\begin{center}
	\subfigure{
	\includegraphics[width=9.0cm, bb=0 0 1024 768]{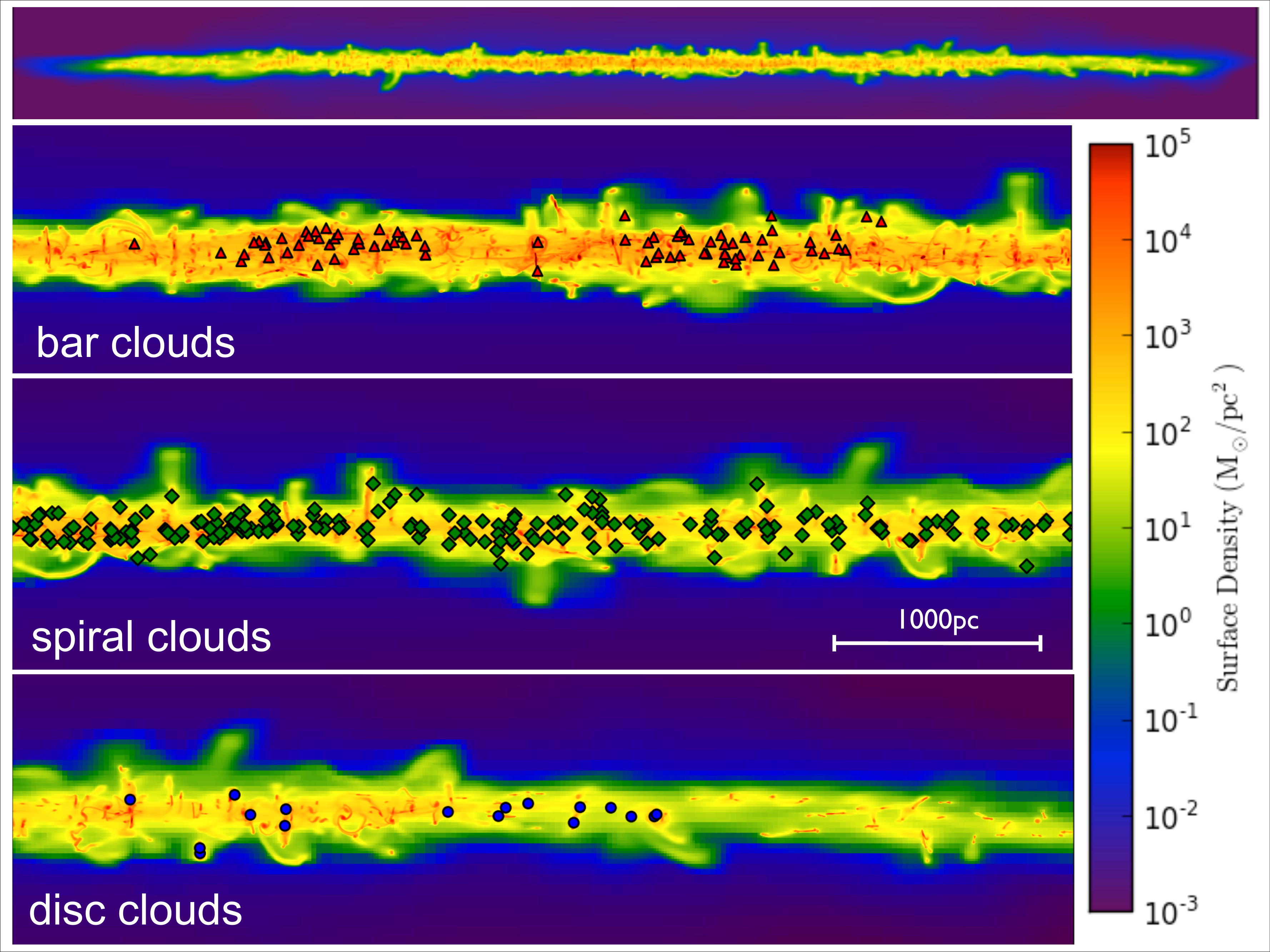}}
	\hspace{-20pt}
	\subfigure{
	\includegraphics[width=90mm]{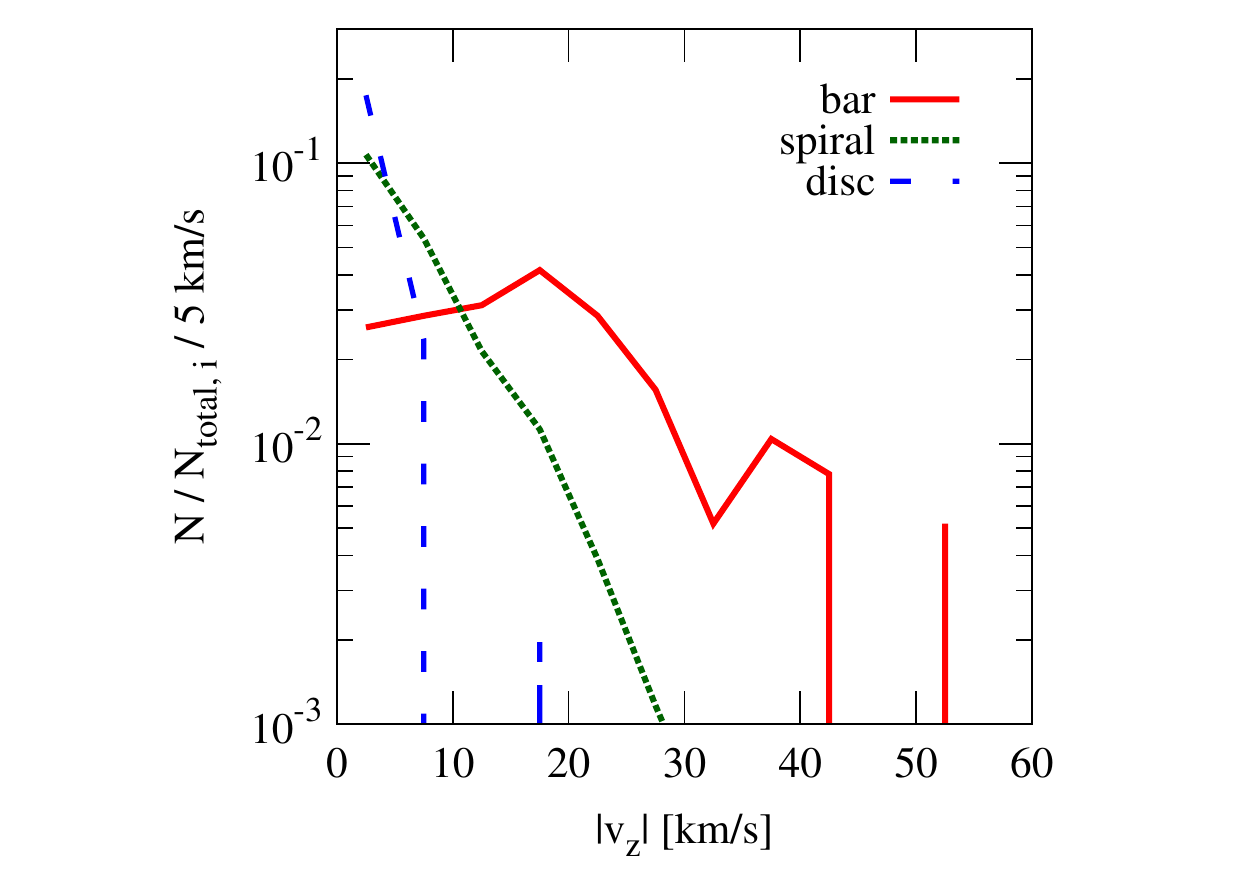}}
	\caption{Galactic disc vertical gas properties. Left panel shows the surface density of edge-on galactic disc with the cloud positions marked by the red, green and blue symbols (bar, spiral and disc region, respectively). Right panel shows the normalised distributions of the vertical velocity of the clouds in each region at 240\,Myr.}
	\label{vertical gas properties}
\end{center}
\end{figure*}

The left-most panel of Figure~\ref{Galactic disc properties} plots the gas surface density, averaged over a height of $\rm -1\,kpc < z < 1\,kpc$, containing the full extent of the disc's vertical height. Small fluctuations are seen in the surface gas density as the gas fragments interact and are stirred by the stellar potential, but the profile shape remains unchanged. At the very centre of the disc, the surface density does increase with time due to infall both from resolution (it is impossible to maintain perfect circular motion on a Cartesian grid at very small radii) and gas motion induced by the bar's potential. In M83, this bar instability causes a starburst in the central region at radius $\rm r < 300pc$ \citep{Harris2001}. If we allowed star formation and stellar feedback in our model, we would expect the gas ejection from such a burst to suppress this concentration of central gas. Since we do not include this process in our model, we ignore the galaxy's central region ($\rm r < 600pc$) in our analysis of the cloud properties. 

The middle panel of Figure~\ref{Galactic disc properties}  shows the radial profile of the mean circular velocity of the gas. This is calculated as a mass-weighted average over $\rm -1 kpc < z < 1 kpc$. Here, the effect of the rotating stellar potential is clearly visible, with the gas deviating from its initial orbit away from the point of co-rotation between the pattern and initial gas orbital speed at $r \sim 3.5$\, kpc. Beyond this point, the stellar potential rotates faster than the initial gas circular speed, driving the gas to a faster orbit. The circular velocity here is approximately 200\,km/s, agreeing with the rotation curve of M83 \citep{Lundgren2004b}. Within 3.5\,kpc, the gas motion is dominated by the bar which, inside the bar ends at $r= 2.3$\,kpc, forces the gas to flow along elliptical orbits. This produces an average velocity lower than the initial conditions, since the motion is not truly circular any more. 

 The final right-most panel of Figure~\ref{Galactic disc properties}, shows the one-dimensional velocity dispersion as defined by $\sigma_{\rm 1D} = \sqrt{(\vec{v}-\vec{v}_{cir})^2/3}$, where $\vec{v}$ is the velocity of the gas and $\vec{v}_{cir}$ is its circular velocity at that point. It is also calculated as a mass-weighted average over $\rm -1 kpc < z < 1 kpc$. With the gas motion initially set to follow a circular orbit, the dispersion at t = 0 is zero. As the gas falls into the stellar potential and fragments into self-gravitating clumps that interact, the dispersion increases. As seen in the middle panel, beyond the co-rotation point, gas is accelerated by the spiral arms. Meanwhile, within the bar region, the velocity dispersion is at its highest as the gas is pulled from its circular orbit to follow elongated elliptical paths through the bar. Within this dense region, further cloud interactions also increase the velocity dispersion, a point we will return to when we consider cloud properties. 

In Figure~\ref{vertical gas properties}, we show the vertical gas distribution and motion in the galactic disc. The left-panel shows the surface density of four regions of the edge-on disc at 240\,Myr (two pattern rotation periods and one gas orbital period at 8\,kpc). The top most image shows the complete disc, while the three images below show close-up sections of the bar, spiral and outer disc regions. Overlaid on the lower three images are the positions of the clouds as identified by the contour method in section~\ref{sec:numerics_cloud}. Despite the gas cooling from its initial equilibrium temperature at $10^4$\,K and the lack of stellar feedback to inject energy, the gas scale height remains around its initial value of 100\,pc, but with marked differences between the regions. Within the bar region, the scale height is 115\,pc, the spiral region has a scale height of 105\,pc and the outer disc has a lower value at 80\,pc.

The reason for the regional variations in the scale height can be seen in the right-hand plot of Figure~\ref{vertical gas properties}. This plot shows the normalised vertical velocity distribution of clouds in each of the three bar, spiral and outer disc regions at 240\,Myr. In the bar region, the clouds have a broad velocity profile, with vertical velocities out to 50\,km/s. In the disc, meanwhile, cloud vertical velocities remain more uniformly around 10\,km/s. This difference is due to strong interactions between the clouds. In the bar region, the gas density is high and its flow is highly elongated by the bar potential, decreasing the distance between the clouds. The shear flow at the bar ridge and bar ends also decreases the interaction time, adding to the boost in the velocity dispersion we saw in Figure~\ref{Galactic disc properties}. A lesser but still notable effect is also seen in the spiral region, where the arms also gather material together to increase cloud interactions. In the outer disc however, the density of clouds is much lower as can be seen visually in the left-panel of Figure~\ref{vertical gas properties}. The number of interactions is therefore less, resulting in a lower scale height. 

\subsection{Cloud classification based on galactic location}
\label{sec:results_location}

\begin{figure}
\begin{center}
	\includegraphics[width=8.3cm]{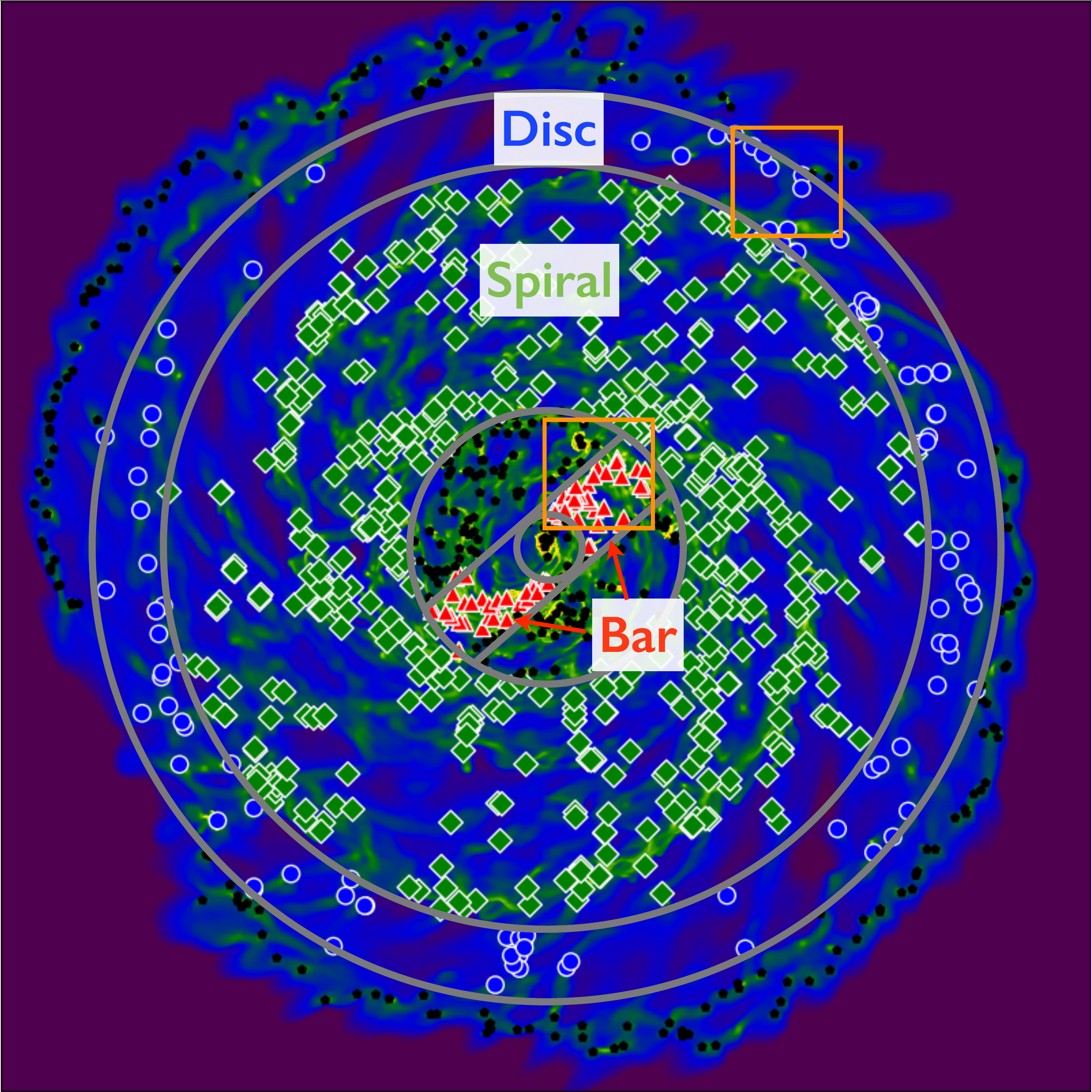}
	\caption{The three different galactic environments: bar, spiral and disc. The surface density of the galactic disc is shown at 240\,Myr, overplotted with markers denoting the cloud type according to location. Blue circles, green diamonds and red triangles show clouds in the disc, spiral and bar respectively. The black markers show clouds not included in our analysis. The width of the image is 20\,kpc, with the two squares marking regions that are shown in close-up in Figure~\ref{figure of three clouds}.}
	\label{Galactic disc with clouds}
\end{center}
\end{figure}

To compare the impact of different galactic environments on GMC properties, we assigned an environment group based on the cloud's physical location within the disc. The boundaries of our three regions, the bar, spiral and disc, are shown in Figure~\ref{Galactic disc with clouds}. If a cloud is found within the radii $2.5 > r > 7.0$\,kpc, we recognise the cloud as a {\it spiral cloud}. Outside $r = 7.0$\,kpc are the {\it disc clouds}, where we intentionally ignore clouds forming on the outer ring instability. This outer dense band of gas is from the Toomre instability \citep{Toomre1964} during the disc's initial fragmentation and therefore not as realistic an environment for cloud formation. {\it Bar clouds} form in a box-like region at the galactic centre, with a length of 5.0\,kpc and width 1.2\,kpc. The nucleus region inside 600\,pc is ignored due to it being very difficult to track clouds in this very high density area and the absence of a star burst degrading the comparison with observed GMC populations, as discussed in section~\ref{sec:results_global}. We also do not distinguish the difference between spiral arm and interarm regions; the number of clouds sitting between the spiral arms is small and hard to identify consistently as the spiral pattern rotates. 

The results of our identification scheme are shown using different coloured markers overlaying the gas surface density at 240\,Myr in Figure~\ref{Galactic disc with clouds}. Blue circles show the position of the disc clouds, green diamonds denote spiral clouds and red triangles mark the bar clouds. Black markers are for clouds identified via our contour method described in section~\ref{sec:numerics_cloud} but which we do not include in our analysis for one of the reasons described above. The number of clouds is roughly constant during our main time period of analysis, from 200 - 280\,Myrs and clouds rarely move environment during their lifetime. In the 240\,Myr snap-shot shown in Figure~\ref{Galactic disc with clouds}, 77 clouds are in the bar region, 515 are in the spiral region and 102 are in the disc.

\subsubsection{Cloud properties in each galactic environment}

\begin{figure*}
\begin{center}
\begin{tabular}{cc}
	\subfigure{
	\includegraphics[width=80mm]{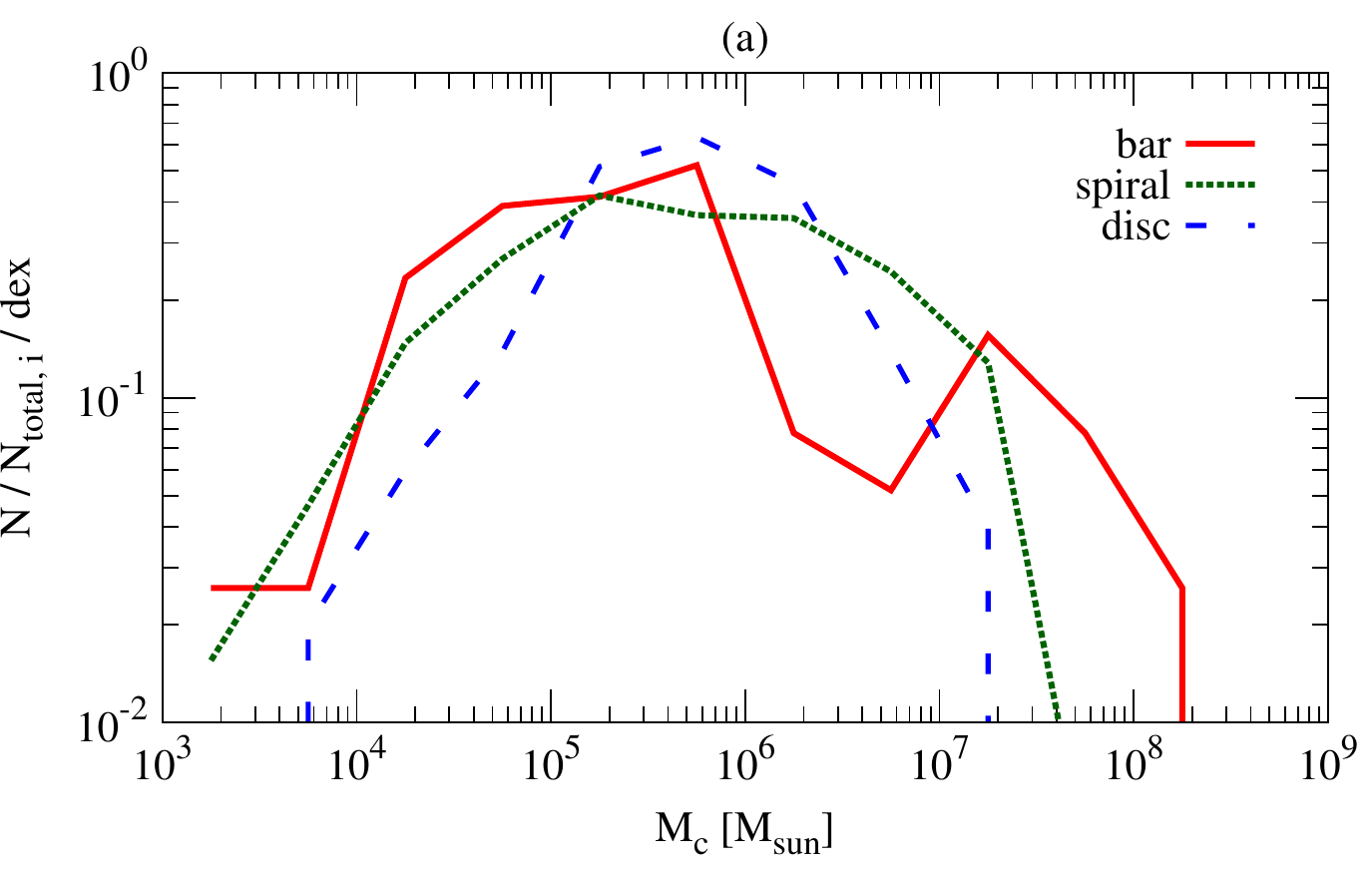}}
	\subfigure{
	\includegraphics[width=80mm]{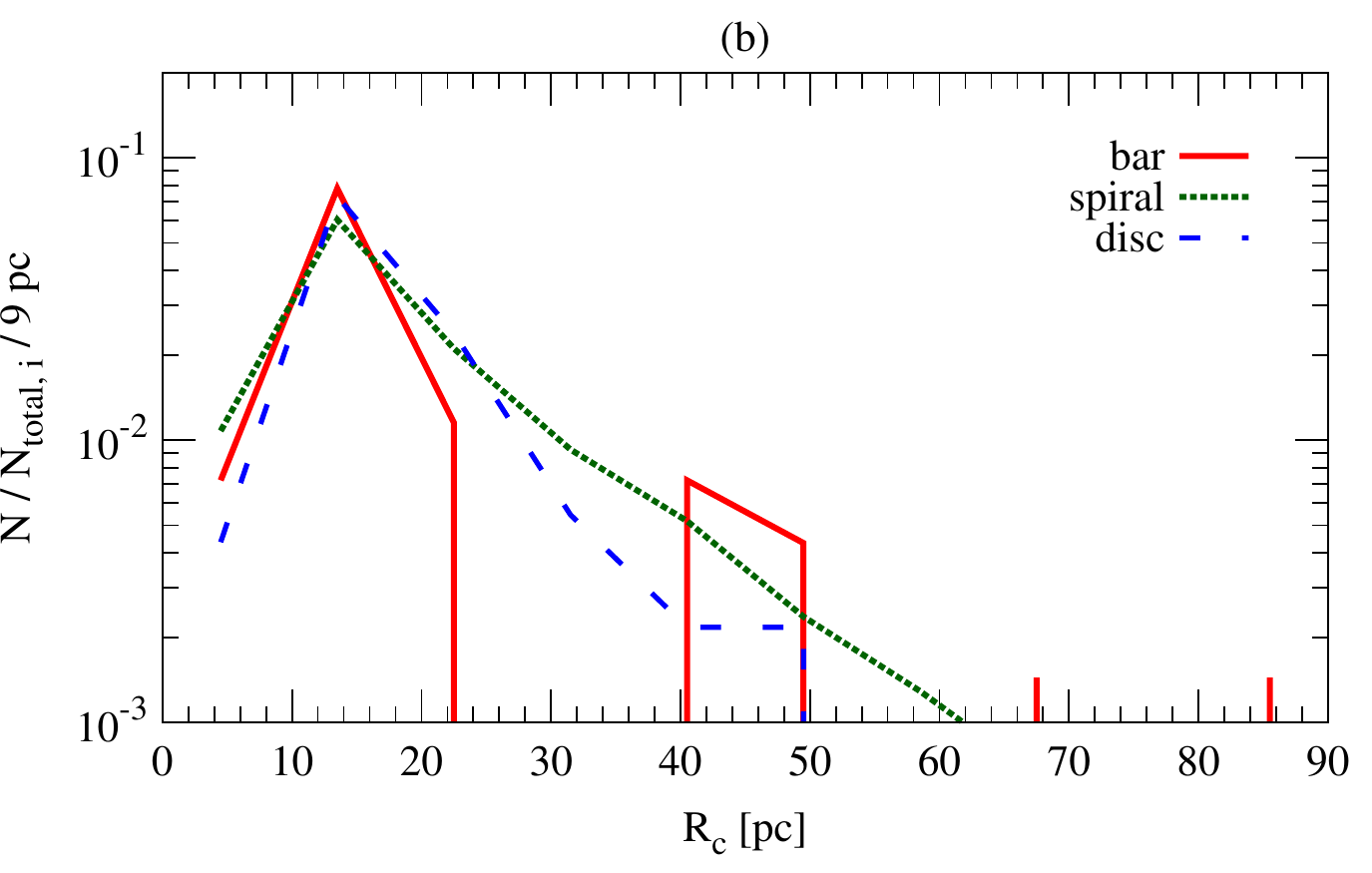}}
\end{tabular}
\begin{tabular}{cc}
	\subfigure{
	\includegraphics[width=80mm]{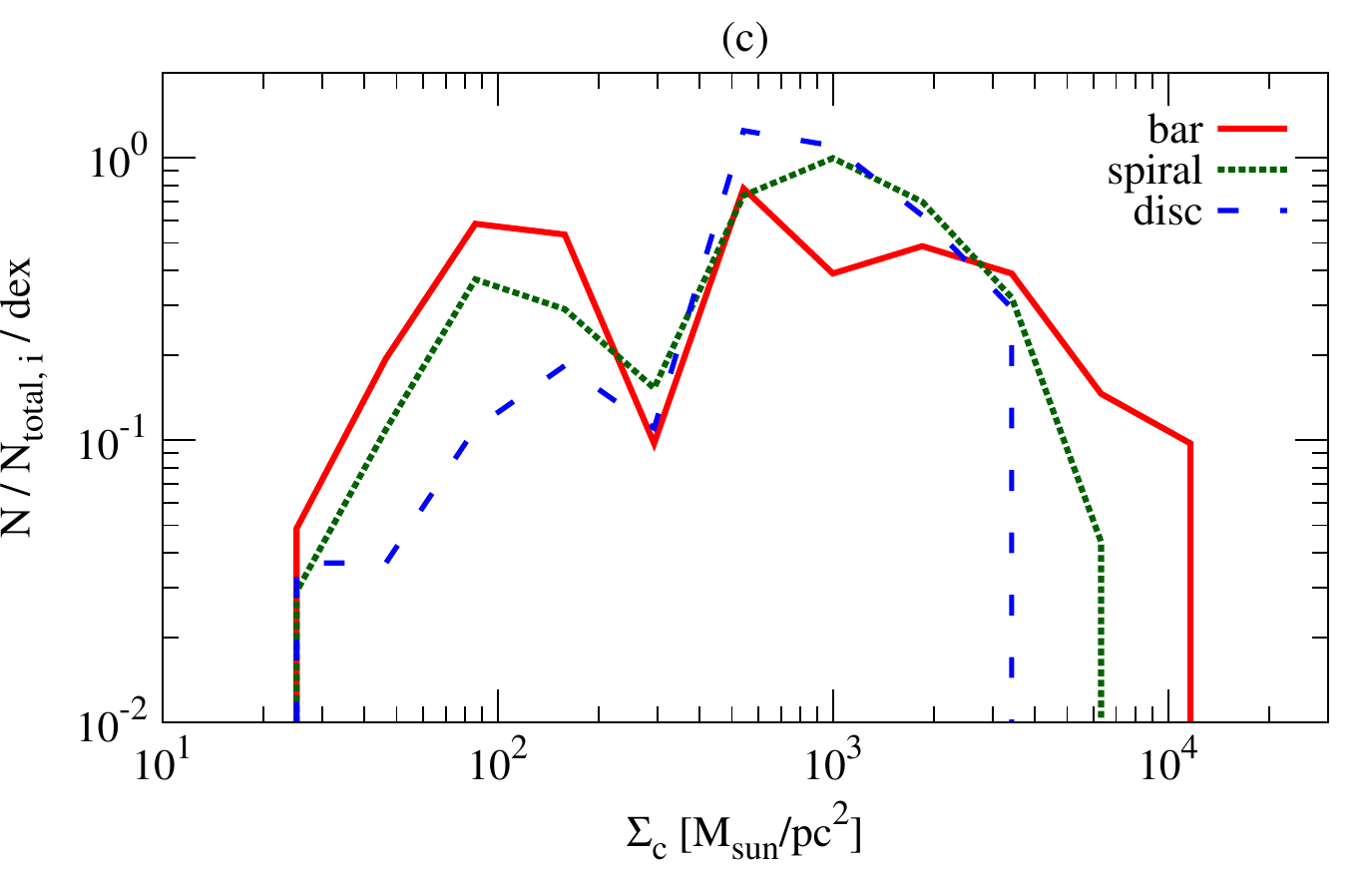}}
	\subfigure{
	\includegraphics[width=80mm]{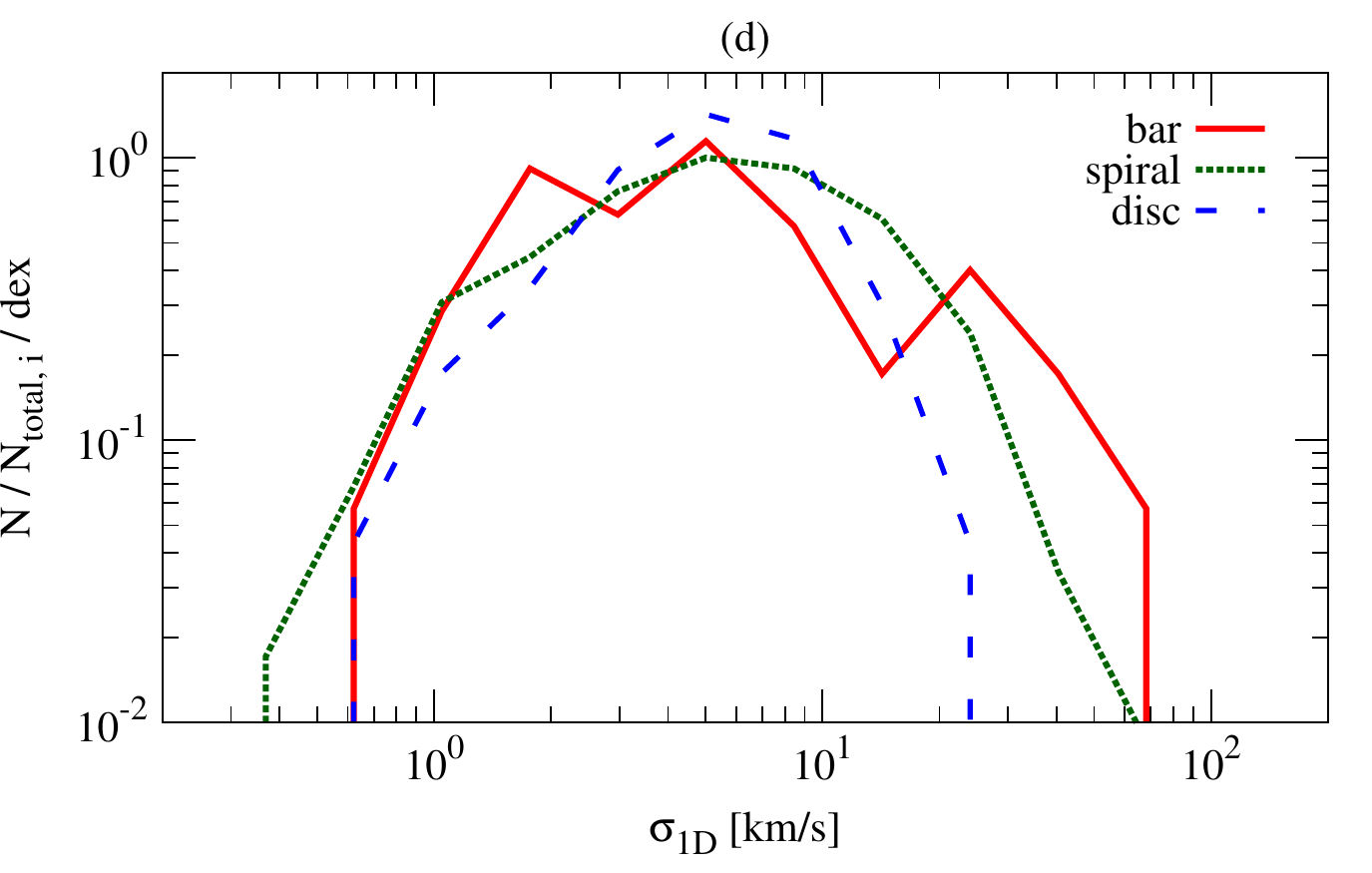}}
	\end{tabular}
\begin{tabular}{cc}
	\subfigure{
	\includegraphics[width=80mm]{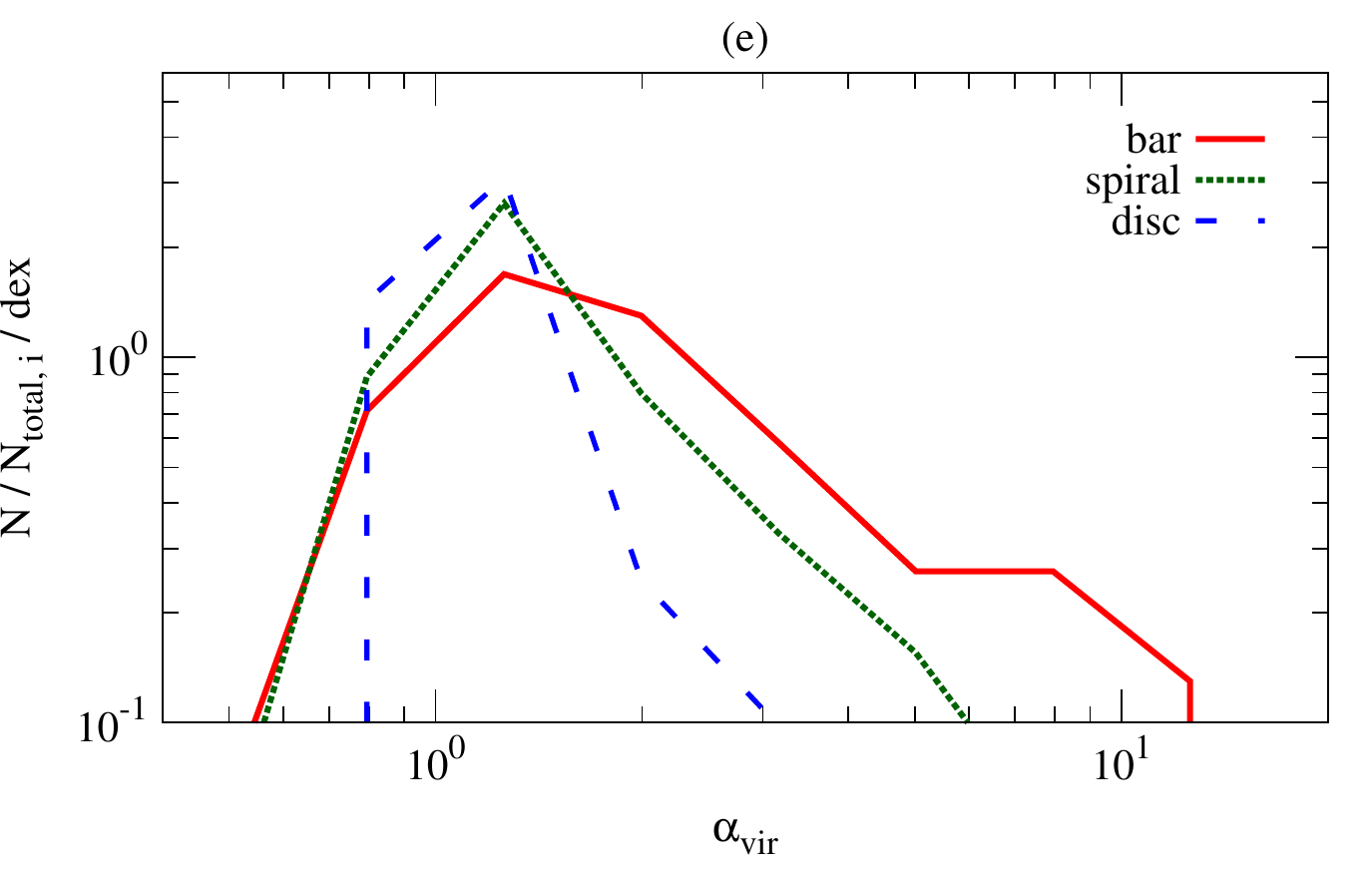}}
	\subfigure{
	\includegraphics[width=80mm]{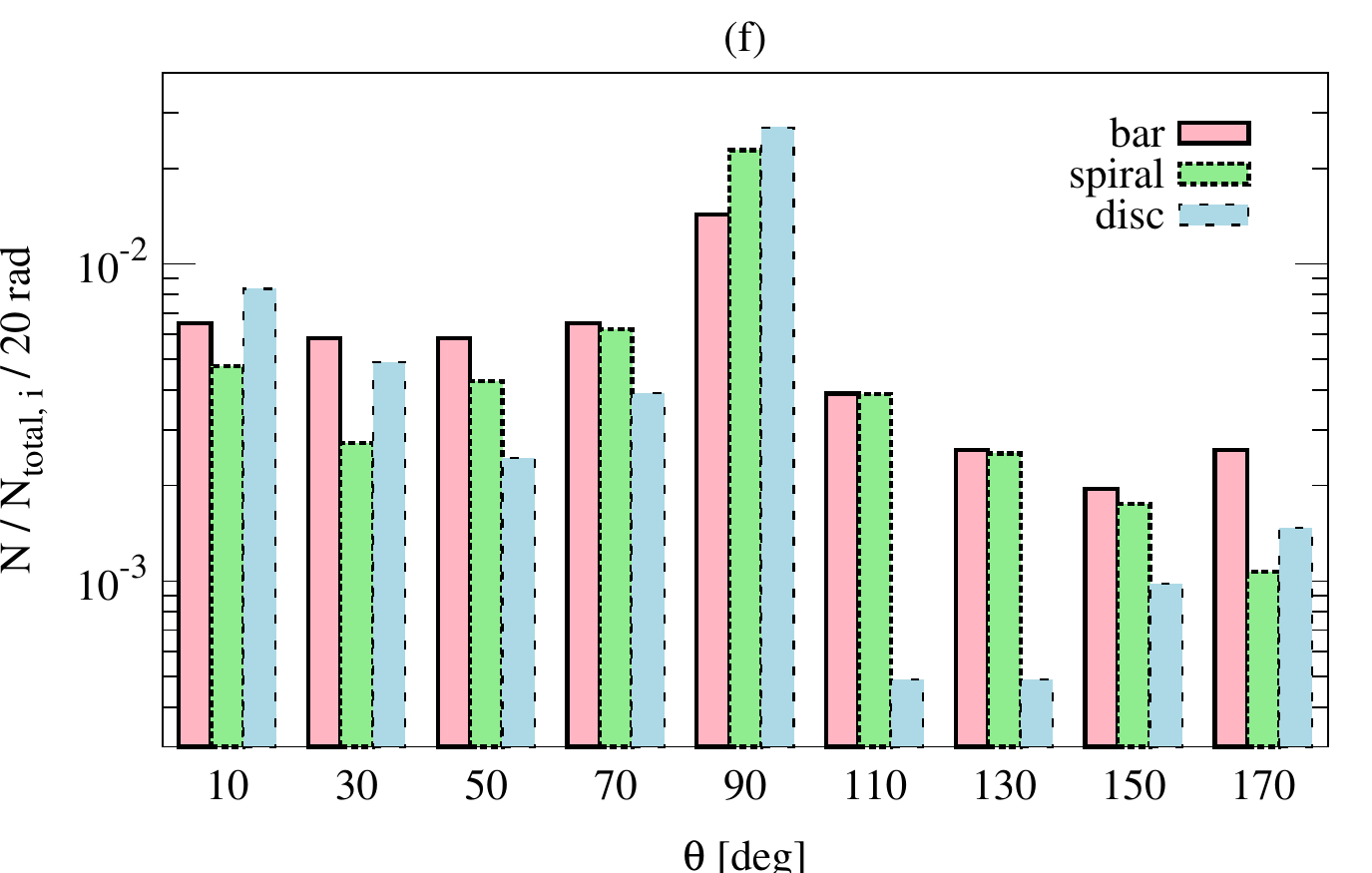}}
\end{tabular}
\caption{Normalised distributions of the cloud properties in the bar region (solid red lines), spiral region (dotted green lines) and disc region (dashed blue line) at 240\,Myr. Plots show: (a) the cloud mass, $M_{\rm c}$, (b) the average cloud radius, $R_{\rm c} = \sqrt{(A_{xy} + A_{yz} + A_{zx})/3\pi}$, where $A_{xy}$ is the projected area of the cloud in the x-y plane, $A_{yz}$ is that in the y-z plane, and $A_{zx}$ is in the z-x plane, (c) the cloud surface density, $\Sigma_{\rm c} = M_{\rm c}/(\pi R_{\rm c}^2)$, (d) the 1D velocity dispersion, $\sigma_{\rm 1D} = \frac{1}{\sqrt 3}\sqrt{[(v_x-v_{cx})^2+(v_y-v_{cy})^2+(v_z-v_{cz})^2]}$, where $(v_{x}, v_{y}, v_{z})$ is the velocity of the gas and $(v_{cx}, v_{cy}, v_{cz})$ is the cloud's centre of mass velocity, (e) the virial parameter, $\alpha_{\rm vir} = 5(\sigma_{\rm 1D}^2 + {c_{\rm s}}^2)R_{\rm c}/(GM_{\rm c})$, where $c_{\rm s}$ is a sound speed. The virial parameter is a measure of gravitational binding; a value greater than 2 indicates that the cloud is gravitationally unbound, (f) the angle $\theta$ between the cloud angular momentum vector and the galactic rotation axis.}
\label{distribution of cloud properties}
\end{center}
\end{figure*}

\begin{figure*}
\begin{center}
\begin{tabular}{cc}
	\hspace{-60pt}
	\subfigure{
	\includegraphics[width=125mm]{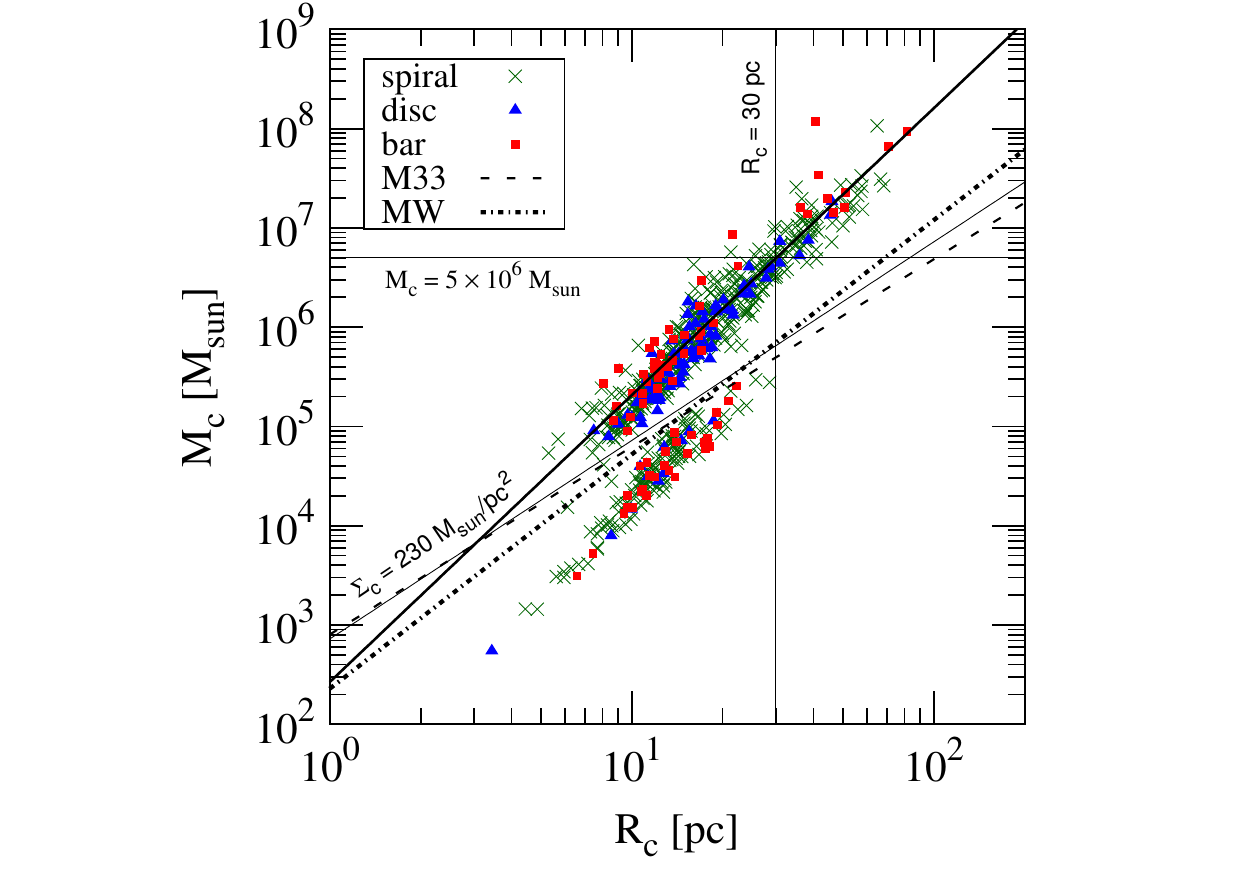}}
	\hspace{-110pt}
	\subfigure{
	\includegraphics[width=125mm]{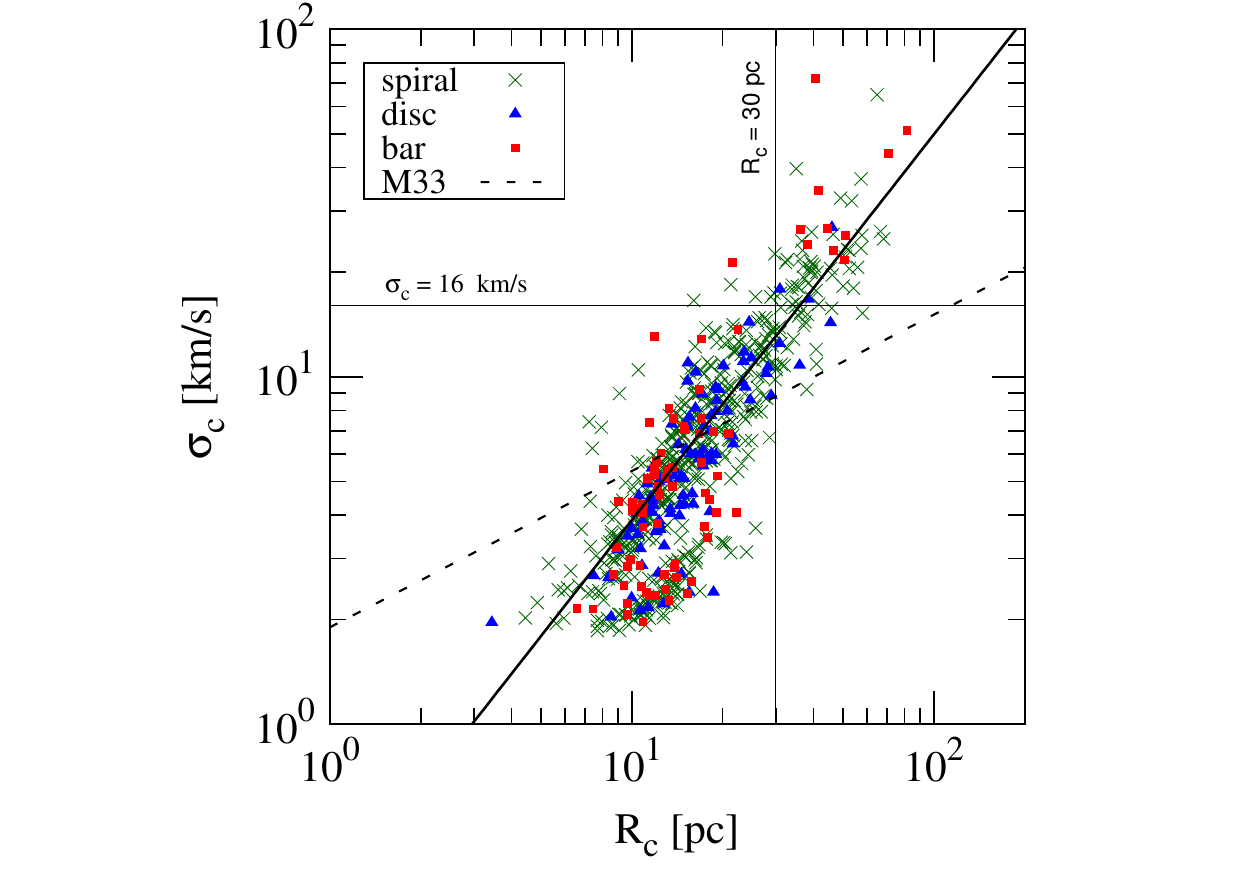}}
\end{tabular}
	\caption{Mass versus radius relation (left) and velocity dispersion versus radius relation (right) for clouds at 240\,Myr. Coloured markers denote clouds in different galactic regions: green crosses show spiral clouds, blue triangles are disc clouds and red squares are bar clouds. The thin black solid lines show the splitting points of the distributions in Figure~\ref{distribution of cloud properties}. The black dashed lines in the both panels show the scaling relation for M33, with $M_{\rm c} = 801\ {R_{\rm c}}^{1.89}$ (left panel) and $\sigma_{\rm c} = 1.9\ {R_{\rm c}}^{0.45}$ (right panel) \citep{Rosolowsky2003}, while the black dotted-dashed line in the left panels shows scaling relation for the Milky Way, with $M_{\rm c} = 228\ {R_{\rm c}}^{2.36}$\citep{Roman-Duval2010}. The fit to the clouds in our simulation are shown as thick solid black lines with power-laws $M_{\rm c} = 260\ {R_{\rm c}}^{2.89}$ (left) and $\sigma_{\rm c} = 0.3\ {R_{\rm c}}^{1.1}$ (right) }
	\label{Larson's law}
\end{center}
\end{figure*}

To see the impact of their environment on the cloud formation, we plot the cloud property distributions for clouds in each of our defined regions at 240\,Myr in Figure~\ref{distribution of cloud properties}. In each plot, the red solid line denotes the distributions for clouds found in the bar region, the green dotted line is for clouds in the spiral region and the blue dashed line is the disc clouds. When describing our results below, we have compared most extensively with observational GMC catalogues from the Milky Way \citep{Roman-Duval2010} and M33 \citep{Rosolowsky2003}. These comparisons have limits, since as \citet{Hughes2013} notes, GMC populations between galaxies have systematic differences and the surveys have been performed at different resolutions. At present, there is no survey of GMCs in M83 due to the low resolution of the molecular line observations; a situation that should change with the results from ALMA cycle 0 and 1. In the meantime, the observations from M33 and the Milky Way provide a guide to assess our simulated cloud results. 

Figure~\ref{distribution of cloud properties}(a) shows the mass distribution for these three environments, where the cloud mass is calculated the sum of the gas mass in each cell belonging to the cloud, as identified by the contour method described in section~\ref{sec:numerics_cloud}. In all three cases, the peak value for the cloud mass lies at around M$_{\rm c} \simeq 5 \times 10^5$\,M$_{\odot}$. This is in reasonable agreement with the GMCs observed in M33, where the peak mass was found to be $\simeq 10^5\ \rm M_{\odot}$ and larger than observations of the Milky Way, where the peak weighs in at $\simeq 5 \times 10^4$\,M$_\odot$. The Milky Way, however, has an average surface density than is almost one-eighth that of M83 \citep{Lundgren2004a,SparkeGallagher2000}, likely aiding the production of smaller clouds. 

While the typical mass for the clouds does not appear to depend on galactic environment, there are two clear differences between the mass distributions in Figure~\ref{distribution of cloud properties}(a) . Firstly, there is a trend in the broadness of the mass distribution, with the bar clouds having the larger abundance of the most massive (M$_{\rm c} > 10^7$\,M$_\odot$) and smallest (M$_{\rm c} < 10^5$\,M$_\odot$) clouds. The disc cloud profile has the most limited range, with the maximum cloud mass found in the disc being just under $2\times 10^7$\,M$_\odot$, compared to the bar region's maximum at almost $2\times 10^8$\,M$_\odot$. This high mass end is in keeping with observations performed by \citet{Foyle2013}, who investigated the compact FIR bright sources on the {\it{Herschel}} maps of M83 to estimate the mass of the giant molecular associations. These GMAs had a gas mass in the range of $10^6 - 10^8$\,M$_\odot$, in agreement with our own clouds. It should be noted, however, that \citet{Foyle2013}'s resolution is limited to spatial scales around 200-300\,pc and our own simulation lacks star formation and feedback. Both of these may have an influence in increasing maximum cloud size, with similar simulations demonstrating star formation can reduce cloud mass above $10^7$\,M$_\odot$ \citep{Tasker2011}. 

The second notable feature is that the distribution of the bar clouds is bimodal, with two peaks at M$_{\rm c} \simeq 5 \times 10^5$\,M$_{\odot}$ and at M$_{\rm c} \simeq 2.0 \times 10^7$\,M$_{\odot}$. The second of these peaks marks out the high mass clouds which appear as a distinct population, rather than a declining tail as seen in the spiral and disc regions. Similar splits can be seen in the radii and surface density distributions and was an unexpected phenomenon whose origin we will return to below. 

All the trends observed in the mass distribution are repeated in the distribution for the cloud radii, shown in Figure \ref{distribution of cloud properties}(b), which defines the average radius of the cloud as

\begin{equation}
R_{\rm c} \equiv \sqrt{\frac{(A_{xy} + A_{yz} + A_{zx})}{3\pi}},
\end{equation}

\noindent where $A_{xy}$ is the projected area of the cloud in the x-y plane, $A_{yz}$ is that in the y-z plane, and $A_{zx}$ is that in the z-x plane. As with the mass, the peak value for the radius is the same for the bar, spiral and disc regions at $R_{\rm c} \simeq 11$\,pc. This typical size agrees well with observations of the GMCs in both the Milky Way and M33, which show characteristic radii of 9\,pc and 10\,pc, respectively. While the majority of clouds are found at this radius in all environments, there is again a trend to find larger clouds in the spiral and bar, with the largest clouds in the galaxy forming in the bar region. Disc clouds, meanwhile, remain below 50\,pc. 

We also see further evidence for a bimodal bar region population, with a clear deficit of clouds between 20-40\,pc in this environment. The spiral and disc populations appear to show a more gradual decline with the numbers of clouds at higher radii steadily decreasing. 

This situation changes when we look at the cloud surface density in Figure~\ref{distribution of cloud properties}(c). Here, the cloud surface density is defined as

\begin{equation}
\Sigma_{\rm c} \equiv \frac{M_{\rm c}}{\pi R_{\rm c}^2}\ .
\end{equation}

\noindent In this distribution, all three environments show a clear bimodal nature, with peaks at $\Sigma_{\rm c} \simeq 10^2$\,M$_\odot$/pc$^2$ and $\Sigma_{\rm c} \simeq 10^3$\,M$_{\odot}$/pc$^2$. The first of these two peaks corresponds well with the typical surface density of the Milky Way clouds, which peaks around $10^2$\,M$_{\odot}$/pc$^2$.

The two populations in the bimodal distributions are most evenly distributed in the case of the bar clouds, with the peaks taking approximately equal magnitudes. The spiral and disc clouds show a smaller population for the lower surface density peak, which is least marked in the disc clouds. The three environment groups also follow the previous trend, with the bar clouds having the largest fraction of low surface density clouds ($\Sigma_{\rm c} < 3 \times 10^2$\,M$_\odot$/pc$^2$) and high surface density clouds ($\Sigma_{\rm c} > 4 \times 10^3$\,M$_\odot$/pc$^2$), while the disc clouds retain the smallest range. 

Perhaps the most surprising feature of the bimodality in the surface density is that it is seen in all three populations, while the mass and radius distributions only show a split for the bar clouds. Moreover, the divide in the surface density appears to occur in the middle of the distribution, whereas the mass and radial plots suggest a smaller population of larger objects. To explore the origin of this phenomenon, we plotted the mass versus radius in the left-hand panel of Figure~\ref{Larson's law}.

Known as one of Larson (empirical) laws \citep{Larson1981}, the scaling relation between cloud mass and radius has been observed both for GMCs in the Milky Way and for cloud properties in other galaxies. The dashed and dot-dashed line on the left-hand panel of Figure~\ref{Larson's law} show the observational fits, with the former being the M33 relation, $M_{\rm c} = 801\ {R_{\rm c}}^{1.89}$ and the latter, the Milky Way's scaling, $M_{\rm c} = 228\ {R_{\rm c}}^{2.36}$. 
 
 However, instead of seeing a single correlation between the cloud mass and its radii, Figure~\ref{Larson's law} shows two parallel sequences lying either side of the observational results. These two sequences exist within each environment and correspond to the bimodality in the cloud surface density distribution in Figure~\ref{distribution of cloud properties}(c). The upper correlation contains clouds with a surface density higher than $\Sigma_{\rm c} = 230 \rm \ M_{\odot}/pc^2$, while the lower line has clouds below that limit.

If we focus purely on the bar clouds in Figure~\ref{Larson's law} (shown with red square markers), we can see there is another split in the upper, high surface density, sequence. In the plot region around M$_{\rm c} \sim 5 \times 10^6$\,M$_{\odot}$ and $R_{\rm c} \sim 30$\,pc, there is a complete deficit of bar clouds. This then, corresponds to the bimodality of the mass and radii distributions in the bar cloud population in Figure~\ref{distribution of cloud properties}(a) and (b). Clouds above these values are large and massive and typically not found in the disc region. The spiral region has clouds that extend into this area, but the number is low compared to the quantity of clouds of smaller size. The bar clouds, meanwhile, have a significant fraction of their number in this upper region, with the smaller clouds split between the two scaling sequences. Thin black lines on Figure~\ref{Larson's law} show the divisions for these three cloud groups which will be discussed as their own cloud types in the next section.

If we return to Figure~\ref{distribution of cloud properties}(d), the distribution of the one-dimensional velocity dispersion of the clouds in the three environments is shown. The mass weighted, one-dimensional velocity dispersion is defined as

\begin{equation}
\sigma_{\rm 1D} = \sqrt{\frac{(v_x-v_{cx})^2+(v_y-v_{cy})^2+(v_z-v_{cz})^2}{3}},
\end{equation}

\noindent where $(v_{x}, v_{y}, v_{z})$ is the velocity of the gas and $(v_{cx}, v_{cy}, v_{cz})$ is the cloud's centre of mass velocity. As with the previous distributions, the peak dictating the typical cloud velocity dispersion is the same for all regions with a value of $\sigma_{\rm 1D} \simeq 6$\,km/s. This is comparable to the 6\,km/s characteristic velocity of the M33 and slightly larger than the Milky Way's 1\,km/s, which corresponds to the smaller cloud size. 

At the high velocity dispersion end of the plot, the bar clouds have the largest relative population, with the greatest fraction of clouds with $\sigma_{\rm 1D} > 30$\,km/s. This is followed by the spiral and then disc populations, with the majority of the disc clouds having velocity dispersions less than 20\,km/s. The bar population is again bimodal, with peaks at $\sigma_{\rm 1D} \simeq 5$\,km/s and at $\sigma_{\rm 1D} \simeq 23$\,km/s.

We can see this bimodality again by looking at the second Larson scaling relation between velocity dispersion and radius ($\sigma_{\rm c} \propto R^b$), as shown in the right-hand panel of Figure~\ref{Larson's law}. To better compare with the observational measurements, the velocity dispersion used includes the combined thermal component and is defined as: 

\begin{equation}
\sigma_{\rm c} = \sqrt{{\sigma_{\rm 1D}}^2+{c_{\rm s}}^2},
\end{equation}

\noindent where $c_{\rm s}$ is the sound speed of cloud. This addition makes only a small difference, since the cold cloud gas has a typical sound speed of $c_{\rm s} \simeq 1.8$\,km/s, 0.6\% of the non-thermal 16\,km/s motions. The power-law fit for the observational data from M33 is shows as a dashed line, with an index of $b = 0.45$. The upper sequence in the bimodality lies on this observational relation but with a steeper inclination, giving $b = 1.1$. This value sits on the upper observed bound \citep{Shetty2012} but is considerably higher than the typical measurements, which agree with the M33 result of $b \sim 0.5$ \citep{Solomon1987, Bolatto2008, Hughes2010}. This steepening may be due to a sensitivity to the physics not included in this simulation. In particular, the lack of feedback may allow our larger clouds to become more bound (and thereby have a higher velocity dispersion) while our smaller clouds may struggle to resolve the internal motions. 

As with the mass-radius scaling relation in the left panel, the linewidth-radius relation shows two sequences, although the lower sequence is significantly smaller than the upper trend. In the upper sequence of the bar clouds, there is a gap at $\sigma_{\rm c} \sim 16$\,km/s and $R_{\rm c} \sim 30$\,pc, corresponding to the bimodal splits in Figure~\ref{distribution of cloud properties}(d) and (b), the same segregation that is seen for the bar clouds in the mass and radius relations. 

The final two plots in Figure~\ref{distribution of cloud properties} show the virial parameter and the orientation of the GMCs. The virial parameter in Figure~\ref{distribution of cloud properties}(e) is defined at

\begin{equation}
\alpha_{\rm vir} = 5\frac{\sigma_{\rm c}^2R_{\rm c}}{GM_{\rm c}}.
\label{eq:virial}
\end{equation}

\noindent and is a measure of the gravitational binding. A value of $\alpha_{\rm vir} > 2$  indicated that the cloud is gravitationally unbound while $\alpha_{\rm vir} < 2$ suggests a bound system \citep{BertoldiMcKee1992}. The clouds in all three environments show a peak $\alpha_{\rm vir}$ value of $\sim 1$, indicating that the majority of the clouds are virialised but only marginally bound. Clouds in the Milky Way are observed to have a slightly lower $\alpha_{\rm vir}$ value of $\simeq 0.46$.

There is no obvious bimodal split in any of the cloud populations, but at values of $\alpha_{\rm vir} > 2$, the bar region contains a significantly higher fraction of clouds. This is followed by clouds in the spiral and disc region, whose distributions drop off smoothly after $\alpha_{\rm vir} \sim 1$. While the bar clouds also peak at this value, the majority of clouds sit to its right, indicating that most clouds in the bar region are unbound and take on a wide range of virial parameters. By contrast, the range in $\alpha_{\rm vir}$ in the disc is much lower, with most of the populations sitting close to the peak value. This difference in the range of $\alpha_{\rm vir}$ could indicate a more dynamic environment, where clouds have less time to settle to a virialised state. 

The final plot in Figure~\ref{distribution of cloud properties}, (f), shows the distribution of the angle $\theta$, between the cloud angular momentum vector and the galactic rotation axis. The cloud angular momentum is defined as the rotation with respect of the centre of mass of the cloud, with $0^\circ < \theta < 90^\circ$ indicating a prograde rotation in the same sense as the galaxy and $90^\circ < \theta < 180^\circ$ consisting of clouds with retrograde motion. In agreement with previous simulations \citep{TaskerTan2009}, clouds forming during the initial fragmentation of the disc ($t < 10\ \rm Myr$)  are born prograde, inheriting the galactic disc's rotational direction, $\theta \sim 0^\circ$. After one pattern rotation period ($t < 120\ \rm Myr$), when the disc has fully fragmented, the fraction of clouds at different spin orientations begins to increase. The disc clouds show the slowest evolution, with the population of high prograde and retrograde clouds increasing fastest in the bar, followed by clouds in the spiral region. By 240\,Myr, all three regions have clouds with the full range of orientations to the galactic rotation axis. The peak rotation angle actually sits at $\theta = 90^\circ$, suggesting most clouds rotate perpendicular to the disc. The fraction of retrograde rotating clouds is largest in the bar region, with the disc clouds remaining predominantly prograde. 

In their isolated Milky Way model, \citet{TaskerTan2009} suggest that the cloud's initial prograde rotation can be lost during encounters with other clouds, e.g. cloud-cloud collisions or tidal interactions. The faster shift towards a more retrograde population is therefore indicative of a more dynamic environment with many cloud interactions. This ties in with the virial parameter distribution in Figure~\ref{distribution of cloud properties}(e), which shows clouds in the spiral and bar tend to be less bound, consistent with a high number of interactions. 

Observations of M33 shows a range of cloud rotations, with 47\% having a prograde rotation, 32\% having a rotation perpendicular to the disc and 21\% with retrograde rotation. If cloud interactions are a dominant form of higher $\theta$ values, then this suggests that cloud environment (which will dictate such encounters) is a key factor in determining GMC properties. 

A last property that can be extracted from the data in this section concerns the stability of the disc. Traditionally, the resistance of a rotating disc to fragmentation is measured by the Toomre Q parameter for gravitational stability \citep{Toomre1964}. Defined as $Q = \kappa c_s/\pi G \Sigma_g$, where $\kappa$ is the epicycle frequency, $c_s$ is the thermal sound speed in the disc and $\Sigma_g$ is the gas surface density, a value of $Q < 1$ indicates  instability while higher $Q$ values imply the disc will not fragment. Since \citet{Toomre1964}'s original calculation involved a two-dimensional disc, the exact threshold for stability is debatable, with values between 1.5 - 0.7 being suggested via calculation and observation for a three-dimensional system \citep{Goldreich1965, Kennicutt1989, Gammie2001}.

Our disc begins with a steadily rising $Q$ value between 2 and 4 from 1\,kpc to the outer edge. As the gas cools, this drops to between 0 and 2 over the majority of the disc surface, suggesting (correctly) that the disc will fragment. However, when the gas breaks into the objects we identify as the GMCs, the azimuthally averaged $Q$ value increases as cold gas is bound up in the clouds while the larger volume of surrounding warmer gas is stable. This has also been seen in other simulations of galaxy discs such as \citet{TaskerTan2009,Tasker2011}. While the $Q$ value implies the disc is now stable, this conclusion has to be incorrect since regions within the clouds must collapse to form stars. This discrepancy was noticed by \citet{Romeo2010}, who pointed out that the Toomre equation assumes a well defined surface density and velocity dispersion, whereas in fact the Larson scaling relations show these properties are strongly dependent on the size of the region being measured due to turbulence. They argue that turbulence within the GMCs produces a transition between instabilities governed by large-scale gravitational fragmentation (Toomre) and those controlled by small-scale turbulence. 

To assess which of these forces has the upper hand, \citet{Romeo2010} created a stability map based on the indices in the Larson scaling relations, $a$ and $b$: $M \propto R^{a+2}$ and $\sigma \propto R^b$. From our fits in Figure~\ref{Larson's law}, we find $a = 0.9$ and $b = 1.1$, placing our clouds close to the border between Toomre stability and small-scale instability. Such a balance is expected if the clouds are virialised since the pressure balancing the self-gravity would follow the same scaling with size \citep{Romeo2010, Hoffmann2012}. However, our exact values place our clouds just above the line, suggesting that while close to viralisation, turbulence will initiate further instability. We therefore conclude that after the initial fragmentation, the disc is borderline stable to gravitational instabilities but unstable to turbulence.

\subsection{Cloud classification based on cloud properties}
\label{sec:results_types}

\begin{figure*}
\begin{center}
	\begin{tabular}{cc}
	\hspace{-40pt}
	\subfigure{
	\includegraphics[width=110mm]{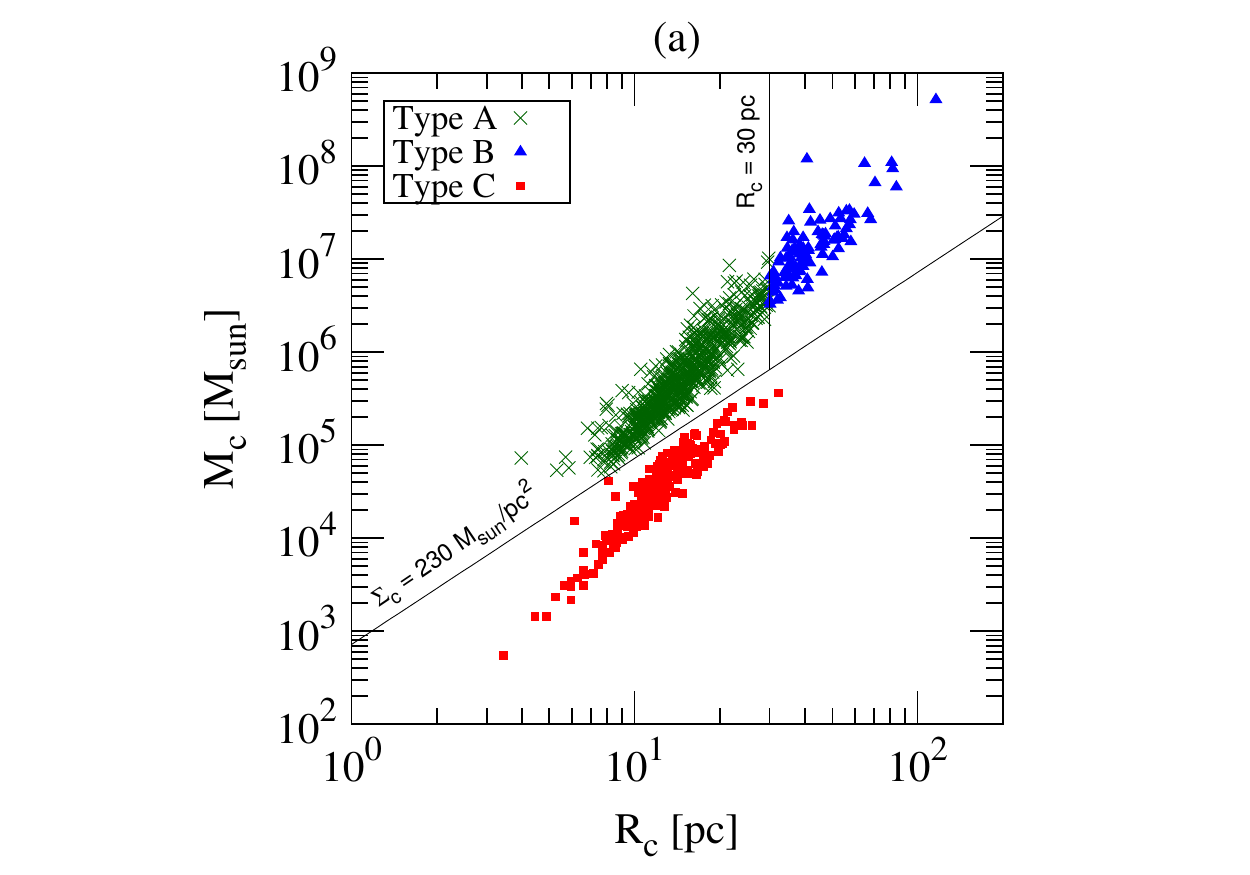}}
	\hspace{-80pt}
	\subfigure{
	\includegraphics[width=110mm]{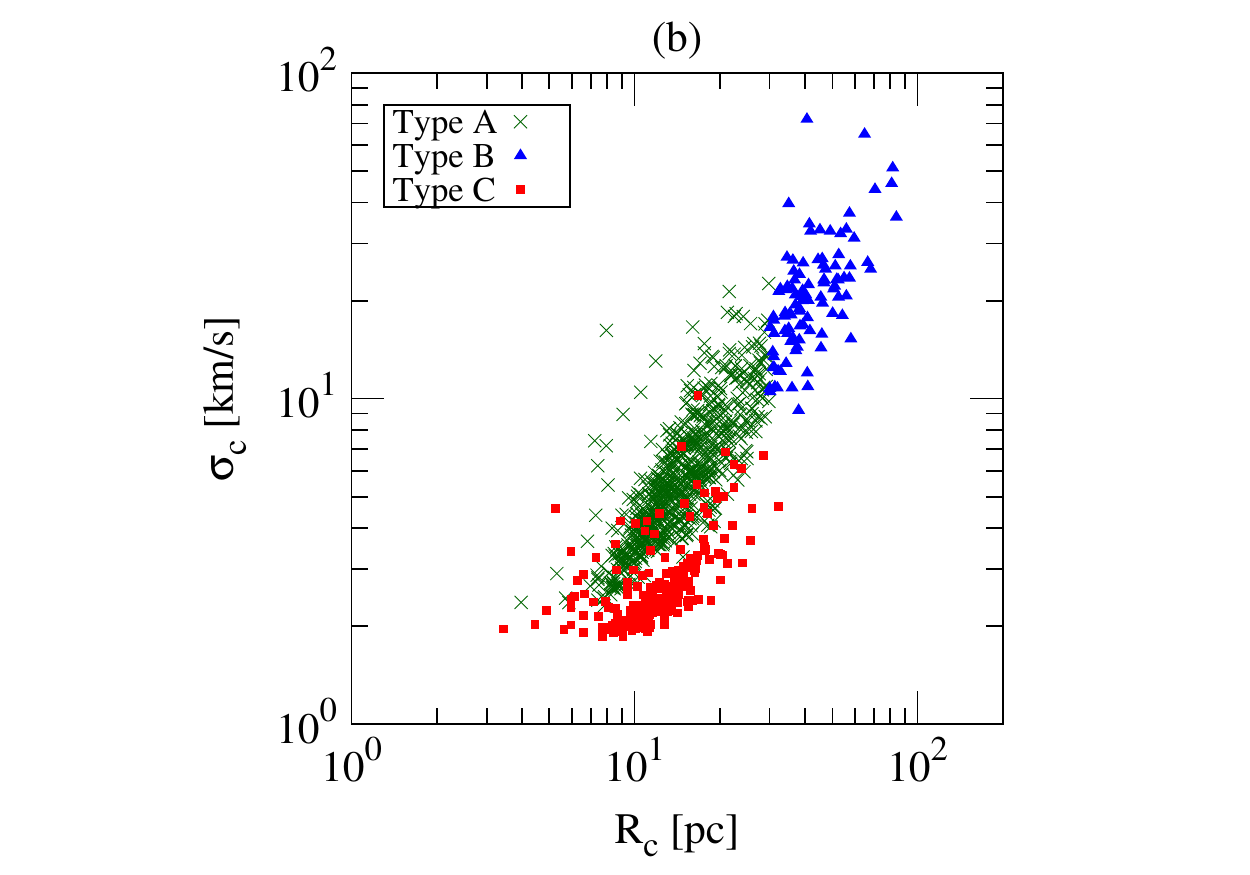}}
	\end{tabular}
	\begin{tabular}{cc}
	\hspace{-40pt}
	\subfigure{
	\includegraphics[width=110mm]{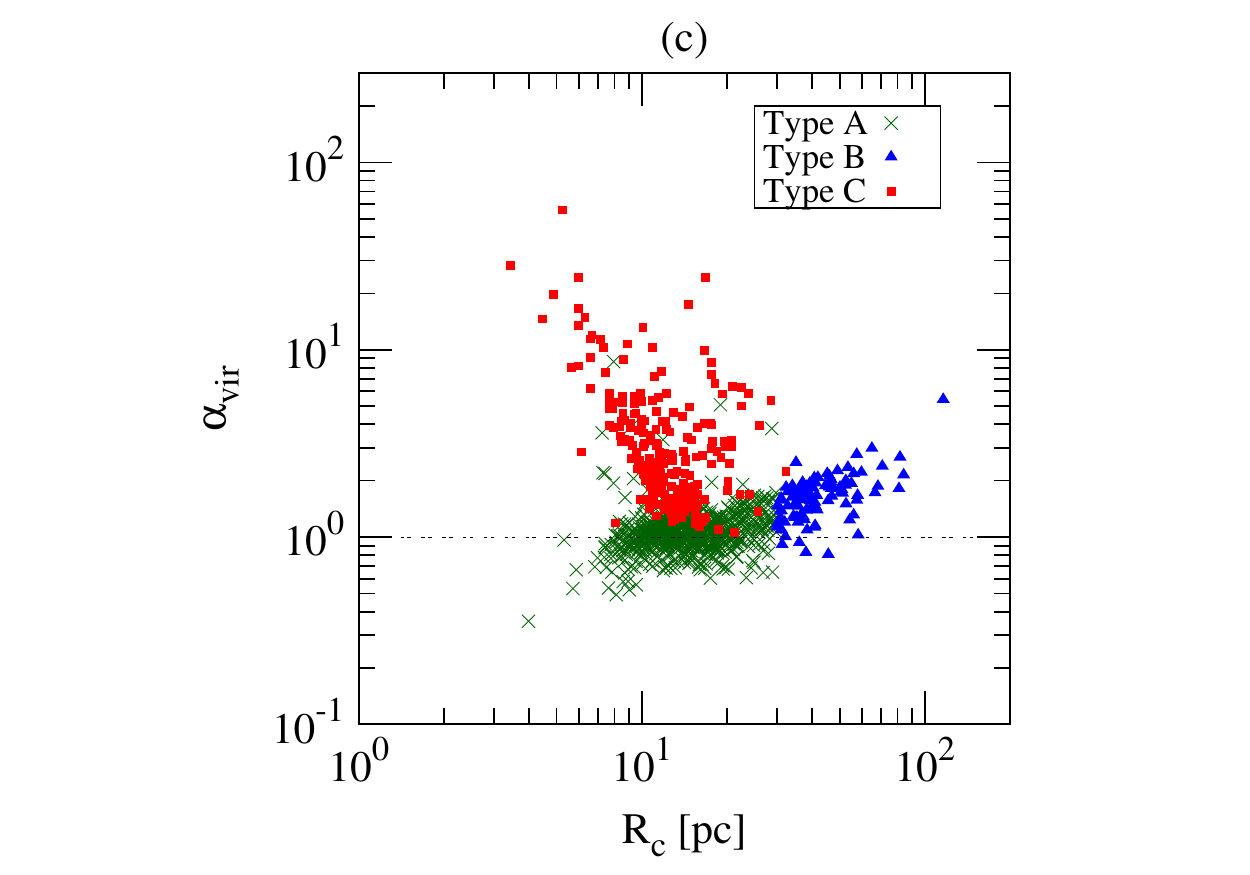}}
	\hspace{-80pt}
	\subfigure{
	\includegraphics[width=110mm]{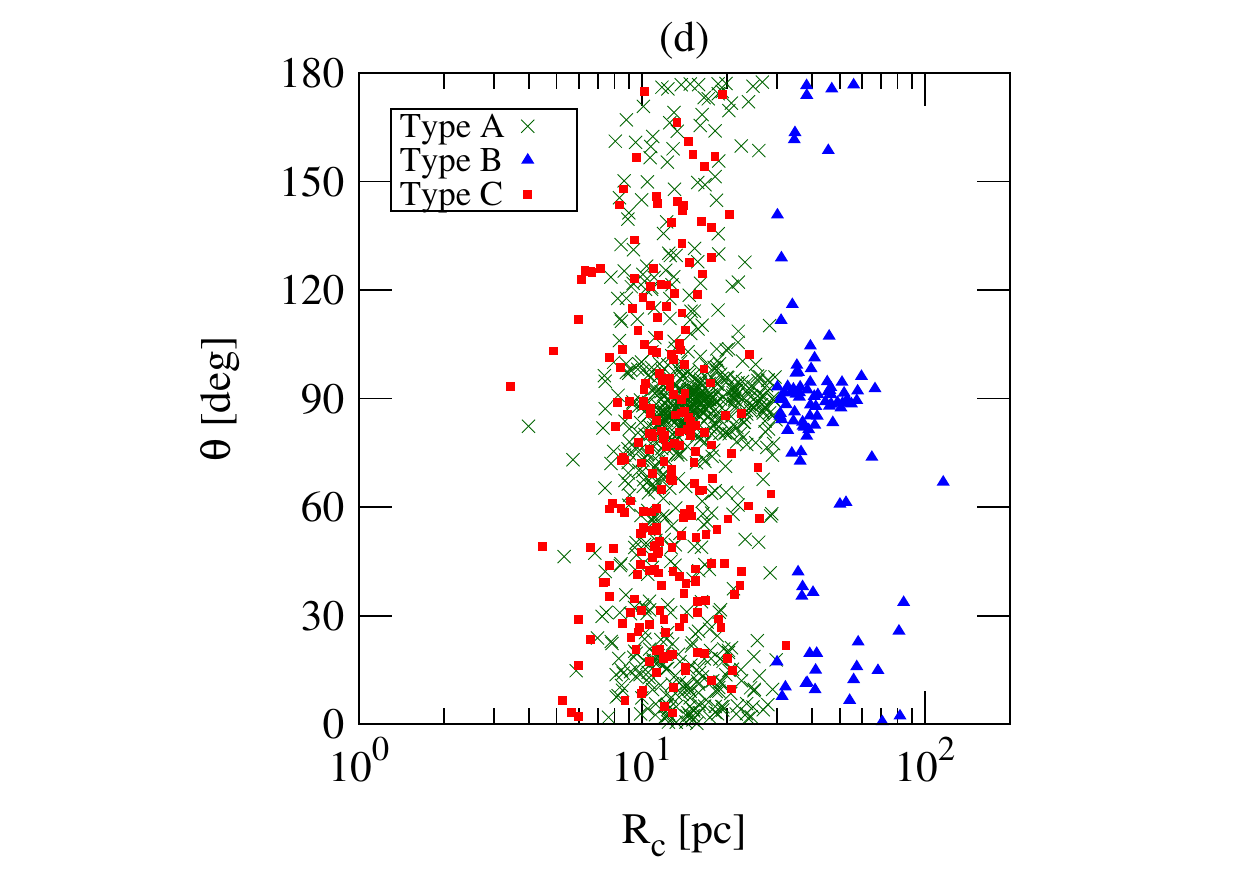}}
	\end{tabular}
	\caption{Scaling relations for our three cloud type categorisations at $t = 240$\,Myr. Top left (a) shows the mass versus cloud radius and defines the three cloud types: {\it Type A} clouds exist on the upper trend of the bimodal split, with surface densities greater than $230$\,M$_\odot$/pc$^2$. {\it Type B} clouds sit at the high end of the same sequence, with radii greater than 30\,pc. {\it Type C} clouds follow the lower trend and have surface densities less than $230$\,M$_\odot$/pc$^2$. The top right plot (b) shows the velocity dispersion versus cloud radius, lower left (c) plots the virial parameter against radius and lower right (d) shows the angle between the cloud's angular momentum axis and that of the disc, plotted against cloud radius.}
	\label{Larson's law of three type}
\end{center}
\end{figure*}

\begin{table}
\begin{center}
\begin{tabular}{|l|c|c|c|} \hline
	& bar & spiral & disc \\ \hline
	Type A & 49.4\% (38/77) & 64.1\% (330/515) & 83.3\% (85/102) \\ \hline
	Type B & 13.0\% (10/77) & 12.8\% (66/515) & 5.9\% (6/102) \\ \hline
	Type C & 37.7\% (29/77) & 23.1\% (119/515) & 10.8\% (11/102) \\ \hline
\end{tabular} 
\caption{The percentage of each cloud type in each galactic region at $t = 240$\,Myr. Bracketed numbers show the actual number of clouds of that type divided by the total cloud number in the region.}
\label{table:cloud_percentage}
\end{center}
\end{table}

To understand with physical reasons for the distinct splits in the cloud properties found in section~\ref{sec:results_location}, we re-classify all clouds according to the two sequences seen on the mass-radius relation in Figure~\ref{Larson's law} and the bimodality in the bar mass and radius distributions in Figure~\ref{distribution of cloud properties}(a) and (b). This is shown in Figure~\ref{Larson's law of three type}(a), where clouds that sit on the upper bimodal trend with surface densities above $230$\,M$_\odot$/pc$^2$ and radii less than 30\,pc form the group of {\it Type A} clouds, clouds along the same sequence but with radii above 30\,pc are declared {\it Type B} and clouds following the lower trend with surface densities below  $230$\,M$_\odot$/pc$^2$ are {\it Type C}.

The split between cloud types in each galactic region is shown in Table~\ref{table:cloud_percentage}. In all galactic environments, the most numerous cloud is {\it Type A}, but this percentage is significantly smaller in the bar regions where 38\% of the cloud population are of {\it Type C} and a further 13\% are {\it Type B}. In contrast, the disc region comprises mainly of {\it Type A} clouds with less than 6\% {\it Type B} and only about 10\% {\it Type C}.

\subsubsection{Properties of the three cloud types}

While these three new cloud classifications are based on their surface density and radius, their other properties also show marked differences. From our initial definition plot in Figure~\ref{Larson's law of three type}(a), it is clear that {\it Type B} clouds are not only extended, they are also massive. This is not surprising, since we see the bimodality in the bar clouds both in the mass and radius distributions, but it rules out the possibility that this cloud type could be dense tidal tails. 

Figure~\ref{Larson's law of three type}(b) shows the scaling relation between velocity dispersion and cloud radii. As was indicated in Figure~\ref{Larson's law}, the velocity dispersion also differs between the three types, with {\it Type C} clouds having a lower velocity dispersion than a {\it Type A} cloud with the same radius. To match their extended structure, {\it Type B} clouds have higher velocity dispersions than either {\it Type A} or {\it Type C} objects, with values above 10\,km/s. 

Another significant difference between the cloud types involves their gravitationally binding. Figure~\ref{Larson's law of three type}(c) plots the virial parameter, $\alpha_{\rm vir}$ as defined in section~\ref{sec:results_location}, against the cloud radius. {\it Type A} clouds are borderline gravitationally bound, with their $\alpha_{\rm vir}$ values clustered around 1.0. The extended {\it Type B} clouds are less bound, fitting in with their larger size and correspondingly higher velocity dispersion. Their values extend between 1 - 5 as the cloud increases in radii. More notable are the {\it Type C} clouds, which for similar radii to the {\it Type A} objects, are far less bound with $\alpha_{\rm vir}$ values extending from 1 to 70 in a reverse trend where the smaller objects are the least gravitationally bound. This implies the {\it Type C} clouds are less compact than the other two populations of GMCs, explaining why they follow the lower trend in Figure~\ref{Larson's law of three type}(a).

In the final panel (d) in Figure~\ref{Larson's law of three type}, we show the variation of the cloud's angular momentum vector with cloud type. There is no correlation between $\theta$ and cloud radius, with clouds at any radii potentially having the full range of spin orientations. However, the massive {\it Type B} clouds have a smaller spread of orientations, with most of the clouds clustered around $\theta \sim 90^\circ$. {\it Type C} clouds meanwhile, appear to have no preferred direction, forming a spread over the full angular range. The {\it Type A} clouds cluster between $\theta = 0 - 90^\circ$, indicating that these clouds may change their orientation during their lifetime. 

\subsubsection{Cloud lifetime and merger rate}

\begin{figure}
\begin{center}
	\subfigure{
	\includegraphics[width=75mm]{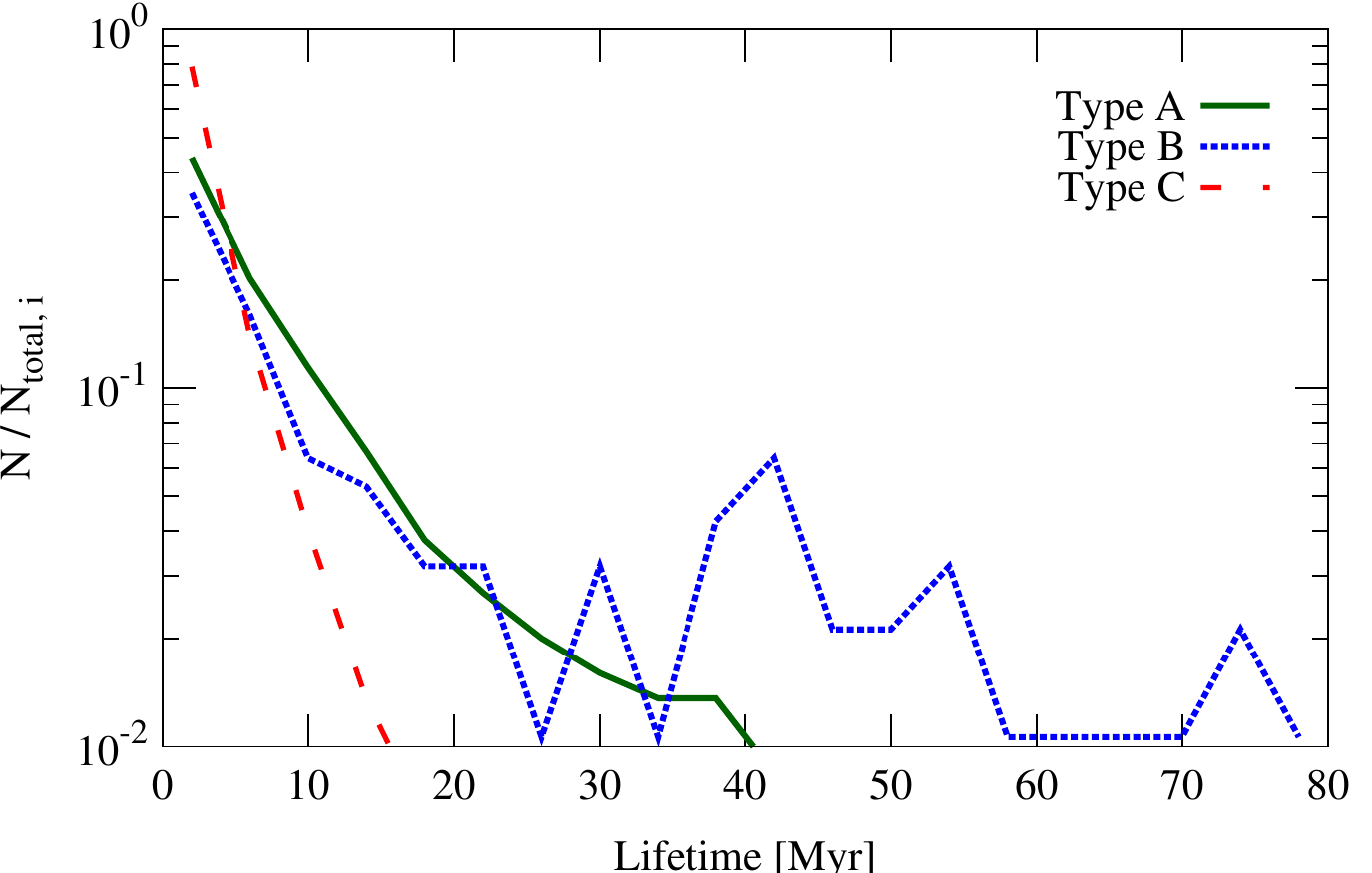}}
	\subfigure{
	\includegraphics[width=75mm]{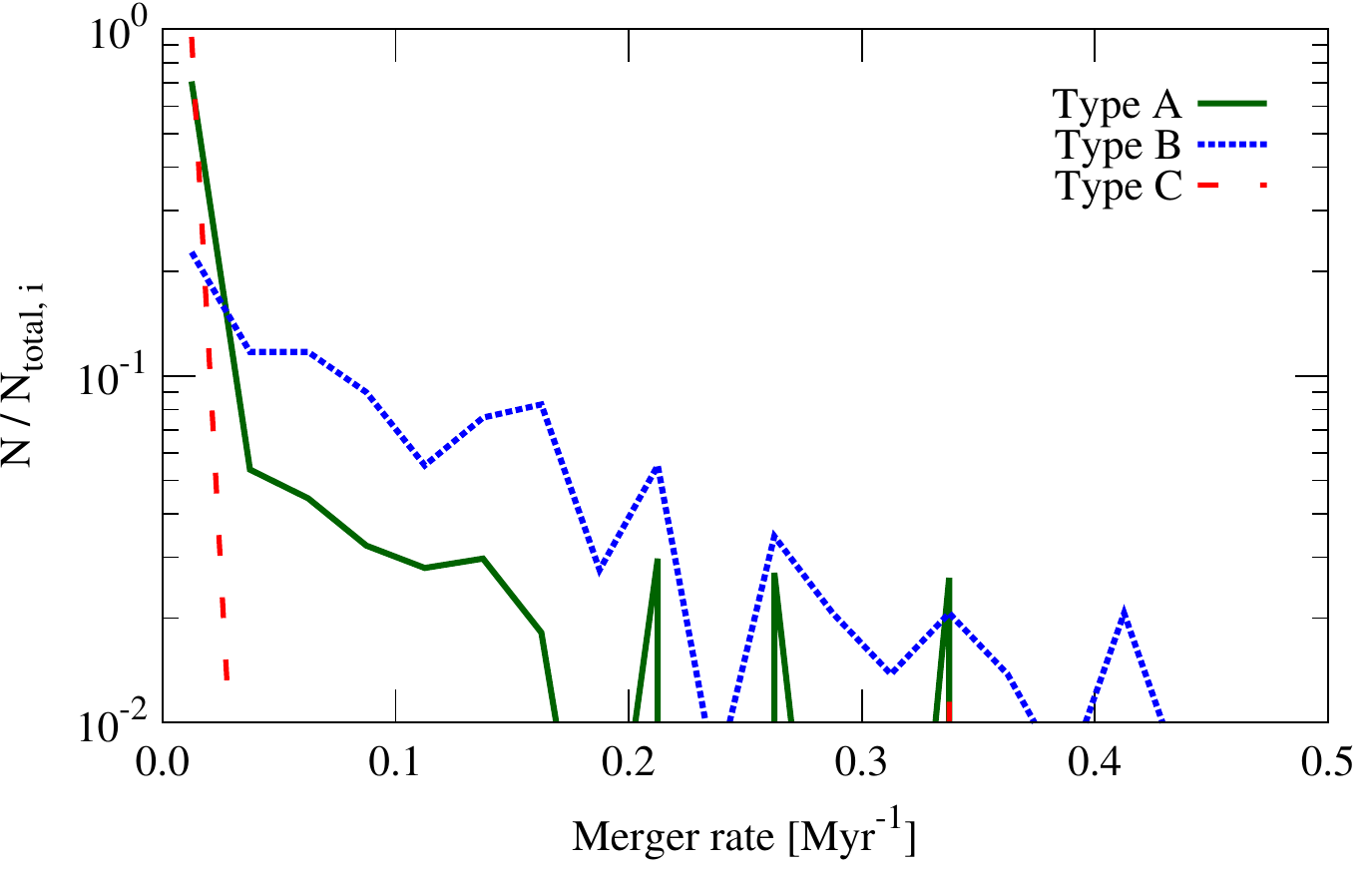}}
	\caption{Normalised distribution of cloud lifetime (top) and the merger rate per Myr of clouds that are born between $t = 200$ - 280\,Myr. }
	\label{lifetime and merger rate}
\end{center}
\end{figure}

Whether there is time during the cloud's life for such a orientation change is considered in Figure~\ref{lifetime and merger rate} where the cloud lifetime distribution (top) and the merger rate distribution are plotted for each of the cloud types. Since the initial fragmentation of the disc is not a realistic environment for the cloud, only clouds that form between $t = 200 - 280$\,Myrs are included in the distributions. Although our model does not have stellar feedback which is believed to aid cloud destruction, clouds can die in our simulation through merger events, tidal stripping leading to dissipation or simply dissipation due to the cloud being unbound and perturbed by nearby structures. 

All clouds, regardless of type, have a typical lifetime of less than 20\,Myr. While the age of GMCs is a heavily debated subject, this agrees well with estimates that suggest clouds live 1-2 dynamical times with ages in the range 5-30\,Myr \citep{Blitz2007, Kawamura2009, Miura2012}. The fact we get such good agreement with observational estimates of cloud lifetimes without any feedback processes is notable; this could imply that feedback has a small impact on the majority of the cloud's evolution. A possible reason for why this could be was noted by \citet{Renaud2013}, who found that in simulations of the Milky Way, the stars have moved out of the gas cloud before they go supernovae, resulting in minimal impact on their host GMC.

While the typical value for cloud lifetime is shorter than 10\,Myr, the range in the distribution differs greatly between the cloud types. {\it Type A} clouds agree most closely with the observed lifetimes, ranging up to 50\,Myr, with 95\% living for less than 20\,Myr. By contrast, {\it Type B} consist of far more long lived clouds, with over 20\% living longer than 40\,Myr. Note that the true maximum lifetime of {\it Type B} clouds is unknown, since they extend up to 80\,Myr, the maximum possible lifetime in our analysis period. We will see in the next section that this cloud type is very hard to destroy in our simulation. The smallest range of lifetimes is for {\it Type C} clouds, the vast majority of which have a lifetime of only a few Myr. These clouds are therefore low density, transient objects that die in a short period of time. 

\begin{figure*}
\begin{center}
	\subfigure{
	\includegraphics[width=8.0cm,bb=0 0 432 605]{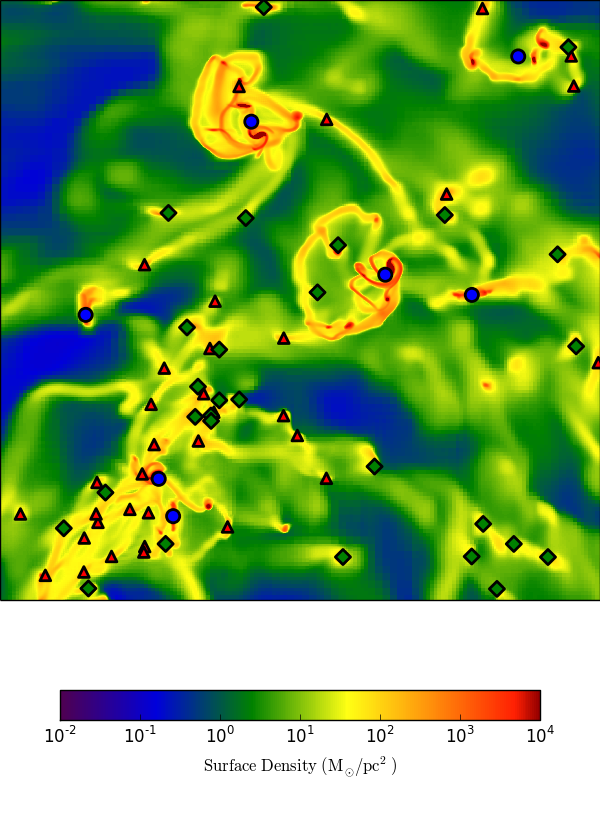}}
	\hspace{20pt}
	\subfigure{
	\includegraphics[width=8.0cm,bb=0 0 432 605]{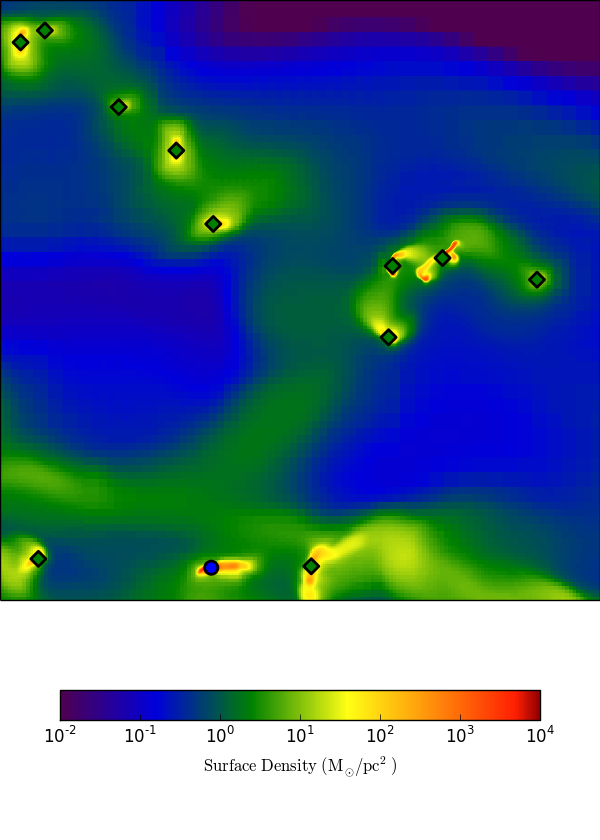}}
	\caption{2\,kpc gas surface density images of regions in the bar 1.5\,kpc from the galactic centre (left) and disc, 8\,kpc from the galactic centre. The position of these two sections is shown on Figure~\ref{Galactic disc with clouds}. Markers show the location of the three different cloud types. Green diamonds label {\it Type A} clouds, blue circles mark {\it Type B} and red triangles are {\it Type C}.}
	\label{figure of three clouds}
\end{center}
\end{figure*}

A measure of the interaction between the clouds during their lifetime can be seen deduced from the lower panel in Figure~\ref{lifetime and merger rate}, which plots the merger rate between clouds. As discussed in section~\ref{sec:numerics_cloud}, a merger is defined where a single cloud appears close to the predicted position of two or more clouds that existed 1\,Myr previously. It is a lower estimate of the true interaction rate, since it does not include tidal shredding where two identifiable objects exist at the end of the encounter. 

The massive {\it Type B} clouds have the highest merger rate, extending up to 1 merger every 2-3\,Myr. This means that these clouds undergo many mergers during their lifetime, accounting for their large mass and size. The transient {\it Type C} clouds have the lowest merger rate, in keeping with their very short lifetimes. Shortly after their birth, they either merge or their low density causes them to dissipate. 

{\it Type A} clouds also experience mergers (although less than for {\it Type B}), with the majority of clouds experiencing 1-2 mergers during their lifetime. This likely accounts for the range in the angular momentum angle distribution, $\theta$, see in Figure~\ref{Larson's law of three type}(d), where clouds were found predominately between  $\theta \sim 0$ and $90^\circ$. {\it Type A} clouds are therefore born prograde but gain the higher $\theta$ value through cloud collisions. This is not true for {\it Type B} who, while undergoing many mergers, are too massive to have their angular momentum greatly affected. The transient {\it Type C}s appear to have no preferred direction and do not live long enough to undergo mergers, suggesting they can be born at any orientation. 

\subsubsection{The effect of galactic environment on cloud formation}

The final confirmation of the origin of the properties of the three cloud types comes from visual inspection of the disc. Figure~\ref{figure of three clouds} shows 2\,kpc patches of the gas surface density taken in the bar region and disc region. The image is overlaid with markers showing the centre of mass of the clouds, with {\it Type A} clouds denoted by green diamonds, {\it Type B} clouds shown with blue circles and {\it Type C} with red triangles. 

In the left-hand image, we see a section of the bar with a large number of {\it Type A}, {\it B} and {\it C} clouds visible. The massive {\it Type B} are the most obvious, forming giant molecular associations that drag in surrounding gas and clouds to produce the high merger rate. The bigger mergers are with {\it Type A} clouds that form the gaseous spiral tidal tails as they pass by or merge with the GMA {\it Type B}s. In these tidal tails sit the {\it Type C} clouds. These objects form briefly in the dense filaments, but are swiftly swallowed or dispersed by the plethora of interaction around them. 

This myriad of action occurs in the bar region due to the high density of material gathered by the stellar bar potential and the constrained elliptical motions, bringing clouds into regular contact with one another. These interactions increase the number of tidal tail filaments formed, birthing a high number of {\it Type C} clouds. Without a source of destruction, {\it Type B} clouds continue to collect matter and grow for an indefinite period. While their size would make them difficult to destroy with internal feedback, including such a mechanism would likely reduce the maximum size the GMAs reached. 

On the other hand, the disc region shows a far more quiescent environment. The clouds are more widely spaced, leading to fewer interactions which slows the creation of the massive GMA {\it Type B}s, explaining the 6\% population shown in Table~\ref{table:cloud_percentage}. The vast majority of the clouds are {\it Type A} which, with less interactions, lack filaments to produce the transient {\it Type C} population. This low merger rate is due to the lack of the grand design potential to gather gas and gravitationally confine it to the region. 

The spiral region forms a mid-point between the inactivity in the disc region and the intense interactions in the bar. It therefore as an intermediate population of clouds in each of the three types. 

In our model, therefore, gas typically fragments into a {\it Type A} GMC. These have properties in good agreements with the typical average observed in many galaxies. However, interactions between clouds produce the tails of these properties in the form of {\it Type B} and {\it Type C} clouds. The role of the galactic environment is therefore to drive these interactions, causing these additional populations of clouds to form. 

\subsection{Star formation}
\label{sec:stars}

Despite not including an active star formation recipe in our simulation, we can estimate the galaxy's star formation rate based on the properties of the gas. Even while restricting star formation to the inner regions of GMCs, the exact conditions that control when a star is born remains an area of active research. We therefore consider three different star formation models which make different assumptions about the parameters controlling the star formation rate. 

\subsubsection{Standard star formation model}
\label{sec:stars_standard}

Since the actual collapse of gas into a star cluster is still below our resolution limit, it is reasonable to assume that all GMCs contain Jeans unstable regions that will collapse to form stars. This first star formation model is the simplest product of this assumption, with the star formation rate depending only on the cloud mass and its free-fall time,

\begin{equation}
{\rm SFR}_{\rm c} = \epsilon\ \frac{M_{\rm c}}{t_{\rm ff, c}} = \epsilon\ \frac{M_{\rm c}}{\sqrt{\frac{3\pi}{32G\overline\rho_{\rm c}}}}
\end{equation}

\noindent where $\epsilon = 0.014$, the star formation efficiency per free-fall time \citet{KrumholzMcKee2005}, and $\overline\rho_{\rm cloud}$ is the mean density of the cloud. 

\begin{figure}
\begin{center}
	\subfigure{
	\includegraphics[width=80mm]{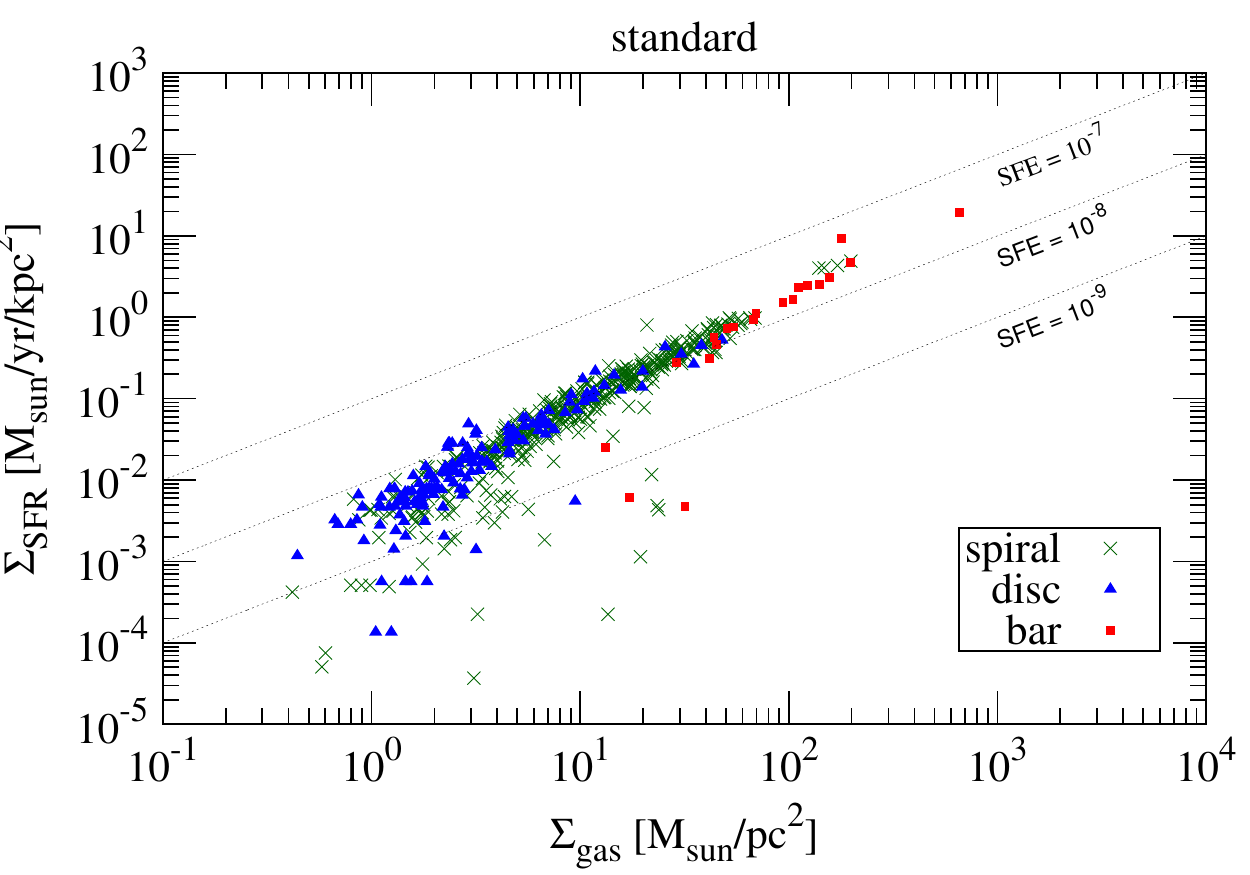}}
	\subfigure{
	\includegraphics[width=80mm]{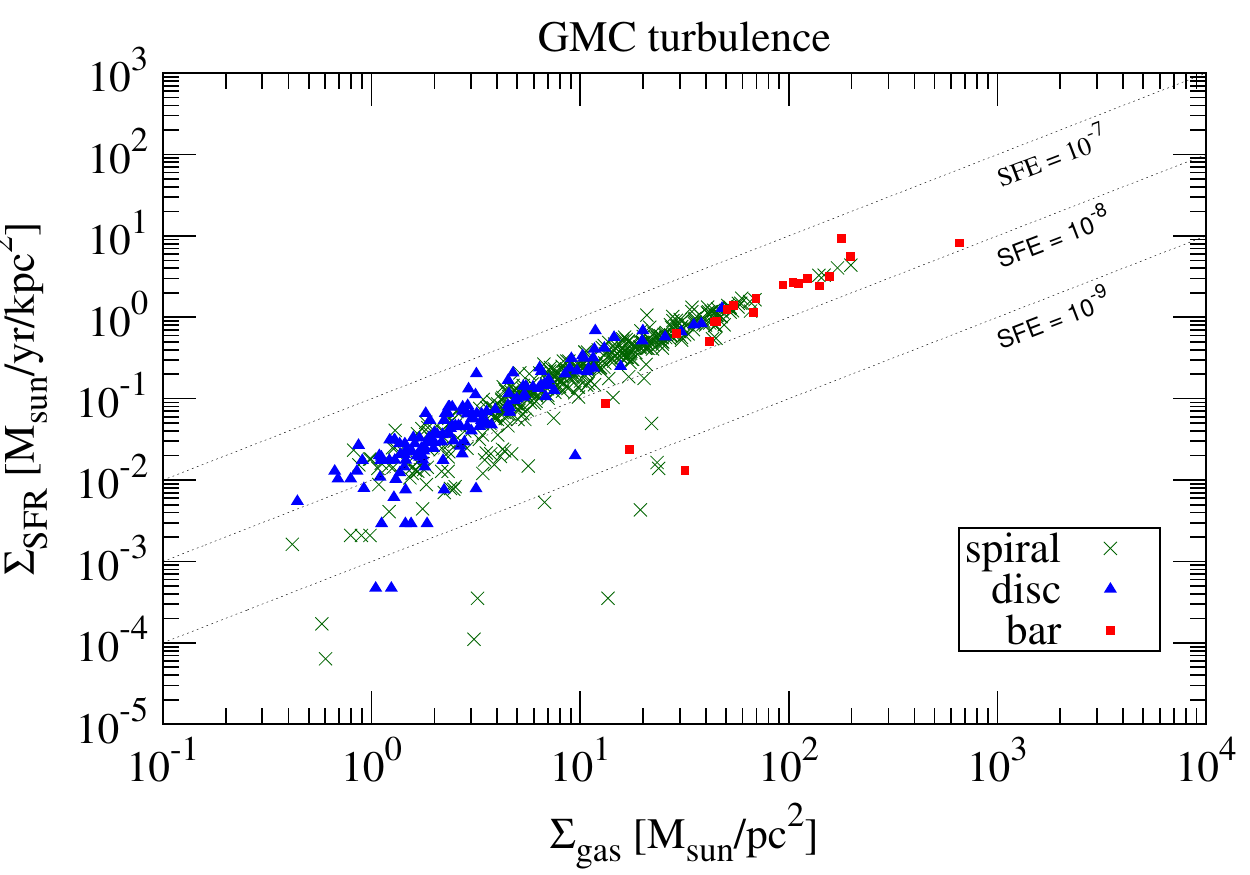}}
	\subfigure{
	\includegraphics[width=80mm]{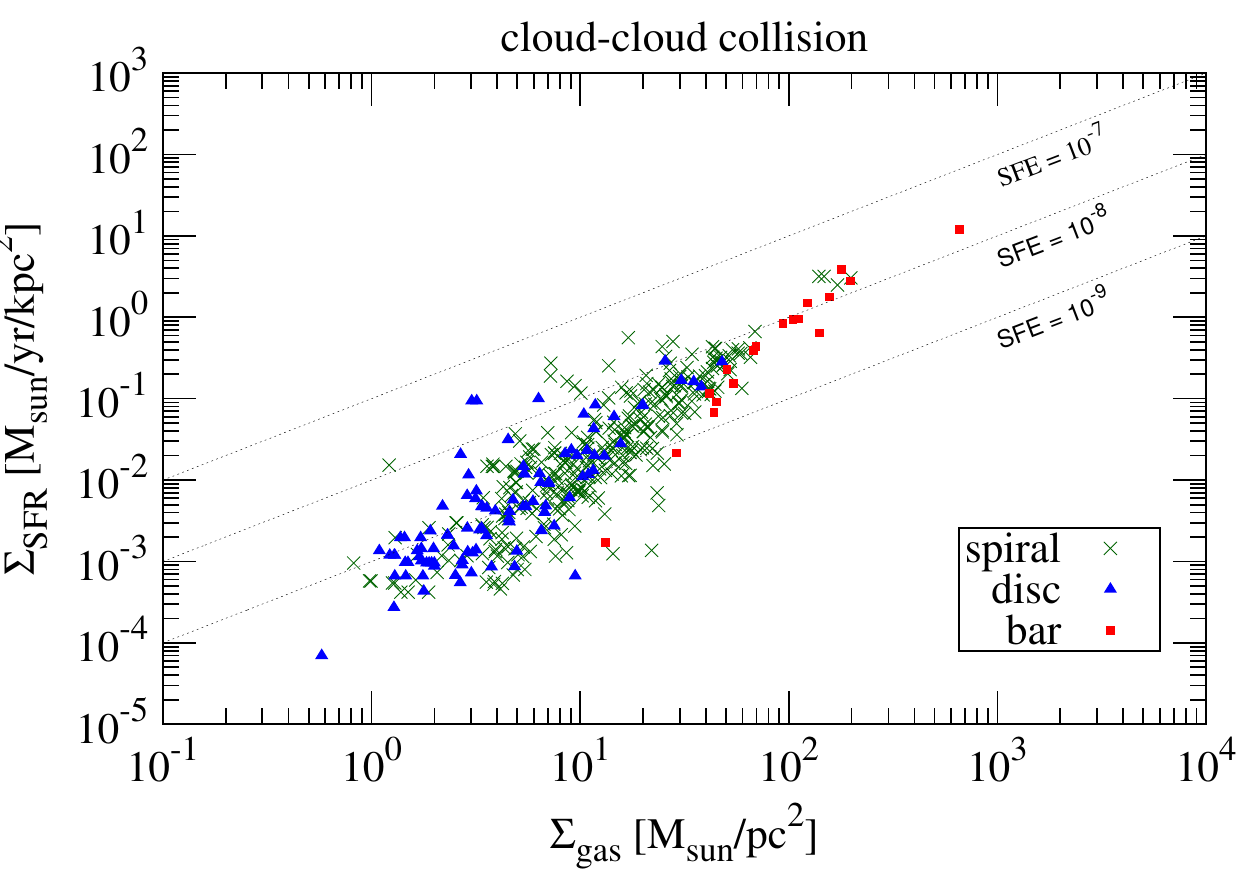}}
	\caption{The Kennicutt-Schmidt relation for three different star formation models. The surface area is calculated in the x-y plane (face-on disc) and the data is averaged over a cylindrical region with radius 500\,pc and 5\,kpc height.  Top panel shows the results from our {\it standard star formation model}, section~\ref{sec:stars_standard}, where the star formation depends only on the mass and free-fall time of the cloud. Middle panel shows the \citet{KrumholzMcKee2005} {\it GMC turbulence model}, section~\ref{sec:stars_turbulence}, where the turbulent motion of the GMCs is considered. The bottom panel is the {\it cloud-cloud collision model} in section~\ref{sec:stars_collision}, proposed by \citet{Tan2000}, where star formation is regulated by GMC interactions. The black dotted lines show constant star formation efficiency: $\rm SFE = 10^{-7},\  10^{-8},\ 10^{-9}\ [yr^{-1}]$.}
	\label{KS Law}
\end{center}
\end{figure}

The top panel in Figure~\ref{KS Law} shows the Kennicutt-Schmidt relation (Equation~\ref{eq:ks}) using this model. Each point on the graph marks the value for a cylindrical region with radius 500\,pc in the galactic plane. This region size was chosen to be comparable to the observational data in nearby galaxies, which finds a near linear relationship between the gas surface density, $\Sigma_{\rm gas}$, and the surface star formation density, $\Sigma_{\rm SFR}$, for densities higher than 10\,M$_\odot$/pc$^2$ \citep{Bigiel2008}.  Since multiple GMCs exist within these regions, the star formation rate is calculated as the sum for each cloud within the cylinder.

In agreement with observations, the gas and star formation rate surface densities follow a nearly linear trend in all three galactic environments. There is a small deviation towards a steeper gradient at densities below $\sim 10$\,M$_\odot$/pc$^2$ and also an increased scatter due to the smaller number of clouds found within our measured region. Note this change has a different origin to the observational results, where the break at the same threshold is due to the transition between atomic and molecular hydrogen. In our simulations, only atomic gas is followed, so we do not expect to observe such a split. It is more likely that clouds in low density regions are less centrally concentrated, due to fewer interactions resulting in tidal stripping. 

The overall star formation rate is approximately a factor of 10 higher than that observed. Such elevation in simulations is usually put down to the absence of localised feedback, which would be expected to dissipate the densest parts of the cloud and thereby reduce the star formation rate regardless of whether the cloud itself was also destroyed \citep{Tasker2011}. In our case, we also lack an actual star formation recipe, meaning that our densest gas is allowed to accumulate inside the cloud without being removed to create a star particle. This adds to the cloud mass and raises the expected star formation rate.  

While there is an overall agreement in the gradient, the difference in the star formation rate in the bar, spiral and disc is also apparent. The bar region contains the highest density of clouds, as well as a larger fraction of the massive {\it Type B} clouds. This produces the upper end of the gas and star formation rate surface densities. The sparser, smaller clouds of the disc region result in correspondingly lower values and the spiral region sits in between. 

\subsubsection{GMC turbulence star formation model}
\label{sec:stars_turbulence}

We can compare the results of the straightforward free-fall collapse with a star formation model that also considers the importance of turbulent motions within the GMCs. Proposed by \citet{KrumholzMcKee2005}, this power-law model assumes that the clouds are supersonically turbulent, producing a log-normal density distribution. By demanding that gas collapses when the gravitational energy exceeds the turbulent energy within a cloud, they find the modified relation,

\begin{equation}
{\rm SFR}_{\rm c} = \epsilon\ \left(\frac{\alpha_{\rm vir}}{1.3}\right)^{-0.68}\left(\frac{\cal{M}}{100}\right)^{-0.32}\frac{M_{\rm c}}{\sqrt{\frac{3\pi}{32G\overline\rho_{\rm c}}}}
\label{eq:stars_turbulence}
\end{equation}

\noindent where $\epsilon = 0.014$ is again the star formation efficiency per free-fall time and $\alpha_{\rm vir}$ is the virial parameter as defined in Equation~\ref{eq:virial}. $\cal{M}$ is the Mach number, defined as the ratio between the cloud's 1D velocity dispersion and sound speed, $\cal{M}$ $\equiv v_c/c_s$.  

The results from this model are shown in the middle panel of Figure~\ref{KS Law}. The surface area and surface star formation rate was calculated as before over a cylindrical region with radius 500\,pc. The addition of turbulence regulation to the star formation rate makes only a small difference to the result, due to the addition terms in Equation~\ref{eq:stars_turbulence} typically multiplying the result by only a factor of 1-2. Clouds in the low density region are affected the most, since these correspond to disc clouds with a lower velocity dispersion. This produces an overall tighter relation throughout the disc, with the gradient of unity.  

The overall trends between the three environments remain unchanged from those observed in section~\ref{sec:stars_standard}. However, the tightening in the Kennicutt-Schmidt relation when environmentally dependent properties such as $\alpha_{\rm vir}$ and the velocity dispersion are included emphasises the importance of the galactic structure in GMC evolution. 

\subsubsection{Cloud-cloud collision star formation model}
\label{sec:stars_collision}

Our final model moves away from a Jeans unstable cloud to a scheme motivated by triggered star formation. In his paper, \citet{Tan2000} suggested that star formation could be initiated by collisions between GMCs, providing a natural connection between the local star formation collapse and the global environment of the disc. Using this method, the star formation rate per unit area becomes,

\begin{equation}
\Sigma_{\rm SFR} = \frac{\epsilon f_{\rm sf} N_{\rm A} M_{\rm c}}{t_{\rm coll}}
\end{equation}

\noindent where $\epsilon = 0.2$, that is the total star formation efficiency, $f_{\rm sf}$ is the fraction of cloud collisions which lead to star formation, $N_{\rm A}$ is the surface number density of clouds and $t_{\rm coll}$ is the time between collision events. The exact value of $f_{\rm sf}$ is not known, so we select $f_{\rm sf} = 0.5$, corresponding to 50\% of collisions leading to star formation.

The Kennicutt-Schmidt relation from using this triggered star formation scheme is plotted in the bottom panel of Figure~\ref{KS Law}. As with the previous two models, each point represents an average within a 500\,pc region, with the star formation rate calculated from the values of $N_{\rm A}$ and $t_{\rm coll}$ within this volume. 

The gradient of the Kennicutt-Schmidt relation is now steeper than unity (index, $N \sim 2$), with a significantly lower star formation rate in the lower density regions. On the one hand, this difference is not surprising, since high density gas leads to many more cloud collisions. However, it is worth noting that taking a constant value for $f_{\rm sf}$ may skew this result; in the bar region, many collisions involve the small {\it Type C} clouds which are unlikely to trigger significant star formation. A value that reflected the differences between cloud types in merger events would lower the surface star formation rate in the bar region more than in the disc, where the majority of cloud mergers are between {\it Type A} clouds and likely more productive. Such a change would support M83 observations by \citet{Hirota2013}, who finds that the star formation rate is elevated in the bar and spiral regions compared to the inter-arm gas, but that the bar region shows a lower star formation rate than the spiral arms. This result is found despite the molecular gas surface density being approximately constant through both the bar and spiral. If collisions drive star formation but are less productive in the bar, this would explain such a result. Additionally, observations of cloud-cloud collisions that result in star formation activity typically have a high relative velocity of 10-20\,km/s \citep{Furukawa2009, Ohama2010}, a result supported in simulations by Takahira et al (ApJ, in prep.) who found that both cloud size and relative velocity were important factors in the formation of stellar cores. Therefore, while \citet{Tan2000}'s model is successful in producing a clear correlation between the surface star formation rate and gas surface density, a more detailed scheme which takes into account cloud differences might yield an even stronger result.

\begin{table}
\begin{center}
\begin{tabular}{|l|c|c|c|} \hline
	& standard & turbulence & cloud collision \\ \hline
	$\rm SFE_{bar}/SFE_{disc}$ & 2.60 & 1.05 & 4.34 \\ \hline
	$\rm SFE_{spiral}/SFE_{disc}$ & 1.42 & 1.05 & 1.73 \\ \hline
\end{tabular} 
\caption{Ration of the star formation efficiency (SFE) in the bar and spiral environments compared to that in the disc. Results from the three star formation models discussed in section~\ref{sec:stars_collision} are shown.\label{table_sfe}}
\end{center}
\end{table}

The star formation efficiency (${\rm SFE} = \Sigma_{\rm gas}/\Sigma_{\rm SFR}$) from each of these three methods is compared in Table~\ref{table_sfe} with respect to the value in the disc. The SFE that is based simply on gas density (standard model, section~\ref{sec:stars_standard}) increases by a factor of 1.5 in the spiral region and 2.6 in the disc. On the other hand, the interaction based SFE (cloud collision, section~\ref{sec:stars_collision}) shows an increase of 1.73 in the spiral and 4.34 in the disc. This simple calculation dramatically shows the main difference between the bar, spiral and disc environments: while the gas density is higher in the bar and spiral and plays a role in shaping the cloud properties, a more major difference is the frequency of the cloud interactions. Notably, when turbulence is included as with the \citet{KrumholzMcKee2005} model (section~\ref{sec:stars_turbulence}), there is no difference in SFE between regions. However, it is unlikely we are resolving the full effect of the cloud interactions on the cloud's internal structure, which is likely to lead to higher compressible turbulent motion.

Since the star formation rate is too high compared to observations, the SFE is likewise above the observed values. \citet{Hirota2013} finds a SFE for M83 between 0.2 - 2\,Gyr$^{-1}$ for the bar and spiral region. The standard star formation model and turbulent model in sections~\ref{sec:stars_standard} and \ref{sec:stars_turbulence} have a SFE almost a factor of 10 too high (in agreement with their star formation rates), with absolute values of 6.1, 8.7 and 16.0\,Gyr$^{-1}$ for the disc, spiral and bar regions respectively in the standard model and roughly 20\,Gyr$^{-1}$ for the turbulent model. The cloud collision model agrees well with \citet{Hirota2013}'s observations in the spiral, with a SFE = 2.7\,Gyr$^{-1}$. However, it is markedly too high in the bar region at 4.34\,Gyr$^{-1}$, due to the reasons discussed above regarding the likely productivity of small cloud collisions.

\section{Numerical dependences}
\subsection{The effect of resolution}
\label{sec:resolution}

\begin{figure*}
\begin{center}
\begin{tabular}{cc}
	\subfigure{
	\includegraphics[width=70mm]{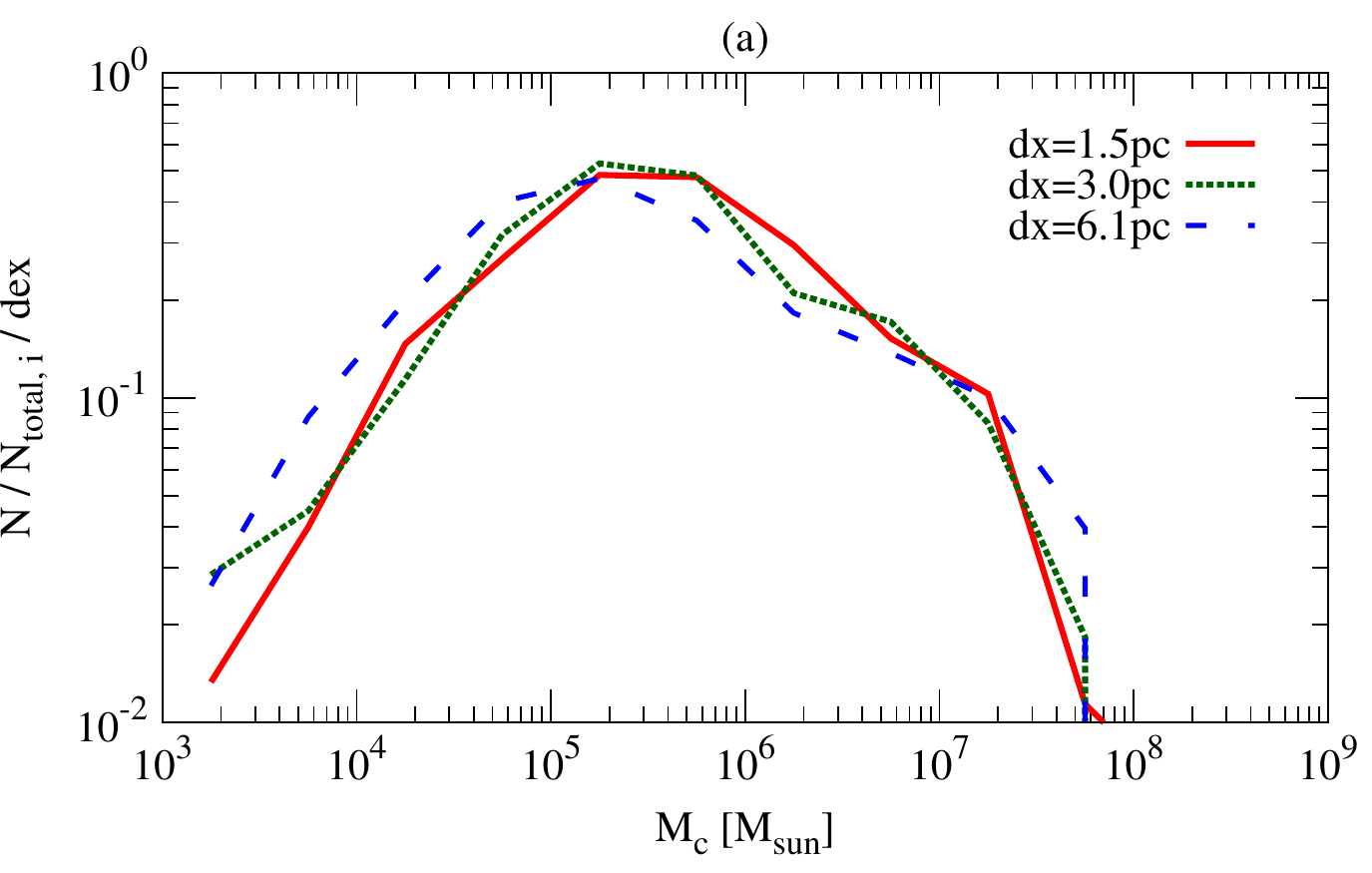}}
	\subfigure{
	\includegraphics[width=70mm]{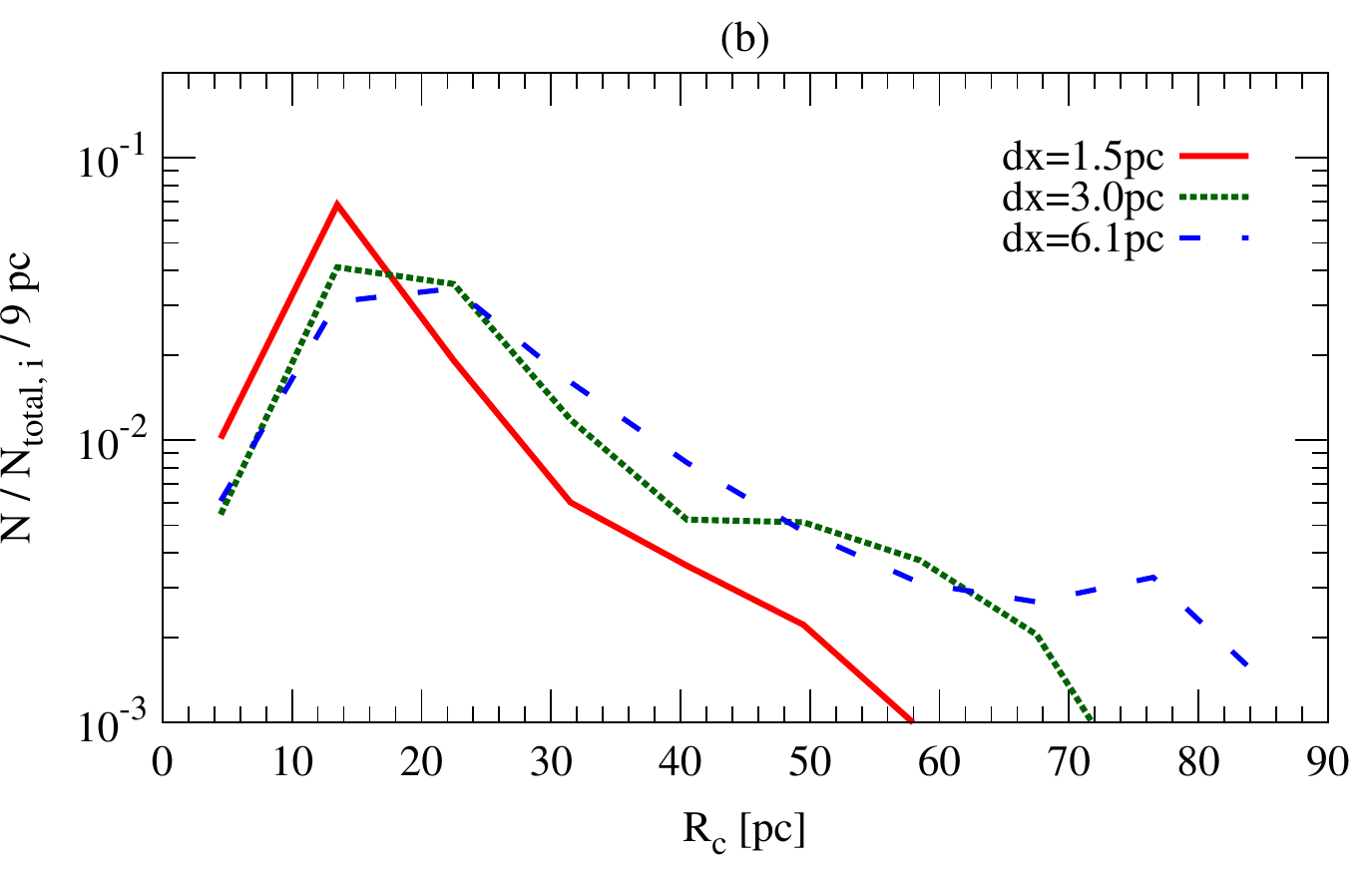}}
\end{tabular}
\begin{tabular}{cc}
	\subfigure{
	\includegraphics[width=70mm]{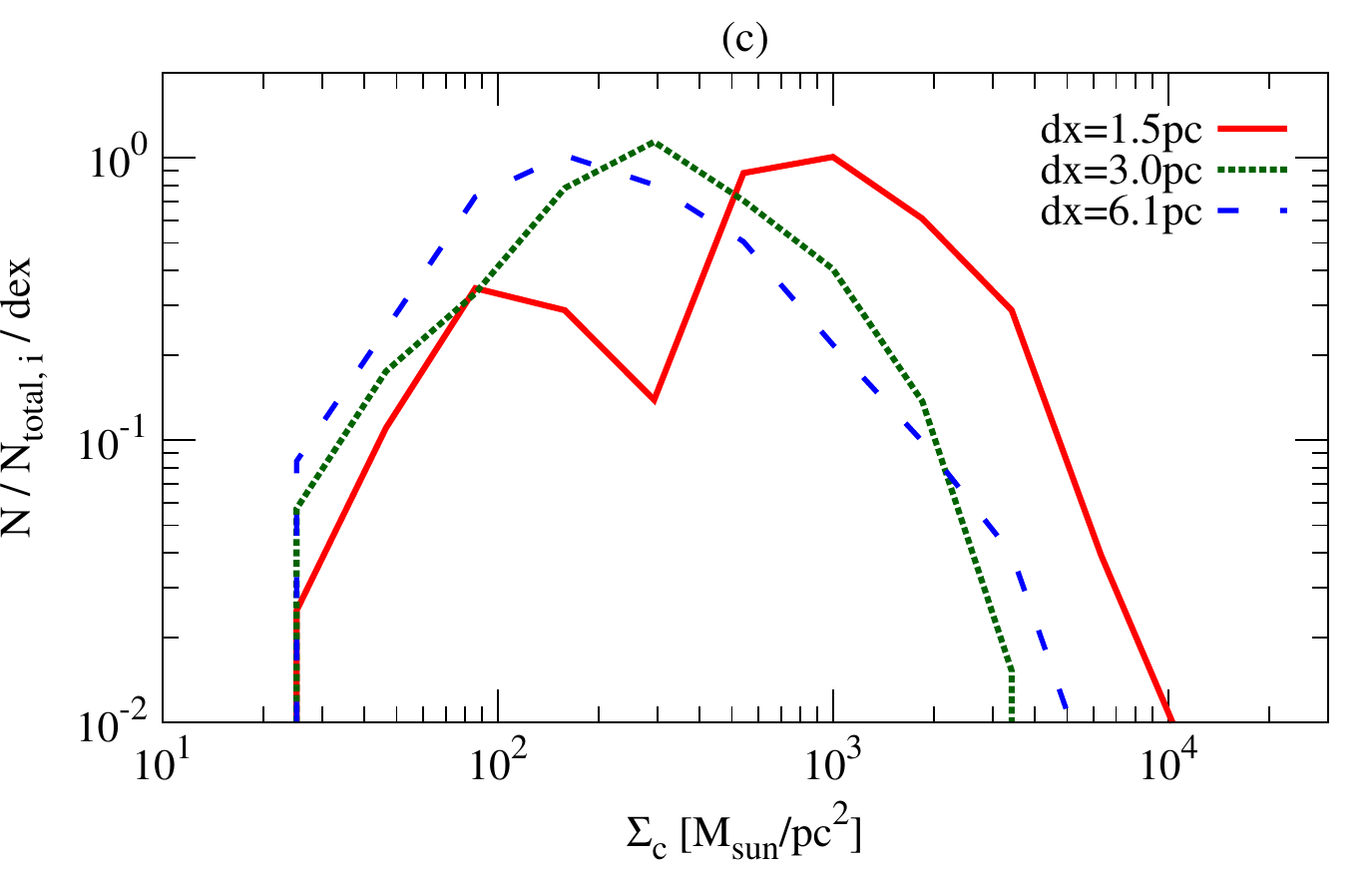}}
	\subfigure{
	\includegraphics[width=70mm]{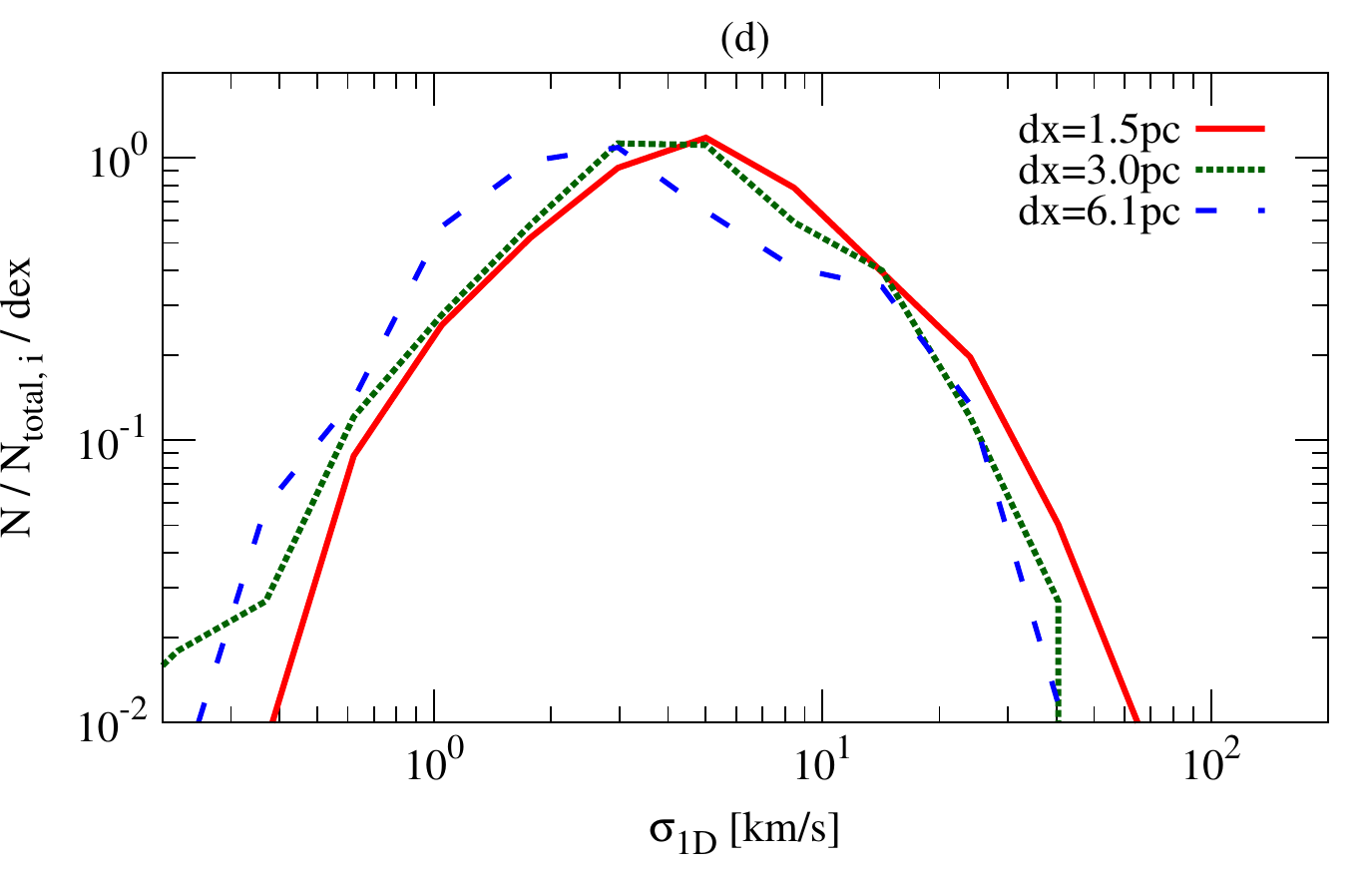}}
	\end{tabular}
\begin{tabular}{cc}
	\subfigure{
	\includegraphics[width=70mm]{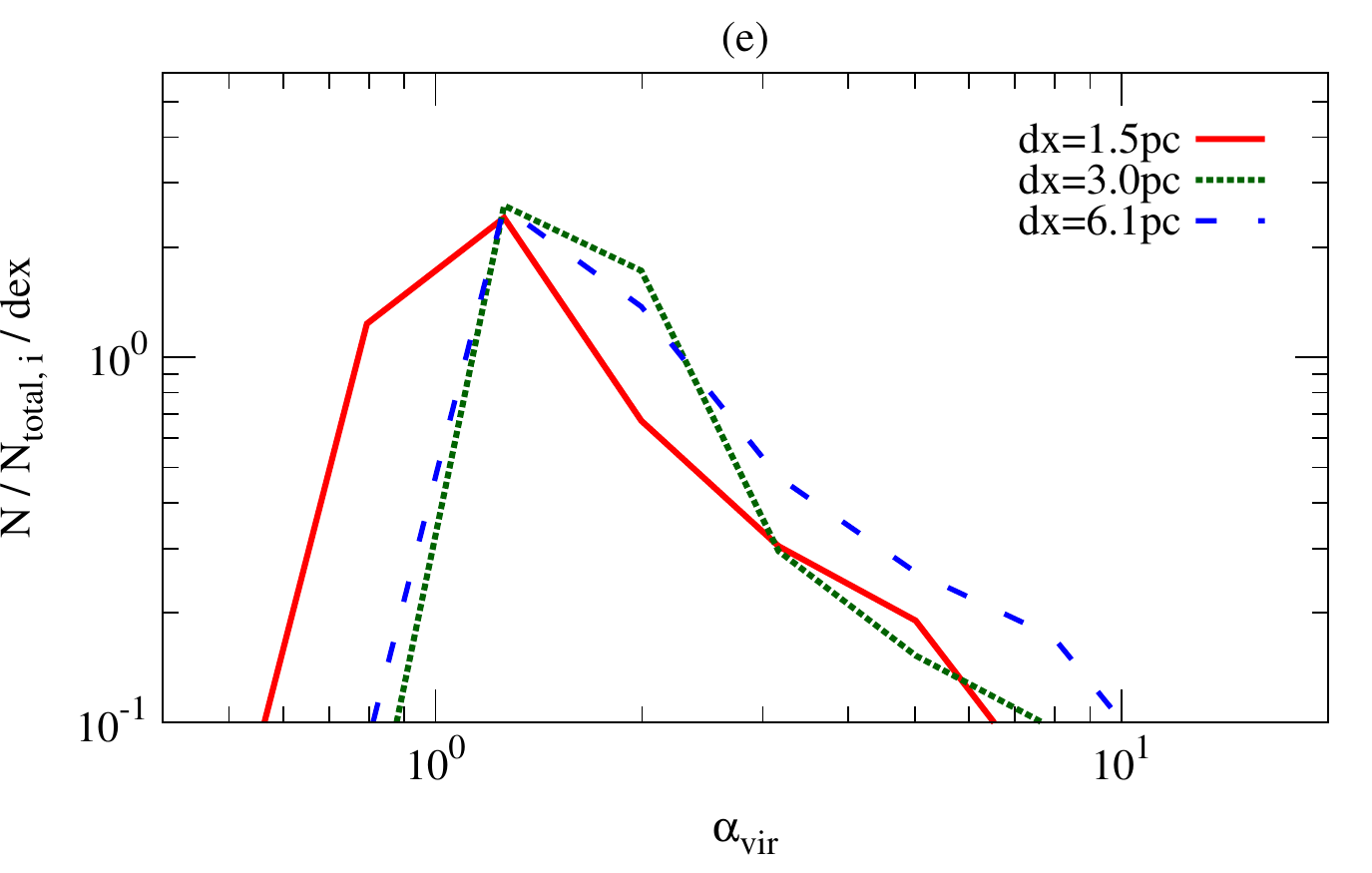}}
	\subfigure{
	\includegraphics[width=70mm]{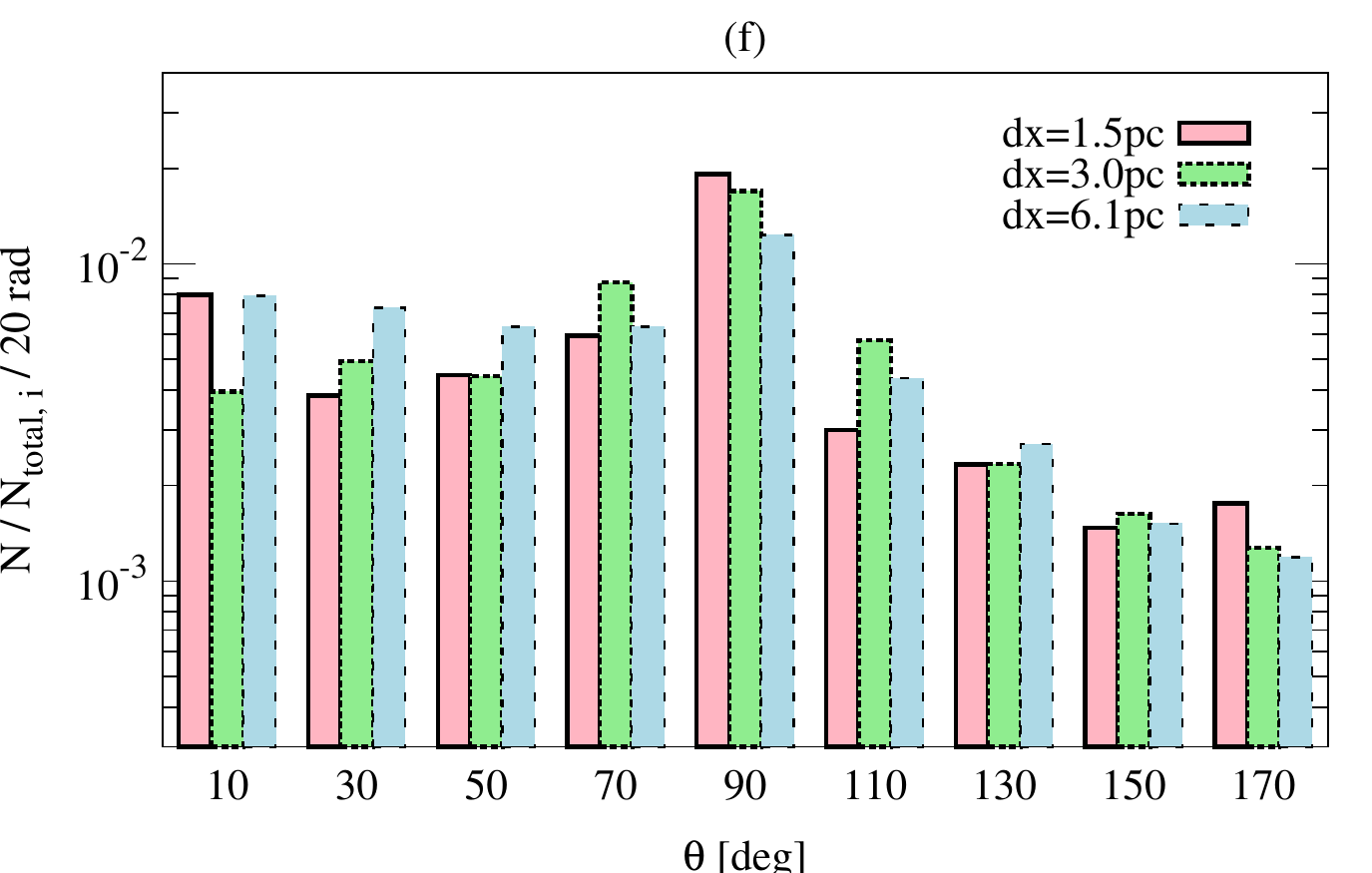}}
\end{tabular}
\caption{Comparison of the cloud property distributions at three different resolutions. For each case, the smallest cell size in the simulation is $\Delta x = 1.5$\,pc (red solid line), $\Delta x = 3.0$\,pc (green dotted line) and $\Delta x = 6.1$\,pc (blue dashed line). The distributions shown are (a) cloud mass, (b) average cloud radius, (c) cloud surface density, (d) cloud 1D velocity dispersion, (e) the virial parameter and (d) the angle of the cloud's angular momentum vector with respect to that of the disc. All properties are calculated as they were for Figure~\ref{distribution of cloud properties}.}
\label{fig:resolution_properties}
\end{center}
\end{figure*}

\begin{figure*}
\begin{center}
\begin{tabular}{ccc}
	\hspace{-40pt}
	\subfigure{
	\includegraphics[width=80mm]{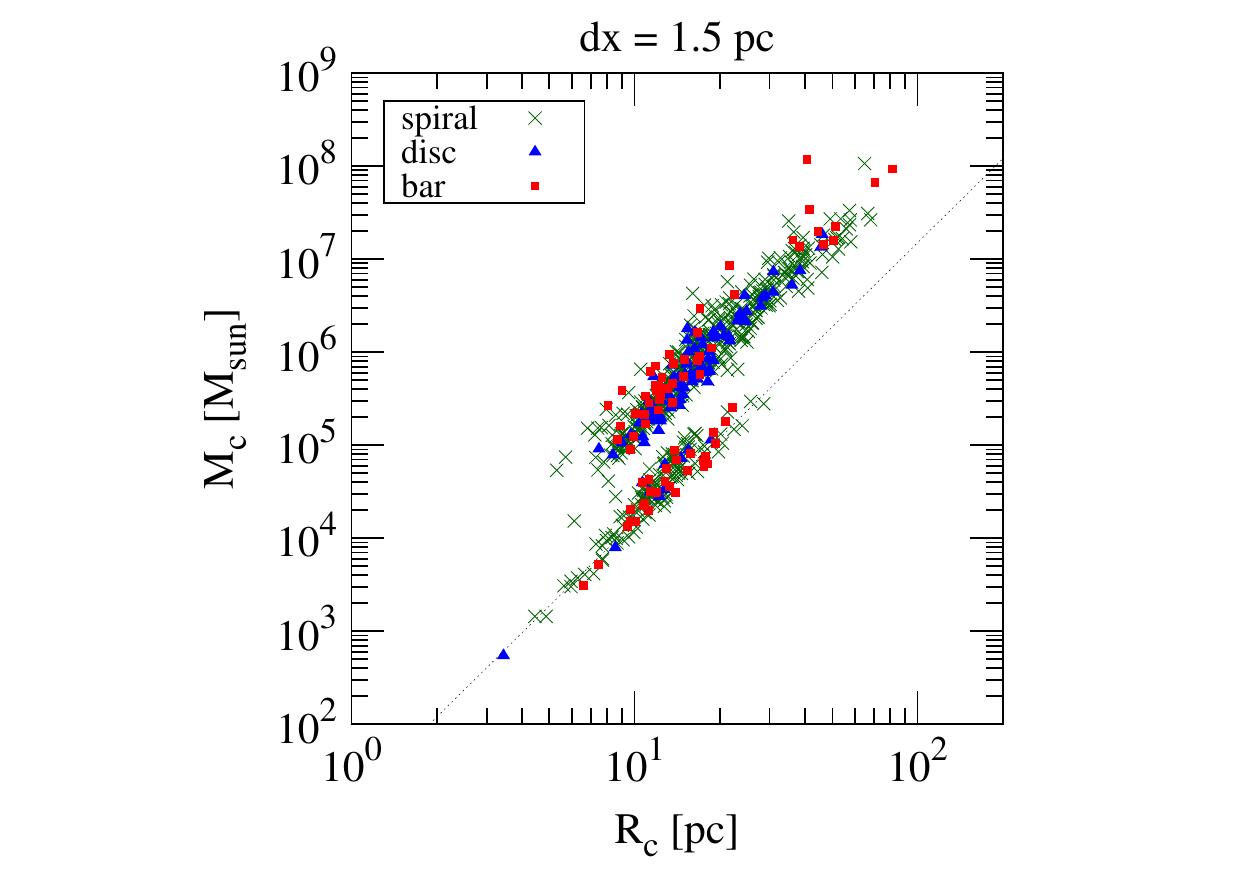}}
	\hspace{-70pt}
	\subfigure{
	\includegraphics[width=80mm]{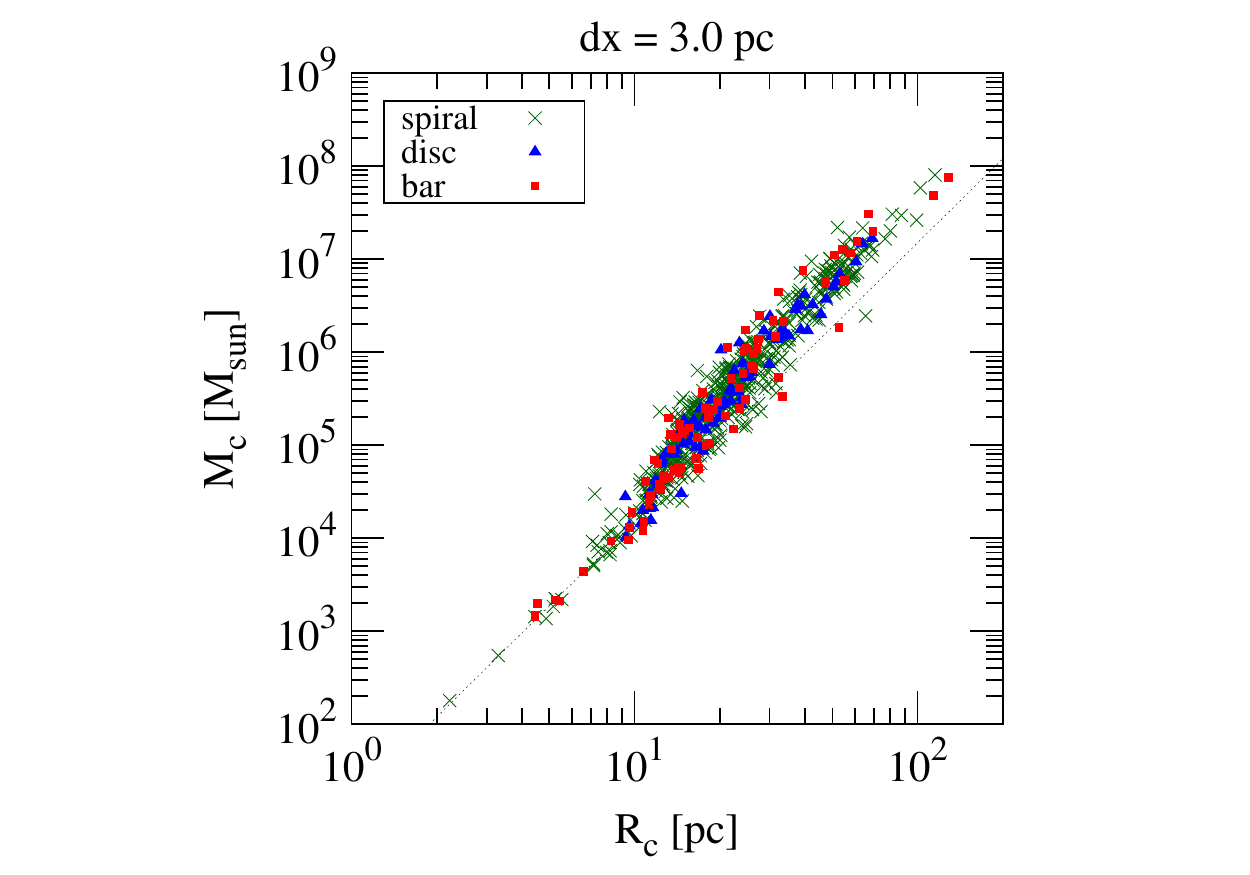}}
	\hspace{-70pt}
	\subfigure{
	\includegraphics[width=80mm]{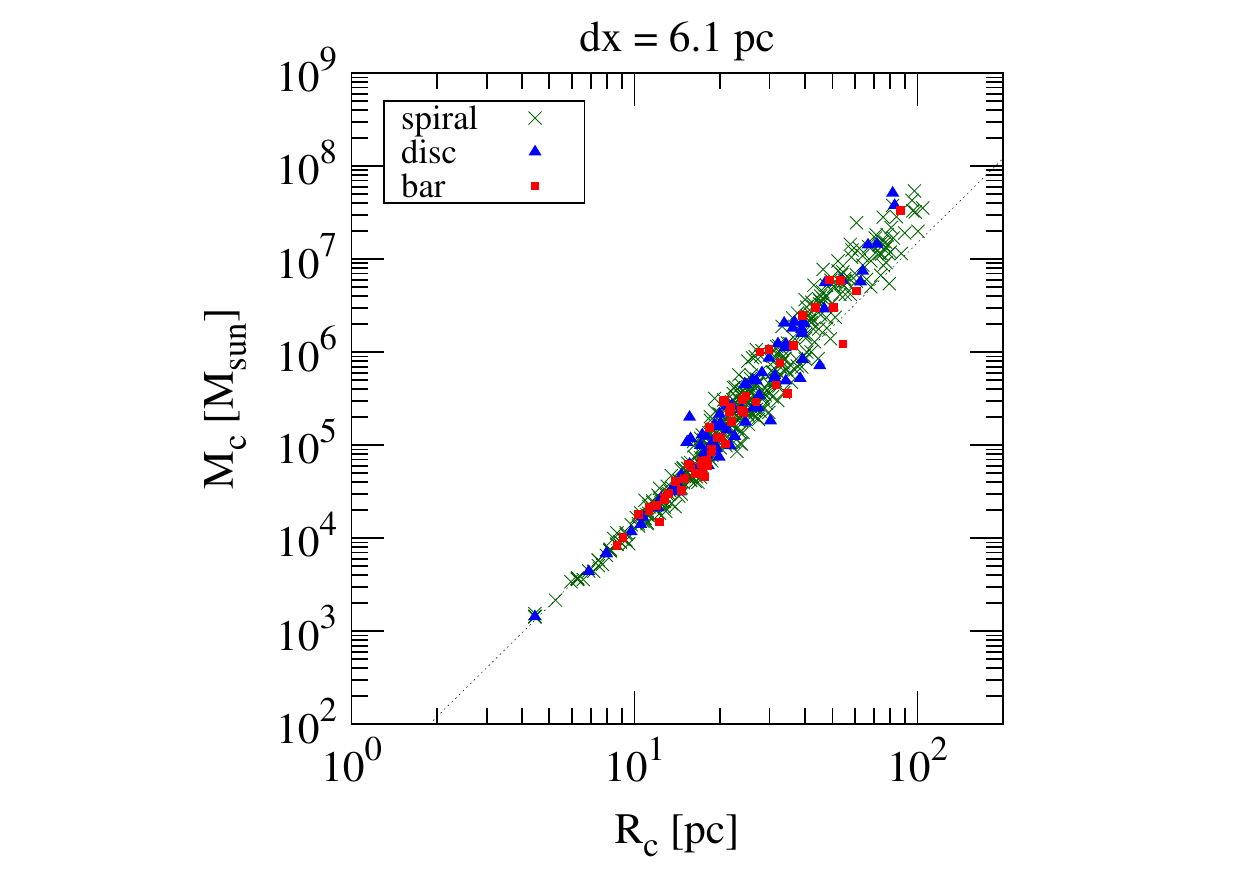}}
\end{tabular}
	\caption{Mass versus radius scaling relation for simulations performed at different resolutions. From left to right, the smallest cell in the simulation volume is $\Delta x = 1.5$\,pc, $\Delta x = 3.0$\,pc and $\Delta x = 6.1$\,pc. The markers designate the same galactic regions as in Figure~\ref{Larson's law}, with red squares showing bar clouds, blue triangles marking disc clouds and green `x' labelling spiral clouds. The dotted line is the fit ($M_{\rm c} = 15\times {R_{\rm c}}^3$) for {\it Type C} clouds which form the lower sequence at our highest resolution ($\Delta x = 1.5 \rm pc$).}
	\label{comparison of mass radius relation between different resolution}
\end{center}
\end{figure*}

The results presented in section~\ref{sec:results} show the existence of three types of GMCs: the regular {\it Type A} clouds, the {\it Type B} giant molecular associations and the transient {\it Type C} clouds. A key question therefore is whether these three different types can be observed in real galaxies. 

One of the controlling factors in both observational and simulation results is that of resolution. We therefore compared the cloud properties presented in section~\ref{sec:results} with those found in two simulations performed at lower resolution. The comparison of the cloud property distributions at the three resolution limits is shown in Figure~\ref{fig:resolution_properties}. The distributions are plotted for all the clouds in the galaxy with the three lines indicating the limiting resolution (smallest cell size in the volume) of the simulation. The red solid line is our main simulation, with a limiting resolution of $\Delta x = 1.5$\,pc. The green dashed line shows the results for a run with one less level of AMR, giving a limiting resolution of $\Delta x = 3.0$\,pc. The blue dashed line marks our lowest resolution simulation with $\Delta x = 6.1$\,pc. 

The cloud mass (Figure~\ref{fig:resolution_properties}(a)) and cloud velocity dispersion (Figure~\ref{fig:resolution_properties}(d)) distributions show very little difference between the three resolutions. This is true even at low masses, where the clouds become more difficult to resolve. However, there is a difference at low cloud radii, where our main highest resolution simulation produces a greater proportion of clouds with $3 < R_c < 15$\,pc. At lower resolutions, the clouds blend to become extended structures with radii out past 60\,pc. This has a very notable effect on the cloud surface density (Figure~\ref{fig:resolution_properties}(c)), where the second population of clouds with $\Sigma_c > 230$\,M$_\odot$/pc$^2$ is entirely missing at the two lower resolutions, removing the bimodality. The larger radii also impacts the virial parameter (Figure~\ref{fig:resolution_properties}(e)), with cloud typical value moving from $1.0\rightarrow 1.5$ for the two lower resolution cases. The effect on the angular momentum angle, $\theta$, in Figure~\ref{fig:resolution_properties}(f) is small overall, showing similar proportions of prograde and retrograde clouds at all resolutions. 

The removal of the bimodality in the surface density profile at lower resolutions can be seen clearly in the mass-radius relation. Figure~\ref{comparison of mass radius relation between different resolution} shows the same plot at the three different resolution limits. Only in our highest resolution case (left) is the upper and lower trend clearly visible. As we progress to lower resolution, the {\it Type A} and {\it Type B} clouds in the upper trend increase in radius, pushing their relation to the right on the plot. The result is a continuous sequence for all cloud types with a more uniform surface density around $\sim \Sigma_{c} = 200$\,M$_\odot$/pc$^2$. This effect could mean that observations are unable to differentiate between {\it Type A} and the transient {\it Type C} clouds unless they are at very high resolution. 

\subsection{Comparison between cloud identification methods}
\label{sec:cloud_comparison}

\begin{figure}
\begin{center}
	\subfigure{
	\includegraphics[width=75mm]{distribution_mass_method2.pdf}}
	\subfigure{
	\includegraphics[width=75mm]{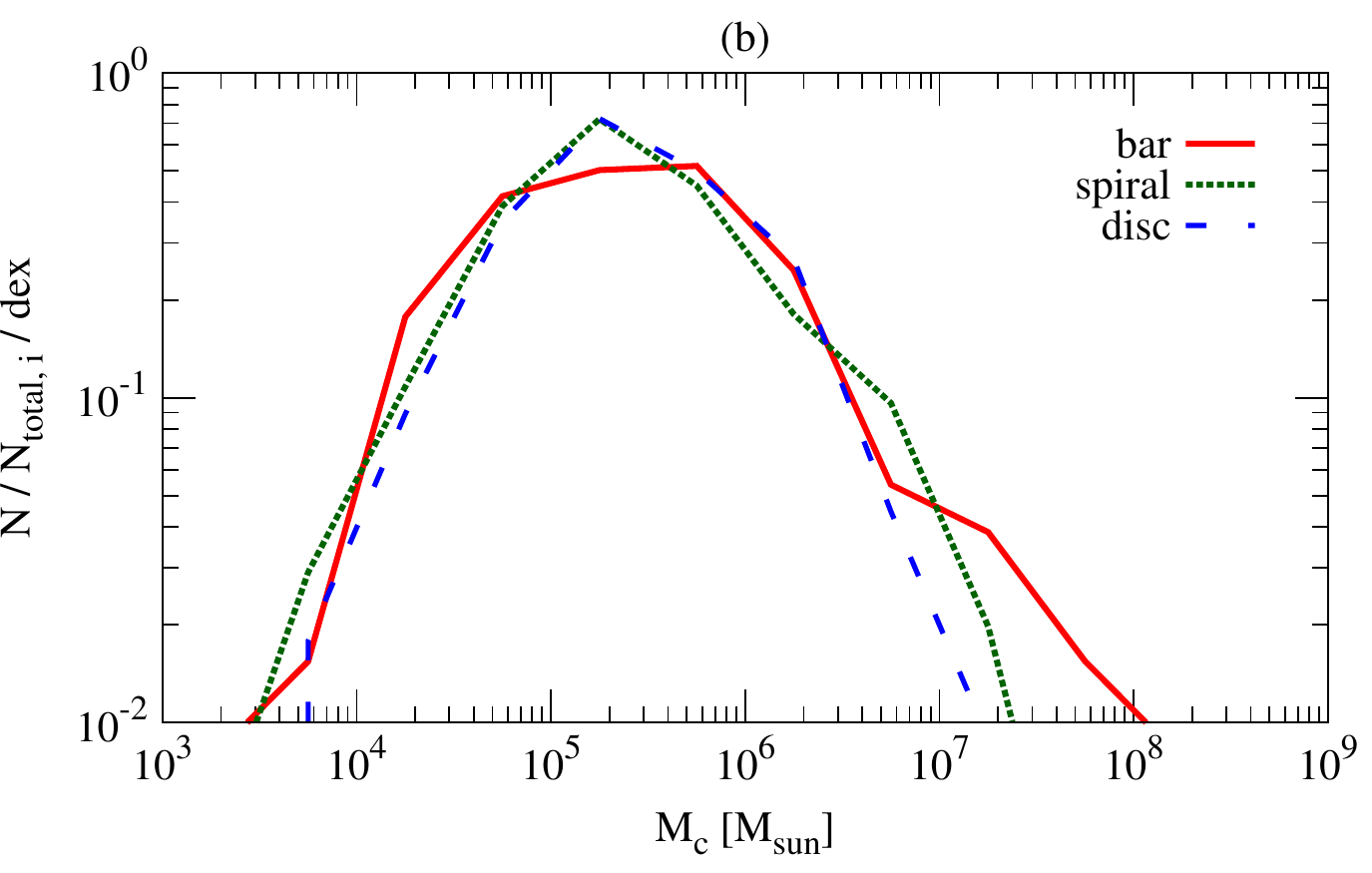}}
	\caption{The cloud mass distribution as calculated by the two different cloud identification schemes mentioned in section~\ref{sec:numerics_cloud}. Top figure shows our original {\it contour method} where clouds are identified as continuous structures within a density contour at $n_{\rm H,c} = 100$\,cm$^{-3}$. The lower figure shows the results from the {\it peak method} where clouds are formed by assigning neighbouring cells to a central peak to build a continuous structure of density above $n_{\rm H,c} > 100$\,cm$^{-3}$.}
	\label{comparison of mass distribution between cloud definitions}
\end{center}
\end{figure}

The exact definition of a GMC is unclear both in theory and observation. Generally, cloud identification schemes use a density threshold to arbitrate where the edge of a cloud should be, but even here there are multiple permeations. Observers cloud find in surface density space or using position-position-velocity data while theorists prefer to use volume density, rather than selecting a viewing angle for their simulation. There is then the question of when an extended body should be considered multiple clouds, with the answer depending both on resolution and the researcher's choice. 

To assess the impact of the choice of cloud definition on our results, we compared cloud properties found using the two identification schemes described in section~\ref{sec:numerics_cloud}. The main difference between the two methods is how peaks within a continuous density structure are treated. Our main {\it contour method} treats these as single cloud, while the {\it peaks method} divides the cloud if the peaks are more than 20\,pc apart (the typical size for a GMC). 

This difference in methodology produces a large variation in the number of clouds. When using the {\it contour method}, we find 77 clouds in the bar region, 515 in the spiral and 102 in the disc at 240\,Myr. The {\it peak method} run on the same output locates 336 clouds in the bar, 1538 in the spiral and 229 in the disc. Unsurprisingly, it is the bar and spiral regions that display the biggest differences in cloud number, with the tidal interactions around the giant {\it Type B} clouds being subdivided by the {\it peak method} into multiple bodies. 

Despite the difference in cloud number, most features in the cloud properties closely coincide. The peak values for the quantities shown in Figure~\ref{distribution of cloud properties} are the same with an overall comparable range of values. The bimodality of the mass-radius scaling relation is also seen with the {\it peak method}, although the number of smaller clouds within a larger body increases the scatter. However, the distinct population of {\it Type B} GMAs is not seen in the mass distribution when using the {\it peak method}. Figure~\ref{comparison of mass distribution between cloud definitions} shows the mass distribution for the {\it contour method} (top) and {\it peak method}. As mentioned above, the peak mass for the clouds is the same in both cases and the range in values is similar, but the bar clouds show no bimodality in the lower distribution. The fact these {\it Type B} GMAs exist in the data is visually seen in Figure~\ref{figure of three clouds}, but their irregular tidal tails produce a multitude of peaks that are broken up into separate smaller clouds by the {\it peak method}.

While neither technique is `right' or `wrong' (since there is no established way to define a GMC), it is harder to discern the environmental differences when using the {\it peak method}, since it tends to produce a more uniformly sized cloud populations in regions of intense interaction. This is also felt to a smaller extent in the comparison of the mass range in the bar and spiral. For the {\it peak method}, the distributions show only a small difference at the high mass end, but the {\it contour method} shows more clearly that the spiral region has a wider spread of cloud masses.  

This comparison suggests that the choice in cloud identification scheme may not be important in determining the broad cloud properties but may make a significant difference when exploring the finer details such as the difference between environmental regions.

\section{Conclusions}

We performed three-dimensional hydrodynamical simulations of a barred spiral galaxy down to a limiting resolution of 1.5\,pc and compared the properties of the GMCs forming in the bar, spiral and disc environments. Our main results are as follows:

\begin{enumerate}

\item The typical (peak) value of the cloud properties such as mass, radius and velocity dispersion, is independent of galactic environment. The values found agree well the GMC observation in the Milky Way and nearby galaxies, having a typical mass of $5 \times 10^5$\,M$_\odot$ and radius 11\,pc (Figure~\ref{distribution of cloud properties}). This is despite having no active star formation or feedback in the simulation.

\item The high-end tail in the mass, radius, surface density, velocity dispersion and virial parameter shows a clear relation to the galactic environment, with the pattern being bar $\rightarrow$ spiral $\rightarrow$ disc for the regions most likely to host clouds with extended structures. 

\item Clouds in the bar region display a bimodality in the mass, radius and velocity dispersion distributions that is not visible in the spiral or disc regions. This is due to the formation of GMAs, which build mass $> 10^7$\,M$_\odot$ through multiple mergers with other clouds. Since the bar is a densely packed region of clouds in close-passing elliptical orbits, the fraction of GMAs is greatest in this environment, producing the bimodal distributions. 

\item All environments show a bimodal surface density distribution. This corresponds to two parallel trends on the mass-radius scaling relation. The lower trend is formed of transient, unbound clouds that are created in the tidal tails and filaments surrounding more massive clouds. (Figures~\ref{distribution of cloud properties}, \ref{Larson's law of three type}, \ref{figure of three clouds}).

\item Based on the distribution results above, clouds can be classified into three types: {\it Type A} are the most common cloud, forming the largest population in all three environments. Their properties agree well with GMCs observed in other galaxies. {\it Type B} are massive GMAs, formed via mergers with smaller clouds and most prominent in high interaction environments. {\it Type C} clouds are unbound, transient objects formed in the dense filaments and tidal tails surrounding other clouds. They usually merge or dissipate within a few Myr, although can live for longer if unperturbed. (Figures~\ref{Larson's law of three type}, \ref{lifetime and merger rate})

\item The main difference between galactic environments is not the properties of a typical cloud born in each region, but the ratio of the above three cloud types. The determining factor in this ratio is the level of interactions between the GMCs. {\it Type B} and {\it C} clouds are formed during cloud collisions, with {\it Type B} being the product of multiple mergers and {\it Type C} forming most frequently in the dense filamentary structures that surround such encounters. The bar region has the highest rate of cloud mergers and also the largest number of {\it Type B} and {\it C} clouds. The spiral region is the next most interactive while the disc is the most quiescent, leading to a cloud population that is predominantly {\it Type A}. (Table~\ref{table:cloud_percentage}, Figure~\ref{lifetime and merger rate}).

\item Lower resolution simulations blur the distinction between the three cloud types, due to {\it Type A} and {\it Type B} clouds having larger radii at lower resolutions. Linked with this, the cloud identification scheme can also affect separating cloud types. The cloud peak properties and range of values are preserved between two different cloud finding algorithms run on our simulation, but the scheme which splits density peaks in close proximity to one another fails to identify the {\it Type B} GMAs. This needs to be considered when comparing populations of clouds in observational and simulation data. 

\end{enumerate}

Although we successfully reproduce many of the properties of observed GMCs, the question of impact from stellar feedback is not addressed in our simulations. On our limited resolution scale of 1.5\,pc, the effect of feedback is especially interesting (with the outcome of feedback affecting our results or not both being of equal importance). This topic will be the subject of future work.

\section*{Acknowledgments}

The authors would like to thank Akihiko Hirota for his assistance with the M83 potential, the yt development team \citep{yt} for many helpful tips during the analysis of these simulations and Alessandro Romeo for his careful reading and suggestions when refereeing this paper. Numerical computations were carried out on the Cray XT4 and Cray XC30 at the Center for Computational Astrophysics (CfCA) of the National Astronomical Observatory of Japan. EJT is funded by the MEXT grant for the Tenure Track System.

\label{lastpage}

\end{document}